\documentclass[12pt,preprint]{aastex}
\usepackage{epsfig}
\usepackage{natbib}
\usepackage{graphicx}
\usepackage{mathrsfs,amssymb}
\usepackage{amsmath}
\usepackage{xcolor}
\usepackage{lineno}
\newcommand	\beq	{\begin{equation}}	
\newcommand	\eeq	{\end{equation}}	
\newcommand       \Angstrom     {\,{\rm \AA}}

\newcommand       \cm           {\,{\rm cm}}

\newcommand       \Gyr      {\,{\rm Gyr}}

\newcommand       \nH           {n_{\rm H}}
\newcommand       \NH           {N_{\rm H}}
\newcommand       \NHI          {N({\rm HI})}
\newcommand       \NHH         {N({\rm H_2})}
\newcommand       \simlt        {\lesssim}
\newcommand       \simgt        {\gtrsim}
\newcommand       \gtsim        {\gtrsim}

\newcommand       \um           {\mu{\rm m}}
\newcommand       \mum          {\,{\rm \mu m}}
\newcommand       \ppm          {\,{\rm ppm}}

\newcommand       \mH           {m_{\rm H}}

\newcommand       \simali       {\sim\,}
\newcommand       \magni        {\,{\rm mag}}
\newcommand       \rmH          {{\rm H}}
\newcommand       \HH           {\,{\rm H}}

\newcommand       \rhosil        {\rho_{\rm sil}}
\newcommand       \rhogra        {\rho_{\rm gra}}

\newcommand	  \xism         {\left[{\rm X/H}\right]_{\rm ISM}}

\newcommand	  \cism         {\left[{\rm C/H}\right]_{\rm ISM}}

\newcommand	  \oism         {\left[{\rm O/H}\right]_{\rm ISM}}
\newcommand	  \feism        {\left[{\rm Fe/H}\right]_{\rm ISM}}
\newcommand	  \mgism        {\left[{\rm Mg/H}\right]_{\rm ISM}}
\newcommand	  \siism        {\left[{\rm Si/H}\right]_{\rm ISM}}
\newcommand	  \xdust        {\left[{\rm X/H}\right]_{\rm dust}}
\newcommand	  \cdust        {\left[{\rm C/H}\right]_{\rm dust}}

\newcommand	  \odust        {\left[{\rm O/H}\right]_{\rm dust}}
\newcommand	  \fedust       {\left[{\rm Fe/H}\right]_{\rm dust}}
\newcommand	  \mgdust       {\left[{\rm Mg/H}\right]_{\rm dust}}
\newcommand	  \sidust       {\left[{\rm Si/H}\right]_{\rm dust}}
\newcommand	  \xgas         {\left[{\rm X/H}\right]_{\rm gas}}
\newcommand	  \cgas        {\left[{\rm C/H}\right]_{\rm gas}}

\newcommand	  \ogas        {\left[{\rm O/H}\right]_{\rm gas}}
\newcommand	  \fegas       {\left[{\rm Fe/H}\right]_{\rm gas}}
\newcommand	  \mggas       {\left[{\rm Mg/H}\right]_{\rm gas}}
\newcommand	  \sigas       {\left[{\rm Si/H}\right]_{\rm gas}}
\newcommand	  \xtot         {\left[{\rm X/H}\right]_{\rm tot}}
\newcommand	  \ctot         {\left[{\rm C/H}\right]_{\rm tot}}
\newcommand	  \otot        {\left[{\rm O/H}\right]_{\rm tot}}
\newcommand	  \mgtot         {\left[{\rm Mg/H}\right]_{\rm tot}}
\newcommand	  \sitot         {\left[{\rm Si/H}\right]_{\rm tot}}
\newcommand	  \fetot        {\left[{\rm Fe/H}\right]_{\rm tot}}
\newcommand       \Bsil          {B_{\rm sil}}
\newcommand       \Bgra          {B_{\rm gra}}

\def    \amin		{a_{\rm min}}
\def    \amax		{a_{\rm max}}

\def    \acC		{a_{c,{\rm C}}}
\def    \acS		{a_{c,{\rm S}}}
\def    \Cext		{C_{\rm ext}}
\def    \alphaC		{\alpha_{\rm C}}
\def    \alphaS		{\alpha_{\rm S}}

\def    \rhosil		{\rho_{\rm sil}}
\def    \rhogra		{\rho_{\rm gra}}

\def    \Bsil		{B_{\rm S}}
\def    \Bgra		{B_{\rm C}}

\def    \xo             {x_{\rm o}}
\def      \AV          {{A_V}}
\def      \RV          {{R_V}}

\def      \lambdaV {\lambda_{V}}

\def    \xo             {x_{\rm o}}

\def \ameansil {{\langle a \rangle}_{\rm sil}}
\def \ameangra {{\langle a \rangle}_{\rm gra}}

\DeclareMathAlphabet{\mathsc}{OT1}{cmr}{m}{sc}
\def\testbx{bx}%
\DeclareRobustCommand{\ion}[2]{%
\relax\ifmmode
\ifx\testbx\f@series
{\mathbf{#1\,\mathsc{#2}}}\else
{\mathrm{#1\,\mathsc{#2}}}\fi
\else\textup{#1\,{\mdseries\textsc{#2}}}%
\fi}
\countdef\decade=200
\advance\decade by \year
\countdef\hours=201
\advance\hours by \time
\divide\hours by 60
\countdef\mins=202
\advance\mins by \hours
\multiply\mins by 60
\multiply\hours by 100
\countdef\miltime=203
\advance\miltime by \hours
\advance\miltime by \time
\advance\miltime by -\mins
\def\today{\number\decade.\number\month.\number\day.\number\miltime}
\shorttitle{Extinction and Abundances of Individual Sight Lines}
\title{Interstellar Extinction and Elemental Abundances:
                 Individual Sight Lines       
\\{\small DRAFT: \today ~~}}

\author{Wenbo Zuo\altaffilmark{1,2,3}, 
        Aigen Li\altaffilmark{3} 
        and Gang Zhao\altaffilmark{1,2,3}}
\altaffiltext{1}{CAS Key Laboratory of Optical Astronomy, 
       National Astronomical Observatories, 
       Chinese Academy of Sciences, 
       Beijing 100101, China;
       {\sf gzhao@nao.cas.cn}}
\altaffiltext{2}{School of Astronomy and Space Science, 
       University of Chinese Academy of Sciences, 
       Beijing 100049, China}
\altaffiltext{3}{Department of Physics and Astronomy,
                  University of Missouri,
                  Columbia, MO 65211, USA;
                  {\sf lia@missouri.edu}}

\begin{document}
\begin{abstract}
While it is well recognized that both the Galactic
interstellar extinction curves and the gas-phase
abundances of dust-forming elements exhibit
considerable variations from one sightline to another,
as yet most of the dust extinction modeling efforts
have been directed to the Galactic average extinction curve,
which is obtained by averaging over many clouds of different
gas and dust properties. Therefore, any details concerning
the relationship between the dust properties
and the interstellar environments are lost.
Here we utilize the wealth of extinction
and elemental abundance data obtained by space
telescopes and explore the dust properties
of a large number of individual sightlines.
We model the observed extinction curve
of each sightline and derive the abundances
of the major dust-forming elements
(i.e., C, O, Si, Mg and Fe) required to
be tied up in dust (i.e., dust depletion).
We then confront the derived dust depletions
with the observed gas-phase abundances of
these elements and investigate the environmental
effects on the dust properties and elemental depletions.
It is found that for the majority of the sightlines
the interstellar oxygen atoms are fully accommodated
by gas and dust and therefore there does not
appear to be a ``missing oxygen'' problem.
For those sightlines with an extinction-to-hydrogen
column density 
$\AV/\NH\simgt4.8\times10^{-22}\magni\cm^2\HH^{-1}$
there are shortages of C, Si, Mg and Fe elements
for making dust to account for the observed extinction,
even if the interstellar C/H, Si/H, Mg/H and Fe/H
abundances are assumed to be protosolar abundances
augmented by Galactic chemical evolution.
\end{abstract}


\section{Introduction}\label{sec:intro}
The interstellar extinction and abundances of metal
elements provide important clues about the size
distribution and chemical makeup of interstellar dust.
While it has been well recognized that the interstellar
extinction curves exhibit considerable variations
from one sight line to another
(e.g., see Witt et al.\ 1984,
Siebenmorgen et al.\ 2018),
most of the extinction modeling efforts
have been so far directed to
the {\it mean} extinction curve of the Galaxy,
obtained by averaging over many clouds of
different gas and dust properties
(e.g., see Mathis et al.\ 1977, 
Draine \& Lee 1984,
D\'esert et al.\ 1990,
Siebenmorgen \& Kr\"ugel 1992,
Mathis 1996,
Li \& Greenberg 1997,
Li \& Draine 2001a,
Weingartner \& Draine 2001,
Zubko et al.\ 2004,
Jones et al.\ 2013).
Cardelli et al. (1989) found that the Galactic average
extinction curve can be approximated by an analytical
formula (known as the CCM formula or parameterization)
characterized by $\RV\approx3.1$,
where $\RV\,\equiv\,\AV/E(B-V)$ is
the optical total-to-selective extinction ratio
and $E(B-V)\equiv A_B-A_V$ is the reddening 
or color excess between the visual extinction
$A_V$ and the $B$-band extinction $A_B$.

Depending on the local physical conditions,
the CCM parameterization of the extinction curves
of various individual sight lines involves a wide
range of $\RV$ values which deviate substantially
from the Galactic average value of $\RV\approx3.1$
(e.g., see Fitzpatrick et al.\ 2019).
More specifically, low-density regions usually
have a smaller $\RV$ for which the extinction curve
is characterized by a strong 2175$\Angstrom$ bump
and a steep far ultraviolet (UV) rise
at $\lambda^{-1}$\,$>$\,6$\mum^{-1}$.
In contrast, sight lines penetrating into dense clouds,
such as the Ophiuchus or Taurus molecular clouds,
usually have $4 < \RV < 6$ and their extinction curves
exhibit a weak 2175$\Angstrom$ bump
and a relatively flat far-UV rise.
Also, the extinction curves toward some sight lines
are ``anomalous'', i.e., they deviate considerably from
that expected from the $\RV$-based CCM parameterization
(e.g., see Cardelli \& Clayton 1991, Mazzei \& Barbaro 2011).
Apparently, the dust size and composition properties of
those ``anomalous'' sight lines or those with $\RV$ values
appreciably smaller or larger the canonical $\RV=3.1$
are naturally expected to differ from that deduced from
the Galactic average extinction curve.
Therefore, by modeling the Galactic average extinction curve,
unavoidably, any details concerning the relationship between
the dust properties and the physical and chemical conditions
of the interstellar environments would have been lost.





Interstellar dust is made of metal elements produced
in stars, especially carbon (C), oxygen (O), silicon (Si),
magnesium (Mg) and iron (Fe).
Space-borne UV spectroscopic observations of
interstellar clouds have provided important data on
the abundances of components of interstellar gas and
have revealed that the gas-phase abundances of heavy
elements are significantly lower, relative to hydrogen (H),
than in the solar photosphere. As those elements ``missing''
from the gas phase must have condensed into dust grains,
the striking abundance deficiencies of heavy elements
such as Si, Mg, Fe, and to a lesser degree, C and O---known
as ``interstellar depletion''---provide useful information
on the possible composition and mass (relative to gas)
of interstellar dust (e.g., see Kimura et al.\ 2003a,b,
Voshchinnikov \& Henning 2010).
Apparently, any viable interstellar dust model
should not contradict the interstellar depletion. 

To quantitatively assess the constraints
on the composition and quantity of interstellar dust placed 
by interstellar depeletion, one has to assume a nominal
reference abundance standard which describes the total
abundance of an element in the combined gas-plus-dust phases.
The dust-phase abundance of an element is then
determined by subtracting off the gas-phase abundance
from the assumed reference abundance.
Apparently, the abundance constraints on interstellar dust
sensitively rely on the knowledge of the gas-phase abundances
of dust-forming elements and the assumption of the reference
abundance (also known as ``interstellar abundance'',  
or ``cosmic abundance'').

However, what might be the most appropriate set of
interstellar reference abundances remains unclear.
The interstellar abundances are often assumed to
be solar (e.g., see Whittet 1984),
subsolar (like that of B stars and young F and G stars;
see Snow \& Witt 1995, 1996, Sofia \& Meyer 2001), and
protosolar (see Lodders 2003).\footnote{%
  The protosolar abundance of an element (except H)
  is the present-day solar photospheric abundance
  of that element {\it increased} by correcting for
  the settling effects. The currently observed solar
  photospheric abundances (relative to H) 
  must be lower than those of the proto-Sun 
  because helium and other heavy elements
  have settled toward the solar interior 
  since the time of its formation 
  $\simali$4.55$\Gyr$ ago (Lodders 2003).
  }
More recently, it has been argued that the interstellar
abundances are better represented by the protosolar
abundances augmented by Galactic chemical enrichment
(GCE; see Zuo et al.\ 2021, Hensley \& Draine 2021).
Nevertheless, a very recent observational study
carried out by De Cia et al.\ (2021) implied that 
the interstellar abundances of refractory elements
in the local Galactic interstellar medium (ISM)
may only be about 55\% solar on average, though
with a high degree of variation. This would place
stringent constraints on dust models.

Also, the overall Galactic average interstellar gas abundances
for the dust-forming elements remain uncertain.
Most dust extinction modeling efforts assume that
the gas-phase abundances of C and O are invariable
with respect to the local interstellar conditions
(e.g., see Cardelli et al.\ 1996, Meyer et al.\ 1998)
and that Si, Mg and Fe are fully depleted from the gas.
However, numerous observational studies carried out
in the past decade have shown that the gas-phase C/H
and O/H abundances vary with the local conditions
(see Zuo et al.\ 2021 and references therein)
and that in many sightlines there are nontrivial 
amounts of gas-phase Si, Mg and Fe 
(e.g., see Jensen et al.\ 2010).
Therefore, in modeling the Galactic average extinction curve,
the assumption of constant gas-phase abundances
for the dust-forming elements would also unavoidably
cause the loss of all the details concerning the abundance
constraints on the dust properties and particularly their
relation to the physical and chemical conditions
of the interstellar environments.

In view of the shortcomings of modeling the Galactic
{\it average} extinction curve and assuming a set of
{\it constant}, {\it environmentally-invariable}
gas-phase abundances for the dust-forming elements,
in this work we are motivated to model the extinction
curves of individual sight lines
from the far-UV to the near infrared (IR),
with the aid of abundance constraints.
To this end, we confine ourselves to those sight lines
for which the gas-phase abundances of 
one or more of the dust-forming elements
have been observationally determined. 
This stands in contrast to the common practice of 
modeling the Galactic {\it average} extinction curve
under the assumption of {\it constant}
gas-phase abundances.
This paper is organized as follows.
We first compile in \S\ref{sec:sample}
a complete sample of all the (81) Galactic sightlines
for which the extinction curves from the far-UV
to the near-IR as well as the gas-phase abundances of 
at least one of the dust-forming elements
(i.e., C, O, Mg, Si, and Fe) have been observationally
measured. We then model in \S\ref{sec:extmod}
the extnction curves of 45 individual sight lines
in terms of the standard silicate-graphite interstellar
grain model since previously 36 sight lines have 
already been modeled by one of us
(Mishra \& Li 2015, 2017).
The results are discussed in \S\ref{sec:discussion}
and summarized in \S\ref{sec:summary}.
%



\section{Extinction Curves of Individual Sight Lines
  \label{sec:sample}}
We first search for in the literature 
an as complete as possible set of individual
interstellar sight lines for which both the extinction curves
have been observationally determined
from the near-IR to the far-UV and 
the gas-phase abundances have been measured 
for at least one of the dust-forming elements 
(i.e., C, O, Mg, Si, and Fe).
As a result, we find 81 such sight lines and tabulate
in Table~\ref{tab:extpara} the extinction parameters
$c_{1}^{\prime}$, $c_{2}^{\prime}$,
$c_{3}^{\prime}$, $c_{4}^{\prime}$,
$\xo$ and $\gamma$ 
as well as $E(B-V)$, $\RV$ and
$A_U$, $A_B$, $\AV$, $A_J$, $A_H$, $A_K$,
the extinction at the U, B, V, J, H, K bands, respectively.
We tabulate in Table~\ref{tab:GasAbund}
the gas-phase abundances of C, O, Mg, Si,
and Fe as well as the column densities
of atomic hydrogen $\NHI$, 
molecular hydrogen $\NHH$, 
the total hydrogen column densities 
$\NH=\NHI+2\NHH$, and the fraction
of H in molecular form $f({\rm H_2})$.

For each sight line, we construct the extinction 
curve as follows.\footnote{%
  We have previously already taken the same
  approach to construct the extinction curves
  for a ``gold'' sample of 10 sight lines
  for which the gas-phase abundances of
  {\it all} the five major dust-forming elements
  C, O, Mg, Si and Fe have been observationally
  measured (see Zuo et al.\ 2021).
  Therefore, we are left with 71 sight lines
  for extinction-curve construction.
  }
First, we make use of the extinction 
parameters $c_{1}^{\prime}$, $c_{2}^{\prime}$, $c_{3}^{\prime}$, 
$c_{4}^{\prime}$, $\xo$, $\gamma$ and $\AV$
tabulated in Table~\ref{tab:extpara}
to represent the UV extinction measured by
the {\it International Ultraviolet Explorer} (IUE)
at $3.3 < \lambda^{-1} < 8.7\mum^{-1}$
as a sum of three components: 
a linear background,
a Drude profile for the 2175$\Angstrom$ extinction bump,
and a far-UV nonlinear rise 
at $\lambda^{-1} > 5.9\mum^{-1}$:
\begin{equation}\label{eq:A2AV1}
  A_\lambda = \AV\times \left\{c_1^{\prime}
              + c_2^{\prime}\,x 
              + c_3^{\prime}\,D(x,\gamma,\xo) 
              + c_4^{\prime}\,F(x) \right\}~~~,
\end{equation}
\begin{equation}\label{eq:A2AV2}
D(x,\gamma,\xo) = \frac{x^2}
  {\left(x^2-\xo^2\right)^2 + x^2\gamma^2} ~~~,
\end{equation}
\begin{equation}\label{eq:A2AV3}
F(x) = \left\{\begin{array}{lr} 
0 ~, & x < 5.9\mum^{-1} ~~~,\\

0.5392\,\left(x-5.9\right)^2 
     + 0.05644\,\left(x-5.9\right)^3 ~, 
 & x \ge 5.9\mum^{-1} ~~~,\\
\end{array}\right.
\end{equation}
where $A_\lambda$ is the extinction 
at wavelength $\lambda$,
$x\equiv 1/\lambda$ 
is the inverse wavelength in $\mu$m$^{-1}$, 
$c_1^{\prime}$ and $c_2^{\prime}$ define 
the linear background, 
$c_3^{\prime}$ defines the strength of 
the 2175$\Angstrom$ extinction bump
which is approximated by $D(x,\gamma,\xo)$,
a Drude function which peaks at 
$\xo\approx4.6\mum^{-1}$ and has 
a full-width-half-maximum (FWHM)
of $\gamma$, and $c_4^{\prime}$ defines 
the nonlinear far-UV rise.
This parameterization was originally introduced
by Fitzpatrick \& Massa (1990; hereafter FM90)
for the interstellar reddening
\begin{equation}\label{eq:E2A}
E(\lambda-V)/E(B-V) = R_V \left(A_\lambda/A_V - 1\right)
              = c_1 + c_2\,x 
              + c_3\,D(x,\gamma,\xo) 
              + c_4\,F(x) ~~~,
\end{equation}
where $E(\lambda-V) \equiv A_\lambda - A_V$.
The extinction parameters $c_j^{\prime}$
listed here in Table~\ref{tab:extpara}
(taken from Valencic et al.\ 2004)
relate to the FM90 parameters through
\begin{equation}
c_j^{\prime} = \left\{\begin{array}{lr} 
c_j/R_V + 1 ~, & j=1 ~~~,\\
c_j/R_V ~, & j=2, 3, 4 ~~~.\\
\end{array}\right.
\end{equation}
In the following, we will refer to the parameterization 
described by Equations (\ref{eq:A2AV1}--\ref{eq:A2AV3})
as the FM parameterization.

For $1.1 < \lambda^{-1} < 3.3\mum^{-1}$,
we compute the extinction $A_\lambda$
from the CCM parameterization which involves
only one parameter (i.e., $\RV$).
As illustrated in Figure~1a of Zuo et al.\ (2021),
there is often a discontinuity 
between the FM parameterization 
at $\lambda^{-1} > 3.3\mum^{-1}$
and the CCM parameterization 
at $\lambda^{-1} < 3.3\mum^{-1}$.
To comply with the observed extinction-to-gas ratio
$\AV/\NH$, we multiply the FM extinction curve
by a factor to smoothly join the CCM curve
(see Figure~1b of Zuo et al.\ 2021).

For $0.9\mum < \lambda < 1\cm$,
we approximate the extinction 
either by the model extinction calculated from
the standard silicate-graphite-PAH model
of Weingartner \& Draine (2001; WD01) for $\RV=3.1$,
or by the model extinction calculated by
Wang, Li, \& Jiang (2015a; WLJ15).
The WLJ15 model is essentially the same
as WD01 but includes an extra population
of very large, micron-sized graphitic grains
which was invoked to account for the observed
flat mid-IR extinction at $3<\lambda< 8\mum$.
Therefore, for each of the 71 sight lines
we construct two extinction curves
(which we refer to as ``WD01'' and ``WLJ15'').

\begin{figure*}
\centering	
\vspace{-10mm}
\includegraphics[width=0.9\textwidth,height=0.8\textheight]{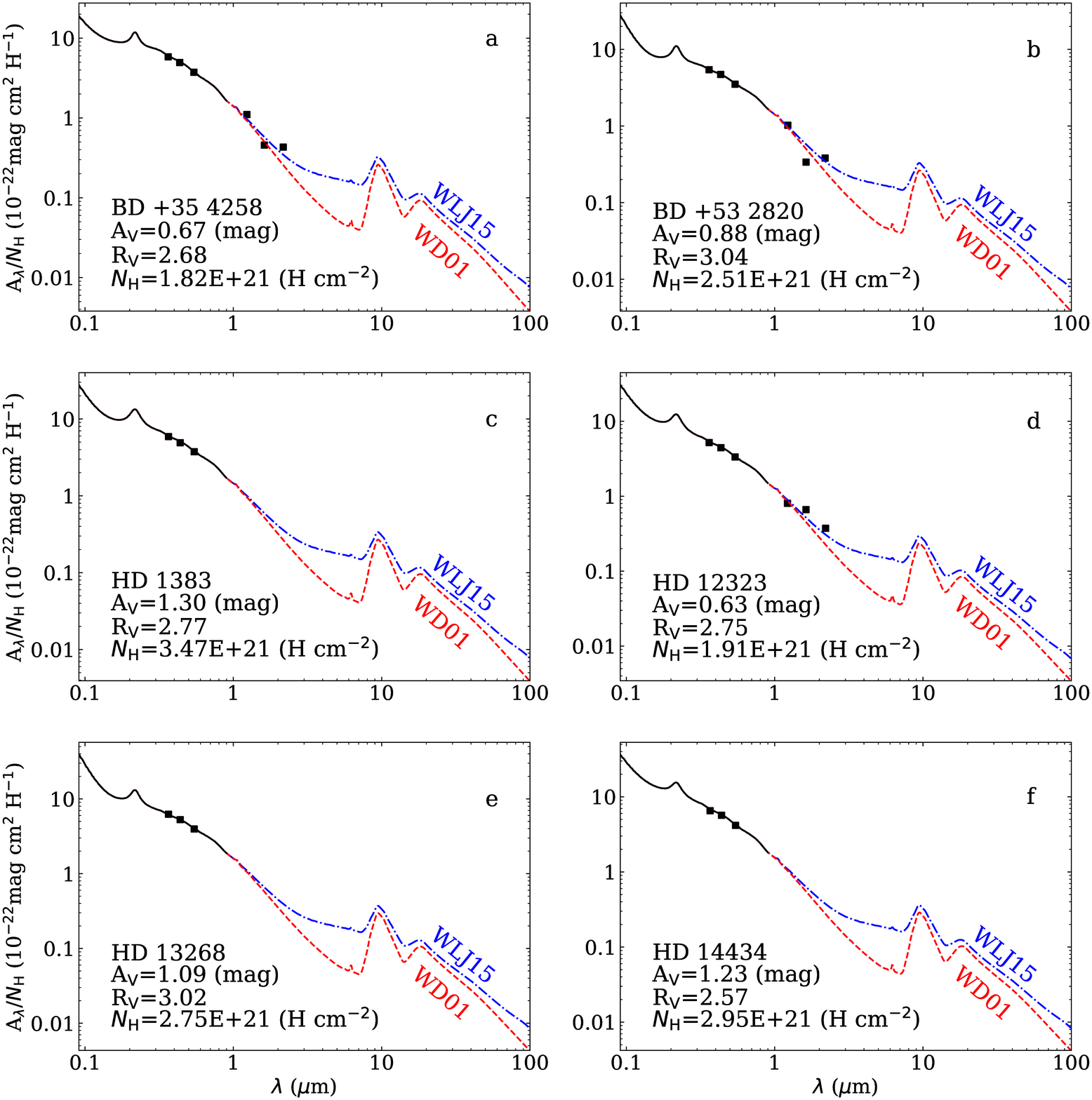}
\vspace{-5mm}
\caption{\footnotesize
  \label{fig:curve1}
         Interstellar extinction curves 
         from the far-UV to the far-IR
         for the lines of sight toward
         BD+35\,4258 (a), BD+53\,2820 (b),
         HD\,1383 (c), HD\,12323 (d),
         HD\,13268 (e) and HD\,14434 (f), 
         with the FM curve for $\lambda^{-1}>3.3\mum^{-1}$,
         the CCM curve for  $1.1\mum^{-1} < \lambda^{-1} < 3.3\mum^{-1}$,
         and the $\RV=3.1$ model curves of 
         Weingartner \& Draine (2001; red dashed line) 
         and Wang, Li \& Jiang (2015a; blue dot-dashed line) 
         for $0.9\mum < \lambda < 1\cm$.
         Whenever available, U, B, V, J, H, and K
         broadband photometric extinction data
         (see Table~\ref{tab:extpara})
         are shown as black squares.
         }
\end{figure*}

\begin{figure*}
\centering	
\vspace{-5mm}
\includegraphics[width=0.9\textwidth,height=0.8\textheight]{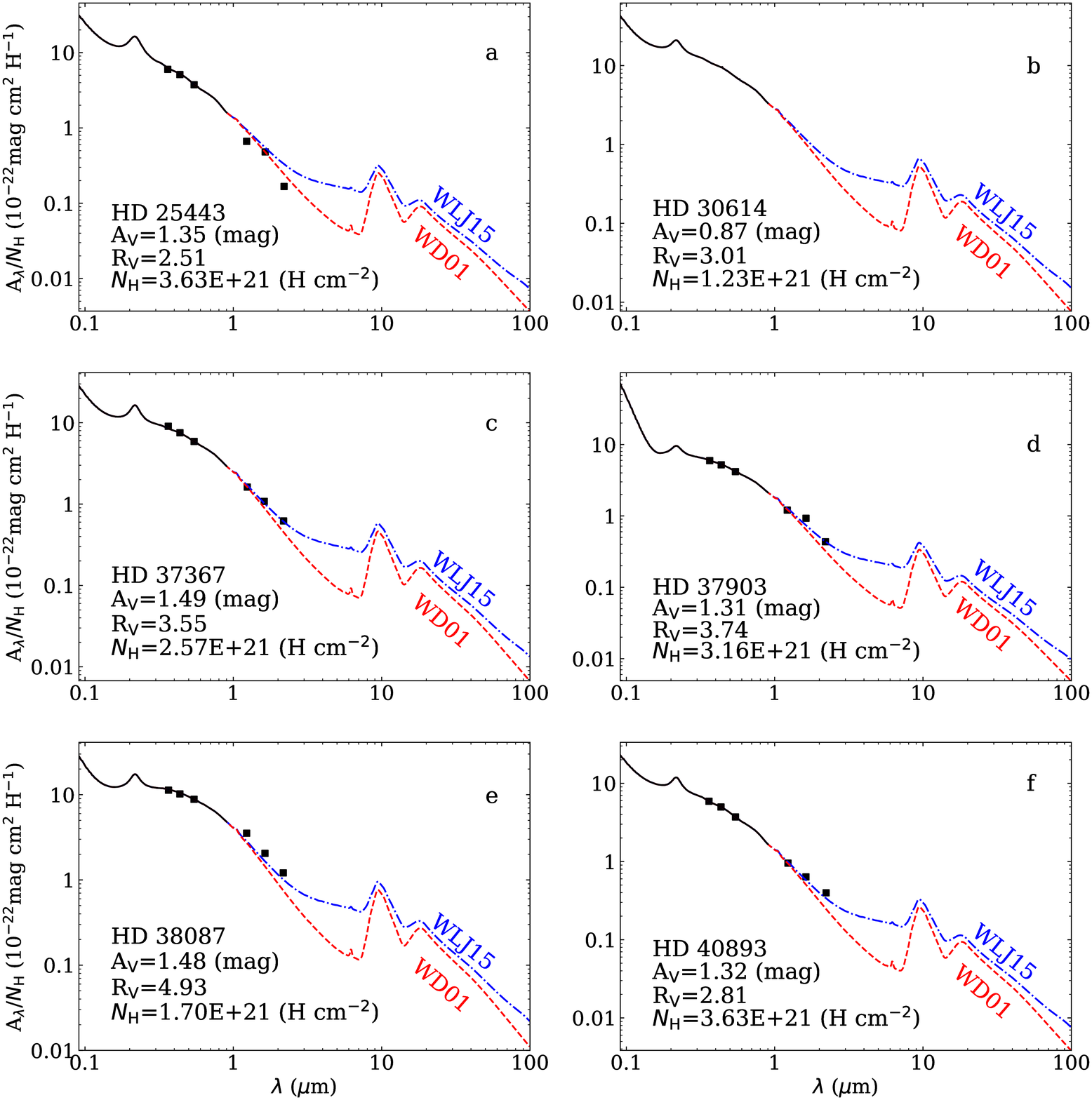}
\vspace{-2mm}
\caption{\footnotesize
        \label{fig:curve2}
        Same as Figure~\ref{fig:curve1} but for HD\,25443 (a),
        HD\,30614 (b), HD\,37367 (c), HD\,37903 (d),
        HD\,38087 (e), and HD 40893 (f).
         }
\end{figure*}

\begin{figure*}
	\centering	
	\vspace{-5mm}
	\includegraphics[width=0.9\textwidth,height=0.8\textheight]{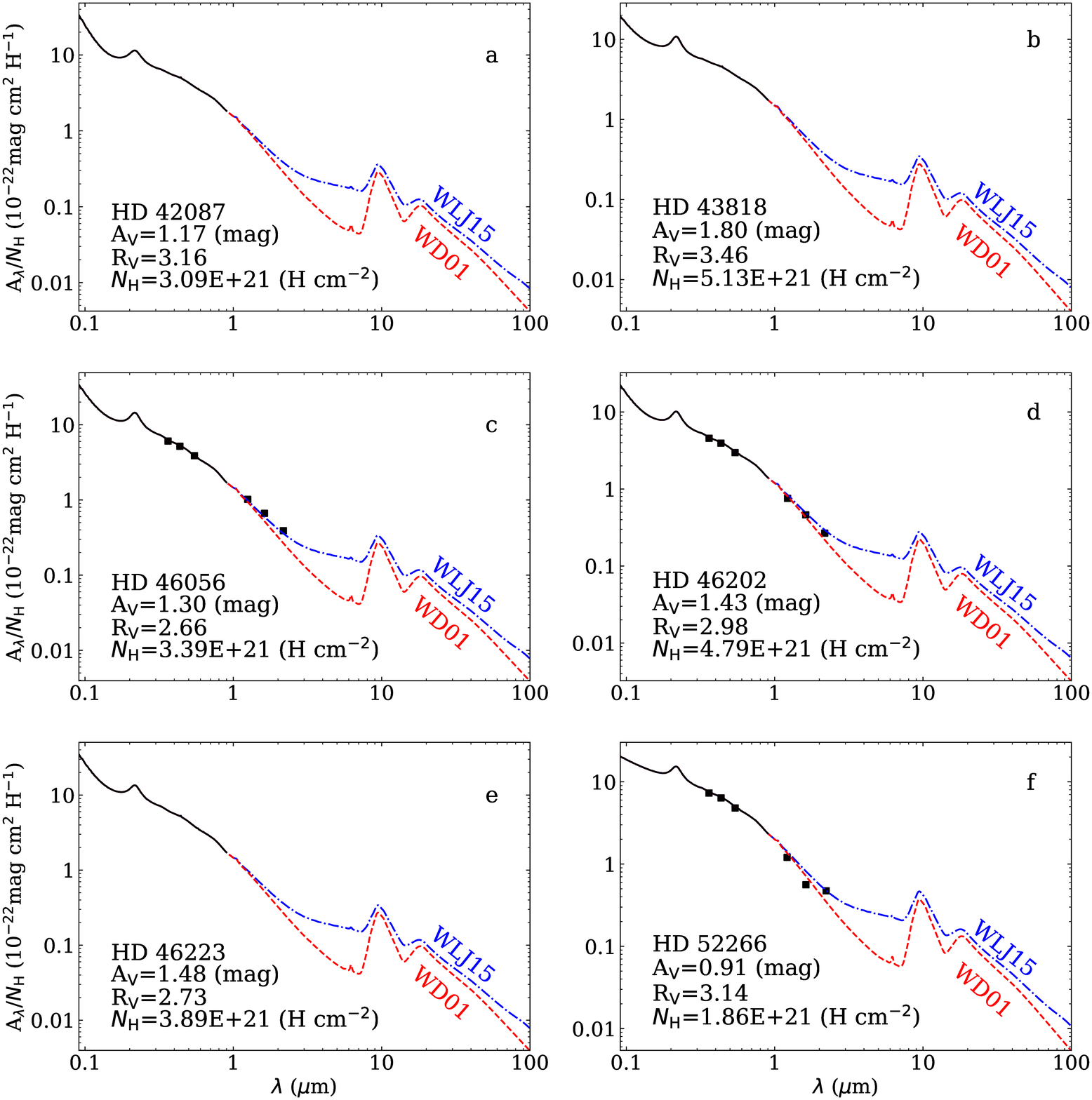}
	\vspace{-2mm}
	\caption{\footnotesize
        \label{fig:curve3}
	Same as Figure~\ref{fig:curve1}
        but for  HD\,42087 (a), HD\,43818 (b),
        HD\,46056 (c), HD\,46202 (d),
        HD\,46223 (e), and HD\,52266 (f).
	}
\end{figure*}

\begin{figure*}
	\centering	
	\includegraphics[width=0.9\textwidth,height=0.8\textheight]{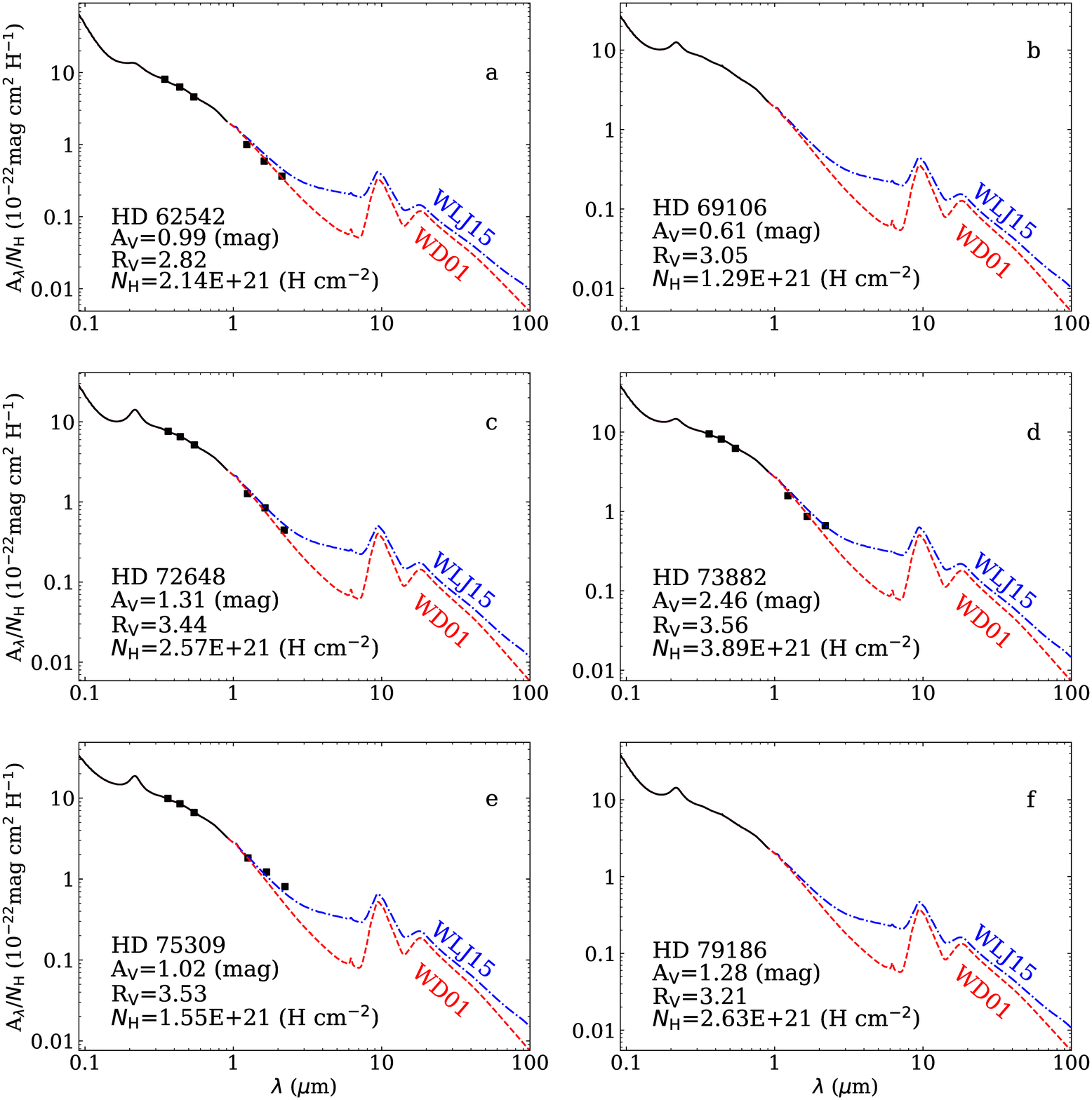}
	\caption{\footnotesize
	\label{fig:curve4}
	Same as Figure~\ref{fig:curve1}
        but for  HD\,62542 (a), HD\,69106 (b),
        HD\,72648 (c), HD\,73882 (d),
        HD\,75309 (e), and HD\,79186 (f).
	}
\end{figure*}

\begin{figure*}
	\centering	
	\includegraphics[width=0.9\textwidth,height=0.8\textheight]{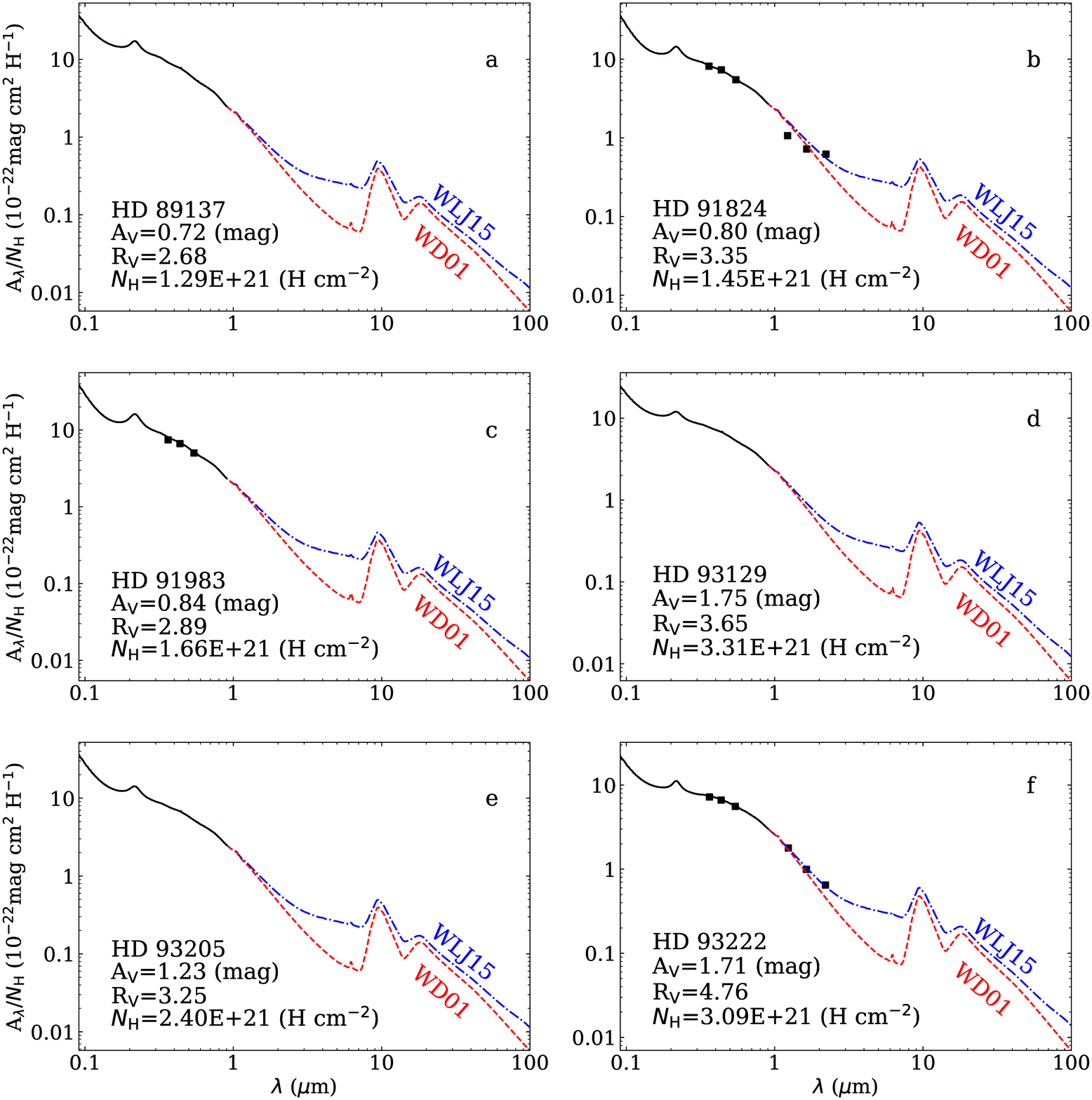}
	\caption{\footnotesize
	\label{fig:curve5}
	Same as Figure~\ref{fig:curve1}
        but for  HD\,89137 (a), HD\,91824 (b),
        HD\,91983 (c), HD\,93129 (d),
        HD\,93205 (e), and HD\,93222 (f).
	}
\end{figure*}

\begin{figure*}
	\centering	
	\includegraphics[width=0.9\textwidth,height=0.8\textheight]{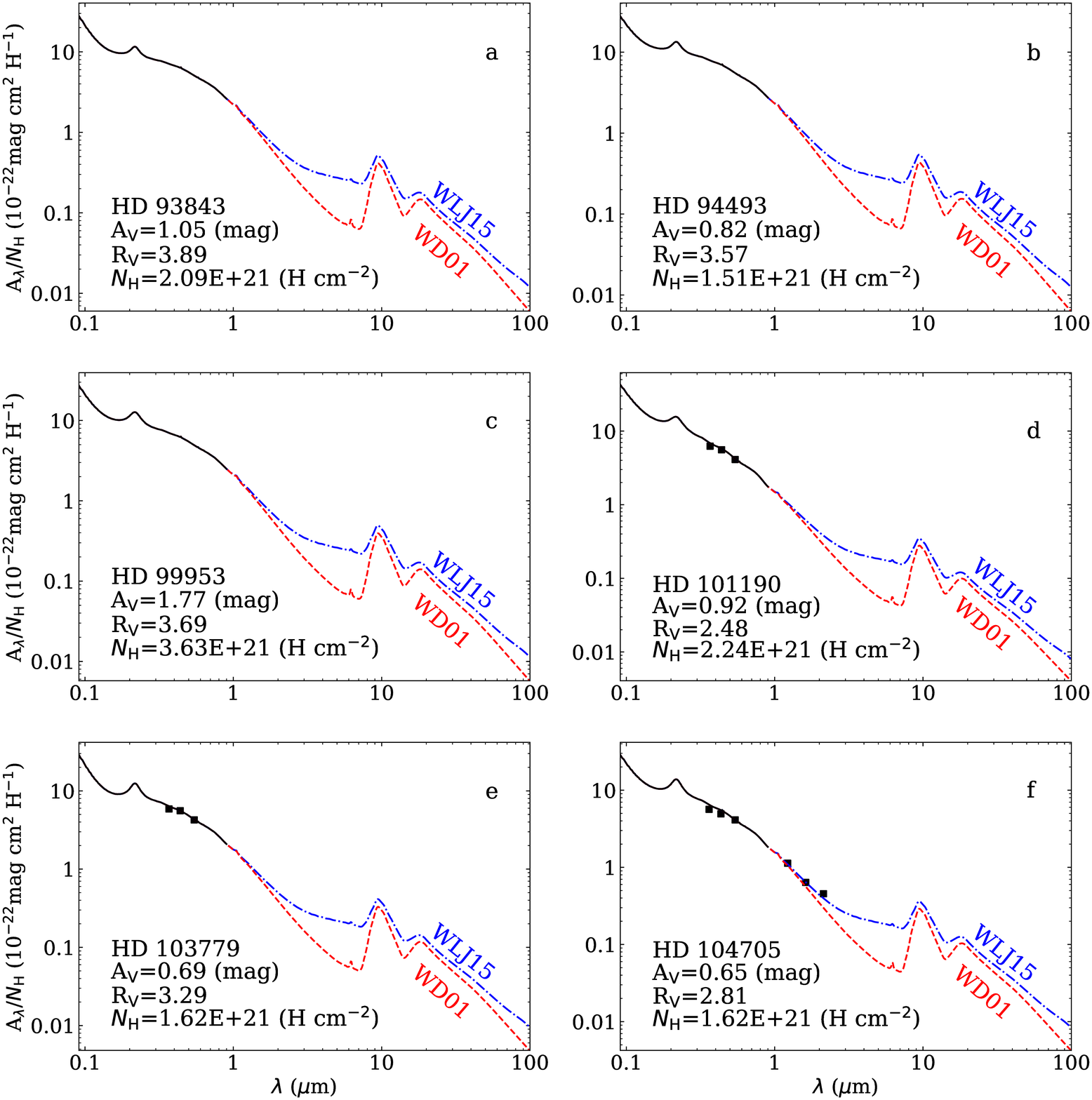}
	\caption{\footnotesize
	\label{fig:curve6}
	Same as Figure~\ref{fig:curve1}
        but for HD\,93843 (a), HD\,94493 (b),
        HD\,99953 (c), HD\,101190 (d),
        HD\,103779 (e), and HD\,104705 (f).
	}
\end{figure*}

\begin{figure*}
	\centering	
	\includegraphics[width=0.9\textwidth,height=0.8\textheight]{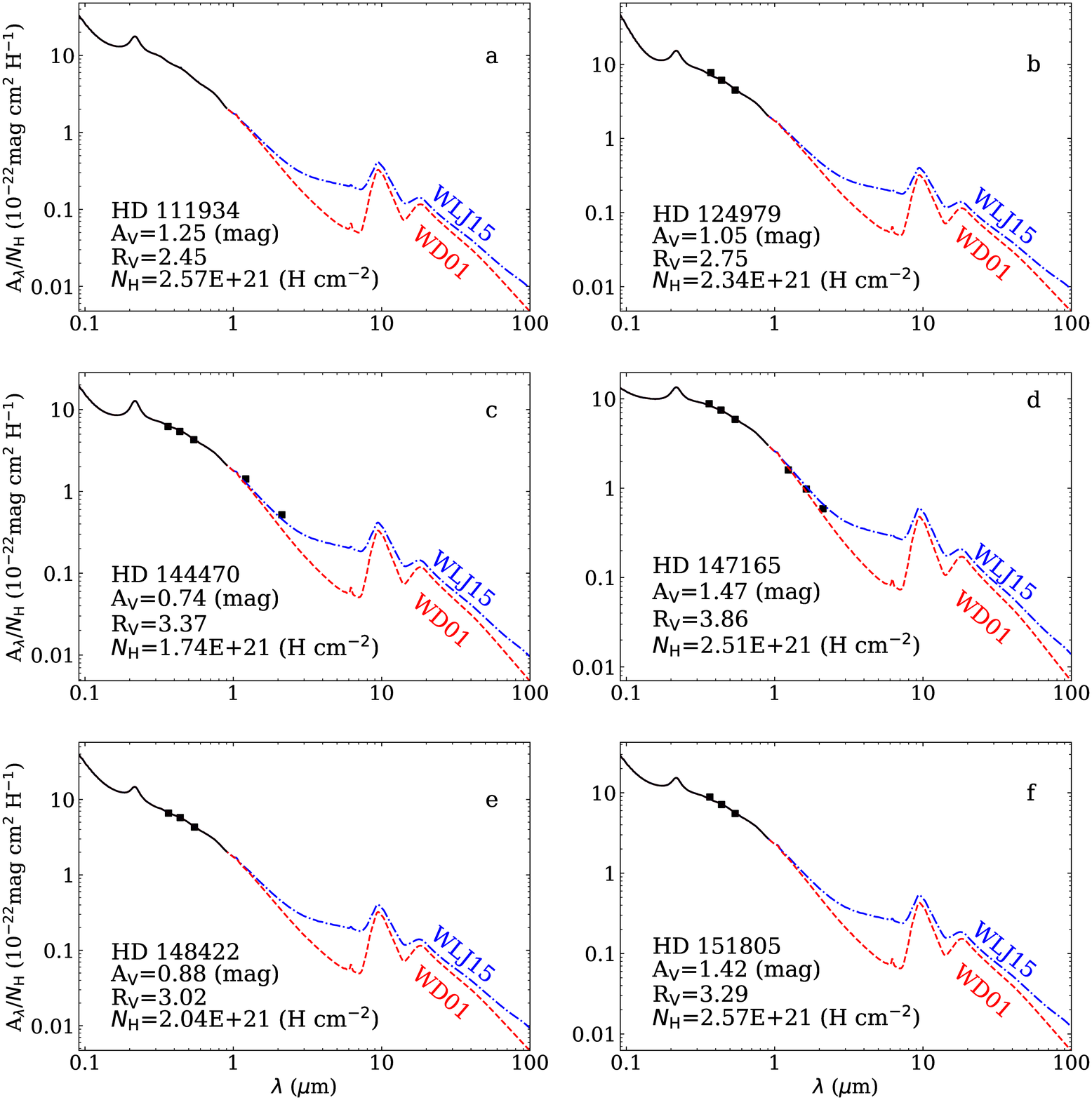}
	\caption{\footnotesize
	\label{fig:curve7}
	Same as Figure~\ref{fig:curve1}
        but for HD\,111934 (a), HD\,124979 (b),
        HD\,144470 (c), HD\,147165 (d),
        HD\,148422 (e), and HD\,151805 (f).
	}
\end{figure*}

\begin{figure*}
	\centering	
	\includegraphics[width=0.9\textwidth,height=0.8\textheight]{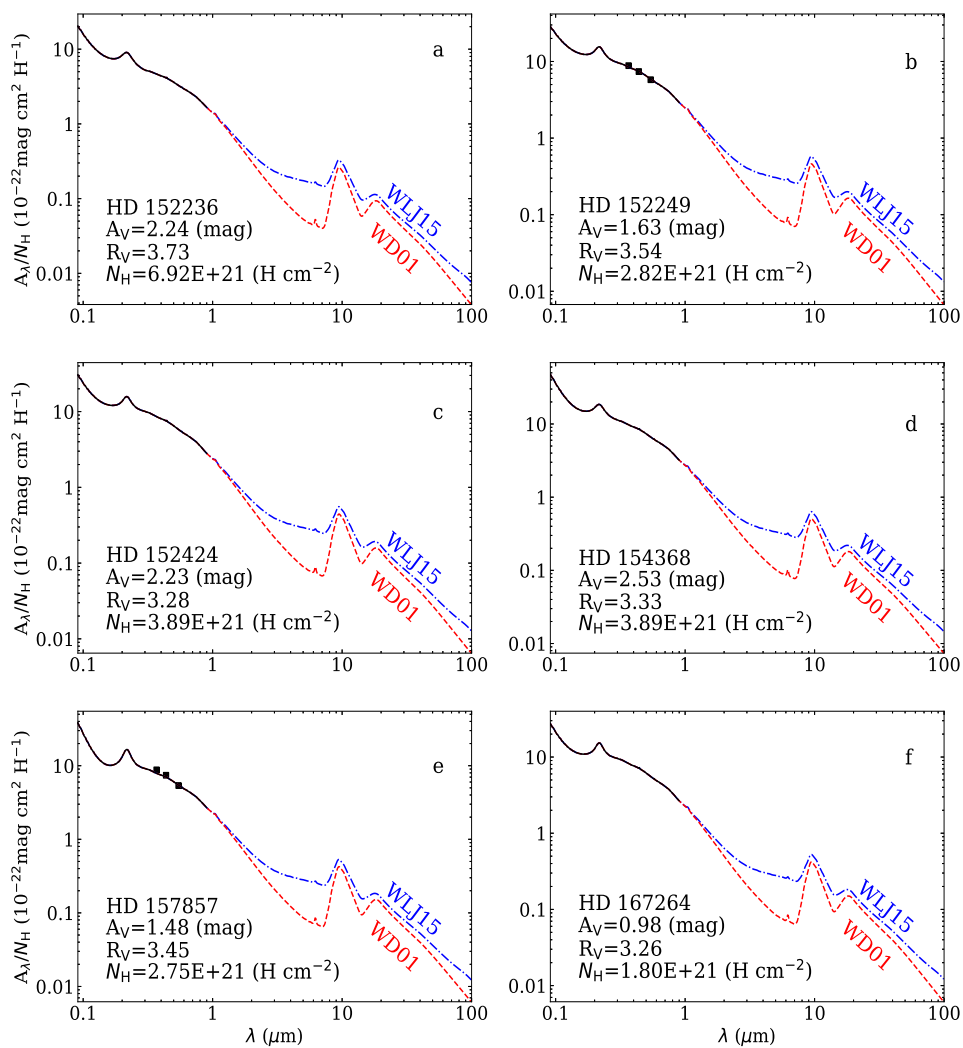}
	\caption{\footnotesize
	\label{fig:curve8}
	Same as Figure~\ref{fig:curve1}
        but for HD\,152236 (a), HD\,152249 (b),
        HD\,152424 (c), HD\,154368 (d),
        HD\,157857 (e), and HD\,167264 (f).
        }
\end{figure*}

\begin{figure*}
	\centering	
	\includegraphics[width=0.9\textwidth,height=0.8\textheight]{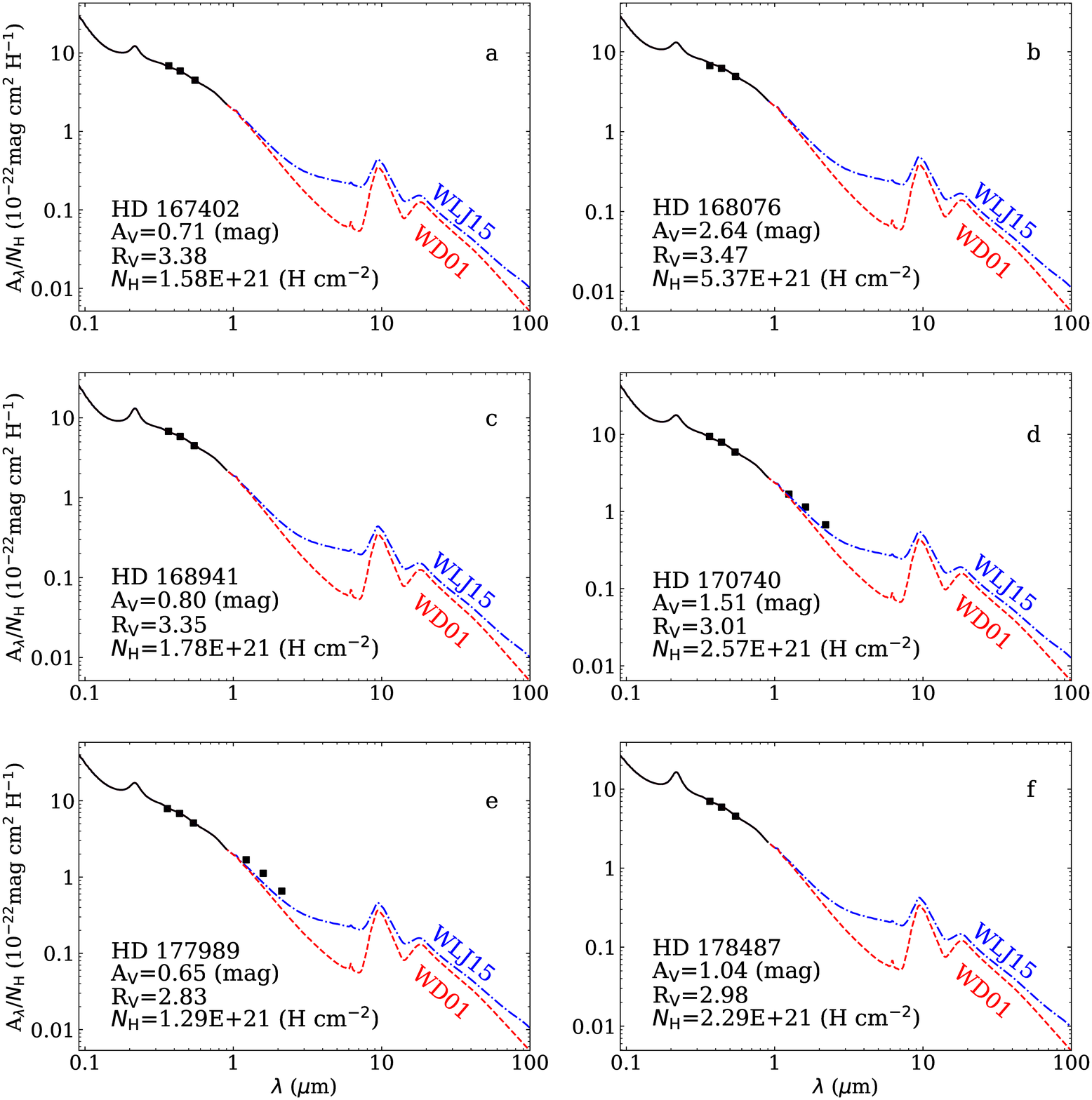}
	\caption{\footnotesize
	\label{fig:curve9}
	Same as Figure~\ref{fig:curve1}
        but for HD\,167402 (a), HD\,168076 (b),
        HD\,168941 (c), HD\,170740 (d),
        HD\,177989 (e), and HD\,178487 (f).
        }
\end{figure*}

\begin{figure*}
	\centering	
	\includegraphics[width=0.9\textwidth,height=0.8\textheight]{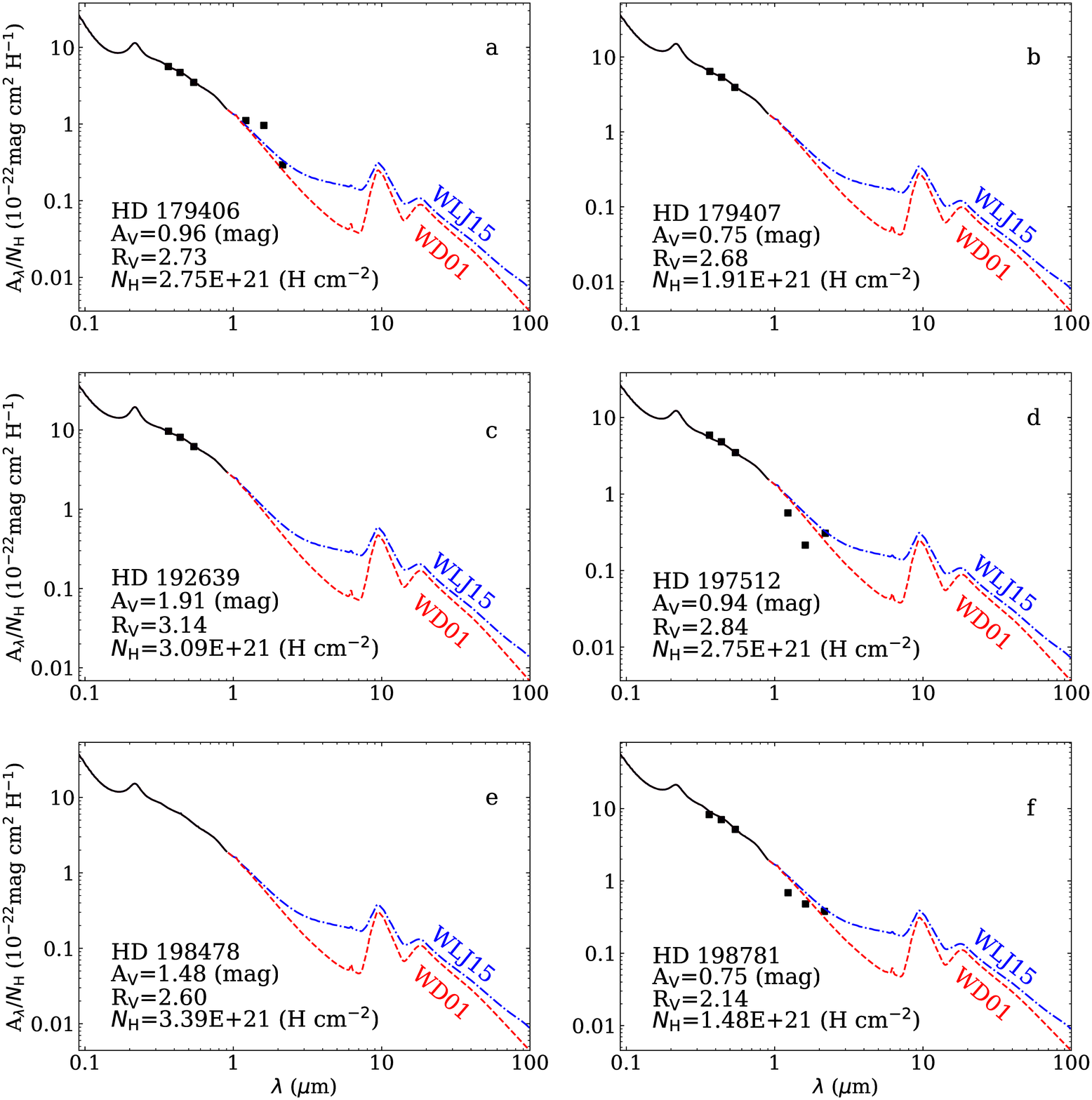}
	\caption{\footnotesize
	\label{fig:curve10}
	Same as Figure~\ref{fig:curve1}
        but for HD\,179406 (a), HD\,179407 (b),
        HD\,192639 (c), HD\,197512 (d),
        HD\,198478 (e), and HD\,198781 (f).
        }
\end{figure*}

\begin{figure*}
	\centering	
	\includegraphics[width=0.9\textwidth,height=0.8\textheight]{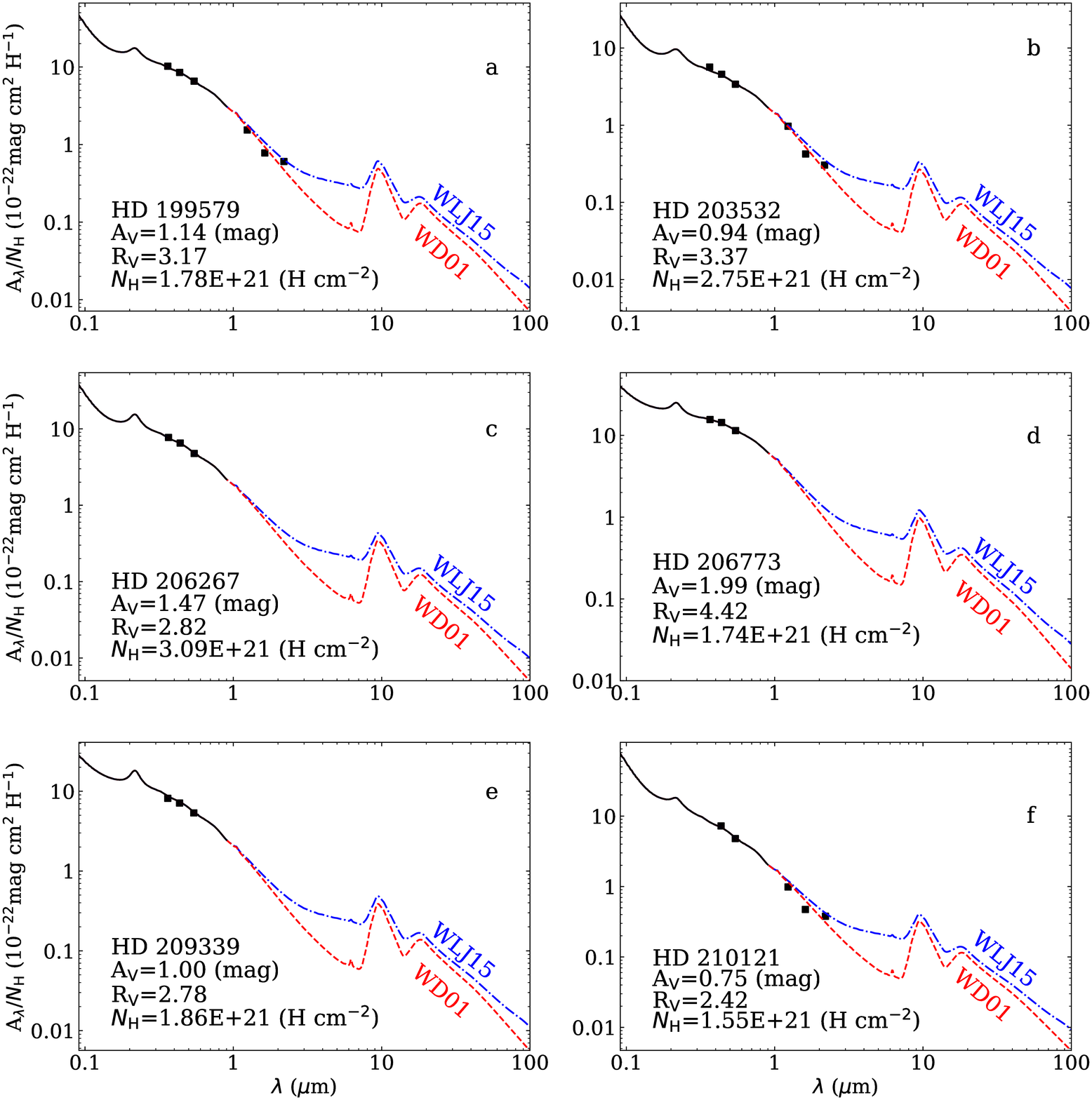}
	\caption{\footnotesize
	\label{fig:curve11}
	Same as Figure~\ref{fig:curve1}
        but for HD\,199579 (a), HD\,203532 (b),
        HD\,206267 (c), HD\,206773 (d),
        HD\,209339 (e), and HD\,210121 (f).
        }
\end{figure*}

\begin{figure*}
	\centering	
	\includegraphics[width=0.9\textwidth,height=0.8\textheight]{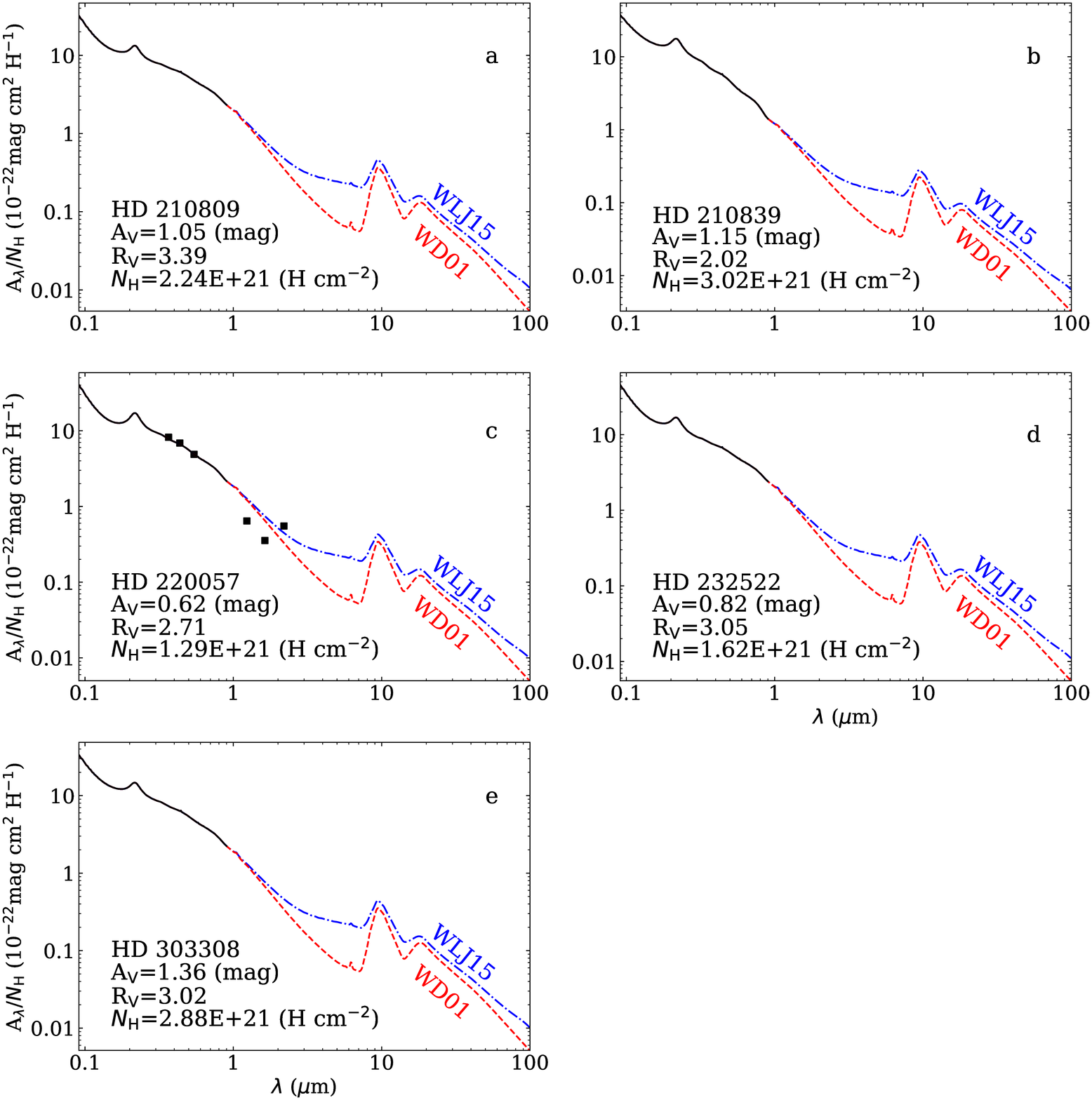}
	\caption{\footnotesize
	\label{fig:curve12}
	Same as Figure~\ref{fig:curve1}
        but for HD\,210809 (a), HD\,210839 (b),
        HD\,220057 (c), HD\,232522 (d),
        and HD\,303308 (e).
	}
\end{figure*}

The synthesized extinction curves for the 71 sight lines
are shown in Figures~\ref{fig:curve1}--\ref{fig:curve12}.
Whenever available, the broadband photometric extinction
data at the U, B, V, J, H and K bands are superimposed
as black squares on the extinction curves.
It is apparent that the synthesized extinction curves
of all 49 sightlines for which the U, B, and V extinction
data are available closely agree with
the observationally-determined U, B and V extinction. 
It is also clear that, for the majorities (24/29) of the sight
lines for which the J, H and K extinction data are available,
the WLJ15 curves closely agree the observationally-determined
J, H and K extinction, suggesting that the WLJ15 model 
may be a realistic representation of
the near- and mid-IR extinction. 
While the WD01 curve approximately agrees with
the J, H, and K extinction data of HD\,25443
(see Figure~\ref{fig:curve2}),
it appreciably exceeds the J and H extinction data
of HD\,197512 (see Figure~\ref{fig:curve10})
and HD\,220057 (see Figure~\ref{fig:curve12}).
On the other hand, the WLJ15 curves of HD\,38087
(see Figure~\ref{fig:curve2})
and HD\,179789 (see Figure~\ref{fig:curve9})
are somewhat lower than their J, H and K extinction data. 

\section{Interstellar Extinction Modeling} \label{sec:extmod}
We now model the extinction curve of each sight line
to derive the dust size distributions and the abundances
of C, O, Si, Mg and Fe required to be depleted in dust. 
We consider the standard silicate-graphite interstellar
grain model which consists of two separate dust
components: amorphous silicate and graphite
(Mathis et al.\ 1977, Draine \& Lee 1984).
For simplicity, we assume the dust to be spherical in shape.
We adopt an exponentially-cutoff power-law size
distribution for both components:
$dn_i/da = \nH B_i a^{-\alpha_i} \exp\left(-a/a_{c,i}\right)$
for the size range of
$50\Angstrom < a < 2.5\um$,
where $a$ is the spherical radius of the dust,
$\nH$ is the number density of H nuclei,
$dn_i$ is the number density of dust of type $i$
with radii in the interval [$a$, $a$\,$+$\,$da$],
$\alpha_i$ and $a_{c,i}$ are respectively
the power index and exponential cutoff size
for dust of type $i$, and
$B_i$ is the constant related to
the total amount of dust of type $i$.
The total extinction per H column
at wavelength $\lambda$ is given by
\begin{equation}\label{eq:Amod}
A_\lambda/\NH = 1.086
            \sum_i \int da \frac{1}{\nH} 
            \frac{dn_i}{da}
            C_{{\rm ext},i}(a,\lambda),
\end{equation}
where the summation is over the two grain types
(i.e., silicate and graphite),
and $C_{{\rm ext},i}(a,\lambda)$
is the extinction cross section of
grain type $i$ of size $a$
at wavelength $\lambda$
which can be calculated 
from Mie theory (Bohren \& Huffman 1983)
using the dielectric functions of
``astronomical'' silicate and graphite 
of Draine \& Lee (1984).

The upper cutoff size of $\amax=2.5\mum$ is
chosen because the amount of large grains
of $a>2.5\mum$ is negligible.
As a matter of fact, we will see later that for
most of the sight lines, the exponential cutoff
sizes ($\acS\simlt0.2\mum$ for silicate,
$\acC\simlt0.5\mum$ for graphite;
see Table~\ref{tab:modpara})
are all much smaller than $\amax=2.5\mum$.
This justifies that the choice of $\amax=2.5\mum$. 
On the other hand,
the presence in the ISM of a population
of angstrom- and nano-sized grains is
revealed by the emission detected by
the {\it Infrared Astronomical Satellite} (IRAS)
broadband photometry at 12 and 25$\mum$
and later confirmed by
the {\it Diffuse Infrared Background Experiment}
(DIRBE) instrument on the
{\it Cosmic Background Explorer} (COBE) satellite
(see Li 2004 and references therein),
as well as by the so-called ``unidentified infrared
emission (UIE)'' bands at 3.3, 6.2, 7.7, 8.6, 11.3
and 12.7$\mum$ which are commonly attributed
to polycyclic aromatic hydrocarbon (PAH) molecules
(Allamandola et al.\ 1985, L\'eger \& Puget 1984, Li 2020).
The lower cutoff size of $\amin=50\Angstrom$
is chosen because angstrom- and nano-sized
grains are in the Rayleigh regime
(i.e., $2\pi a/\lambda\ll1$) in the far-UV and,
on a per unit volume basis,
their extinction cross sections
are independent of grain size $a$
and, therefore, the observed far-UV extinction
is not able to constrain the size distribution of
angstrom- and nano-sized grains. Instead,
it is the near- and mid-IR emission that allows
one to derive the size distribution of grains
of $a<50\Angstrom$ (see Li \& Draine 2001a).

In fitting the extinction curve,
for a given sightline, 
we have six parameters:
the size distribution power indices 
$\alphaS$ and $\alphaC$ 
for silicate and graphite, respectively;
the exponential cutoff sizes 
$\acS$ and $\acC$
for silicate and graphite, respectively;
and $\Bsil$ and $\Bgra$.
We derive the silicon and carbon depletions from
\begin{equation}\label{eq:Si2H}
\sidust = \left(\nH\Bsil/172\mH\right) 
        \int da \left(4\pi/3\right) a^3\,\rhosil
               a^{-\alphaS}
               \exp\left(-a/\acS\right) ~~,
\end{equation}
\begin{equation}\label{eq:C2H}
\cdust = \left(\nH\Bgra/12\mH\right) 
        \int da \left(4\pi/3\right) a^3\,\rhogra
               a^{-\alphaC}
               \exp\left(-a/\acC\right) ~~,
\end{equation}
where we assume a stoichiometric composition of
MgFeSiO$_4$ for amorphous silicate
(of which the molecular weight
is $\mu_{\rm sil}\approx172\mH$)
so that the abundances
of O, Mg and Fe tied up in dust are respectively
$\odust\approx4\times\sidust$,
$\mgdust\approx\sidust$, and
$\fedust\approx\sidust$.
As mentioned earlier, both silicate
and graphitic grains are taken to
be larger than $\amin=50\Angstrom$
and thus nano silicate grains are not
considered here, although an appreciable
amount of nano silicate grains may be
present in the ISM
(Li \& Draine 2001b, Hoang et al.\ 2016,
Hensley \& Draine 2017).
Also, PAHs are not included
in the extinction modeling
and the 2175$\Angstrom$ extinction
bump is attributed to small graphitic grains.

For a given sightline,
we seek the best fit to the extinction 
between $0.3\mum^{-1}$ and $8\mum^{-1}$
by varying the size distribution power indices 
$\alphaS$ and $\alphaC$, and
the upper cutoff size parameters 
$\acS$ and $\acC$.
Following WD01,
we evaluate the extinction at 
100 wavelengths $\lambda_i$, 
equally spaced in $\ln\lambda$.
We use the Levenberg-Marquardt method
(Press et al.\ 1992) to minimize $\chi^2$ 
which gives the error in the extinction fit:
\begin{equation}
\chi^2 = \sum_i \frac{\left( \ln A_{\rm obs} 
       - \ln A_{\rm mod} \right)^2} {\sigma_i^2}~~~,
\end{equation}
where $A_{\rm obs}(\lambda_i)$ 
is the observed extinction at wavelength $\lambda_i$,
$A_{\rm mod}(\lambda_i)$ 
is the extinction computed
for the model at wavelength $\lambda_i$
(see eq.\,\ref{eq:Amod}), 
and the $\sigma_i$ are weights.  
Following WD01,
we take the weights $\sigma_i^{-1} = 1$ 
for $1.1 < \lambda^{-1} < 8\mum^{-1}$ 
and $\sigma_i^{-1} = 1/3$ for 
$\lambda^{-1} < 1.1\mum^{-1}$.

\begin{figure*}
\centering
\vspace{-20mm}
\includegraphics[width=0.9\textwidth,height=0.8\textheight]{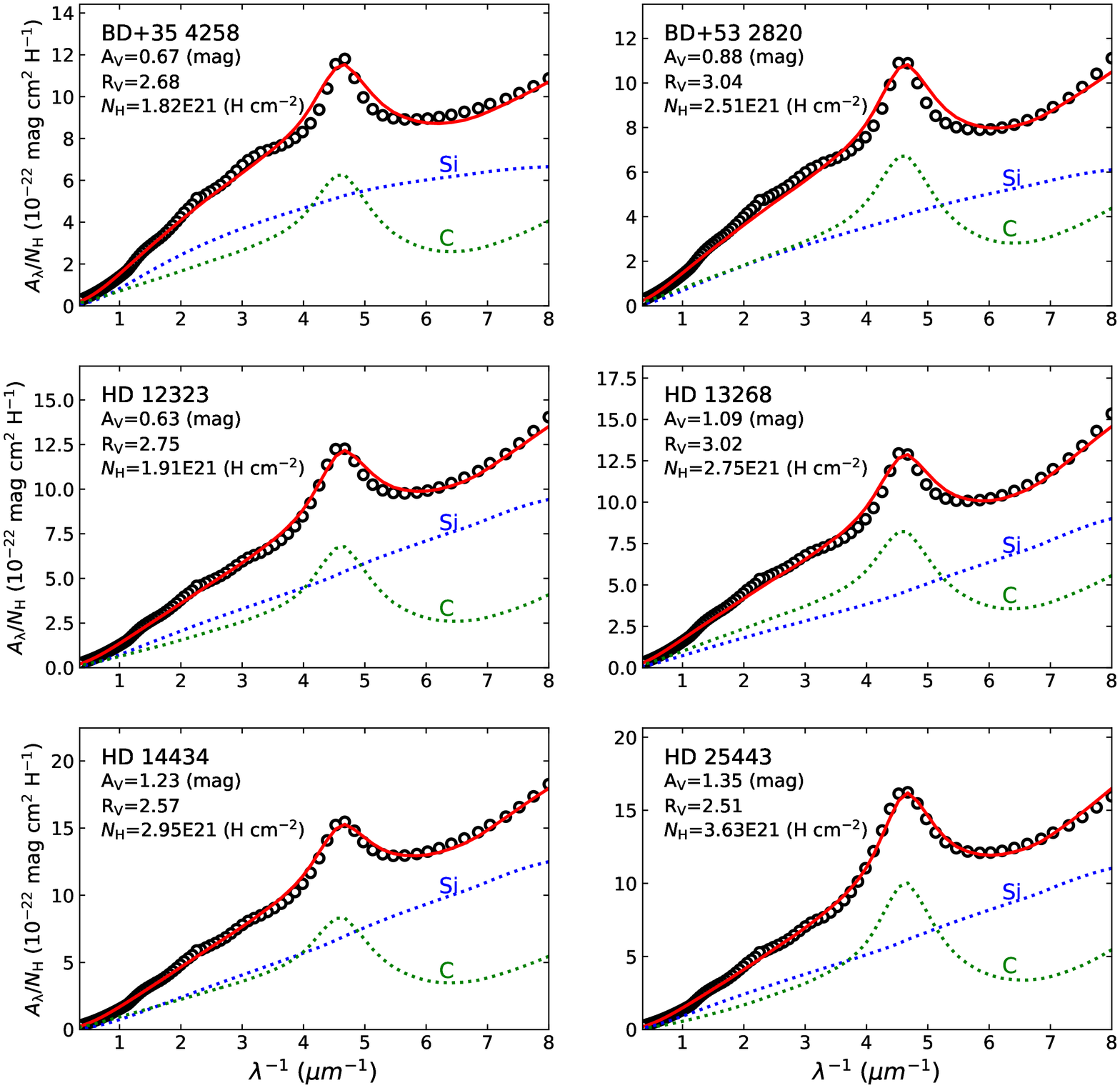}
\vspace{-5mm}
\caption{\footnotesize
              \label{fig:extmod1}
	Fitting the extinction curves of BD+35\,4258,
        BD+53\,2820, HD\,12323, HD\,13268,
        HD\,14434, and HD\,25443.
        The observed extinction curves (open black circles),
        as described in \S\ref{sec:sample} 
        and shown in Figures~\ref{fig:curve1}--\ref{fig:curve12}, 
         are represented by the FM90 parameterization 
         at $\lambda^{-1} > 3.3\mum^{-1}$,
         by the CCM parameterization 
         at $1.1<\lambda^{-1} < 3.3\mum^{-1}$, and
         by the WLJ15 model curve
         at $\lambda^{-1} < 1.1\mum^{-1}$
         (or by the WD01 model curve if it agrees
         better with the J, H, K extinction data).
         The solid red line plots the model
         extinction curve which is a combination
         of amorphous silicate (dotted green line, labeled ``Si'')
         and graphite (dashed blue line, labeled ``C'').
         }
\end{figure*}

\begin{figure}[htbp]
	\centering
	\includegraphics[width=0.9\textwidth,height=0.8\textheight]{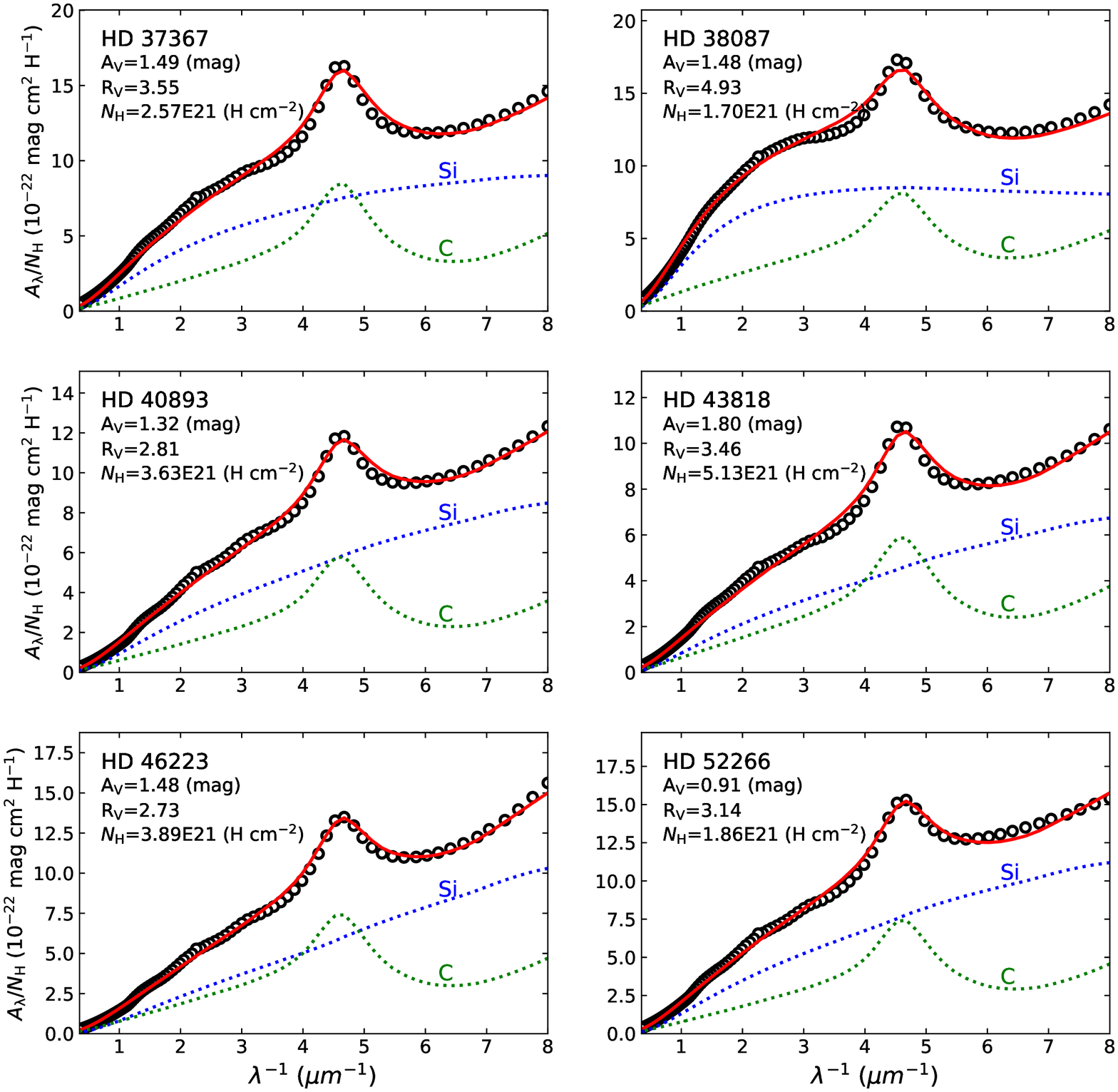}
	\caption{\footnotesize 
	\label{fig:extmod2}
	Same as Figure~\ref{fig:extmod1},
        but for HD\,37367, HD\,38087, HD\,40893,
        HD\,43818, HD\,46223, and HD\,52266.
	}
\end{figure}

\begin{figure}[htbp]
	\centering
	\includegraphics[width=0.9\textwidth,height=0.8\textheight]{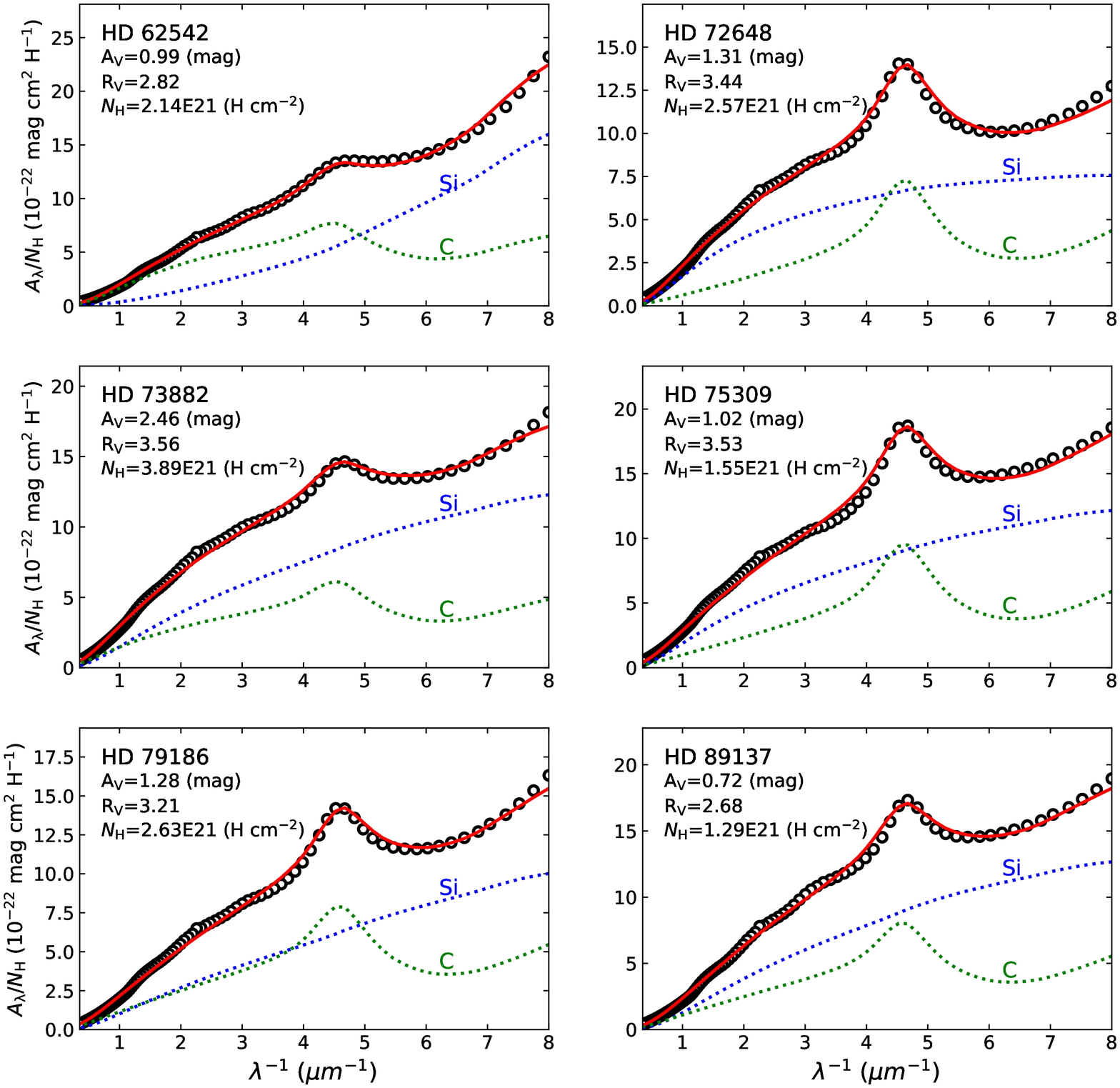}
	\caption{\footnotesize 
	\label{fig:extmod3}
	Same as Figure~\ref{fig:extmod1},
        but for HD\,62542, HD\,72648, HD\,73882,
        HD\,75309, HD\,79186, and HD\,89137.
	}
\end{figure}

\begin{figure}[htbp]
	\centering
	\includegraphics[width=0.9\textwidth,height=0.8\textheight]{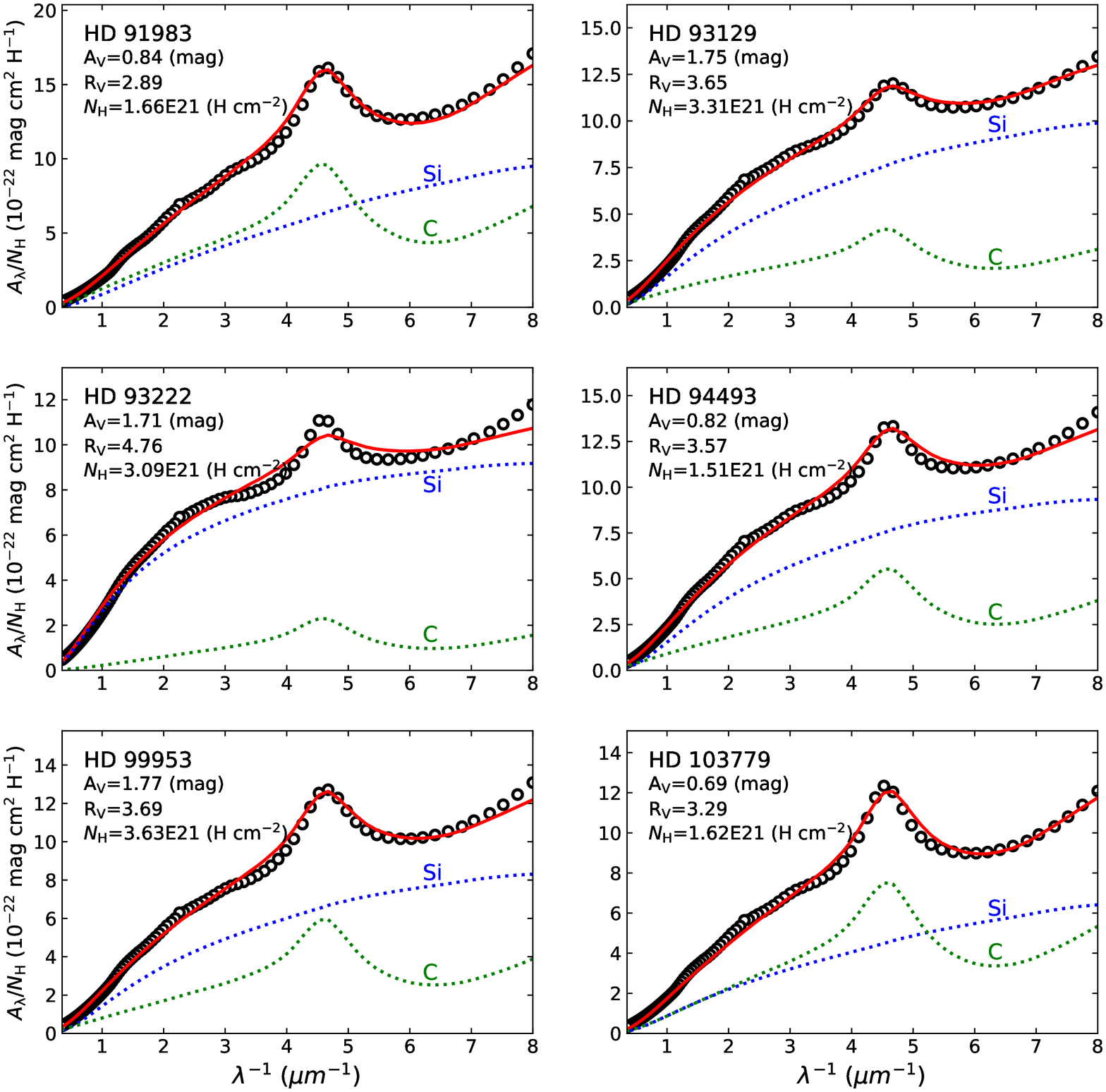}
	\caption{\footnotesize 
	\label{fig:extmod4}
	Same as Figure~\ref{fig:extmod1},
        but for HD\,91983, HD\,93129, HD\,93222,
        HD\,94493, HD\,99953, and HD\,103779.
	}
\end{figure}

\begin{figure}[htbp]
	\centering
	\includegraphics[width=0.9\textwidth,height=0.8\textheight]{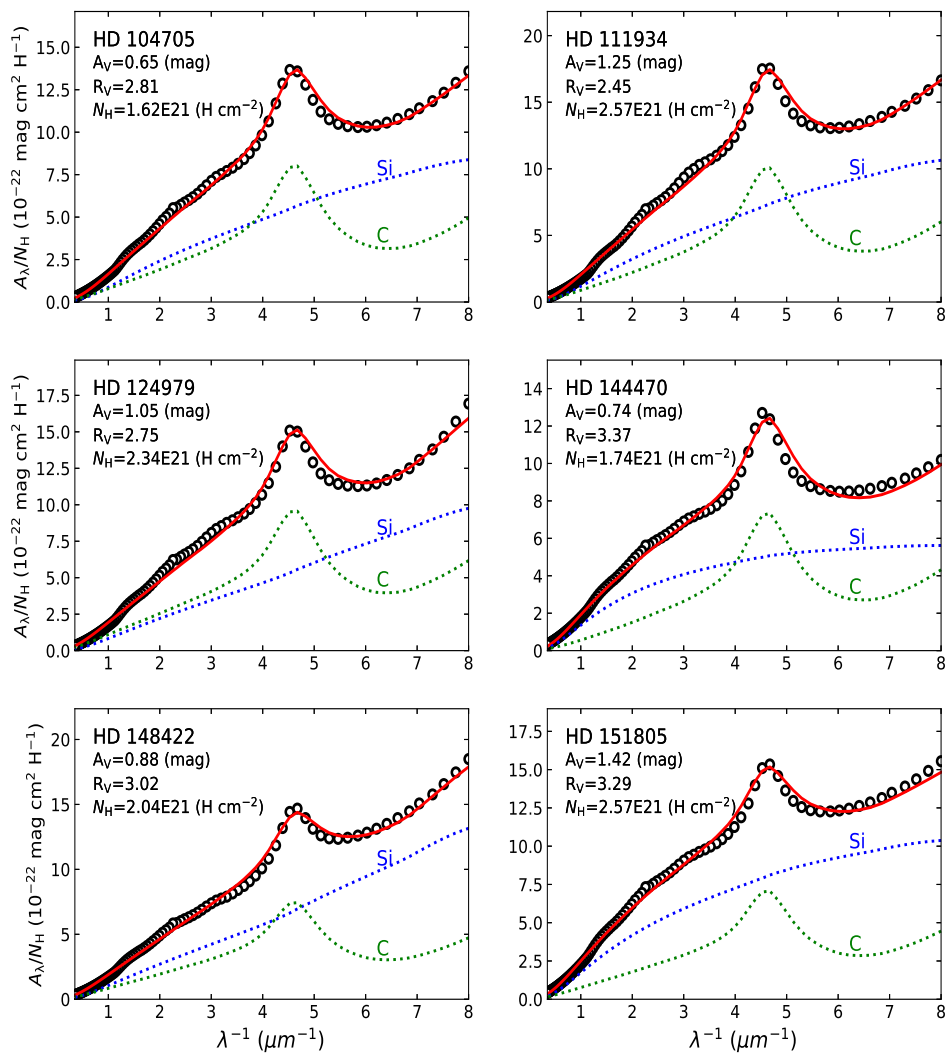}
	\caption{\footnotesize 
	\label{fig:extmod5}
	Same as Figure~\ref{fig:extmod1},
        but for HD\,104705, HD\,111934, HD\,124979,
        HD\,144470, HD\,148422, and HD\,151805.
	}
\end{figure}

\begin{figure}[htbp]
	\centering
	\includegraphics[width=0.9\textwidth,height=0.8\textheight]{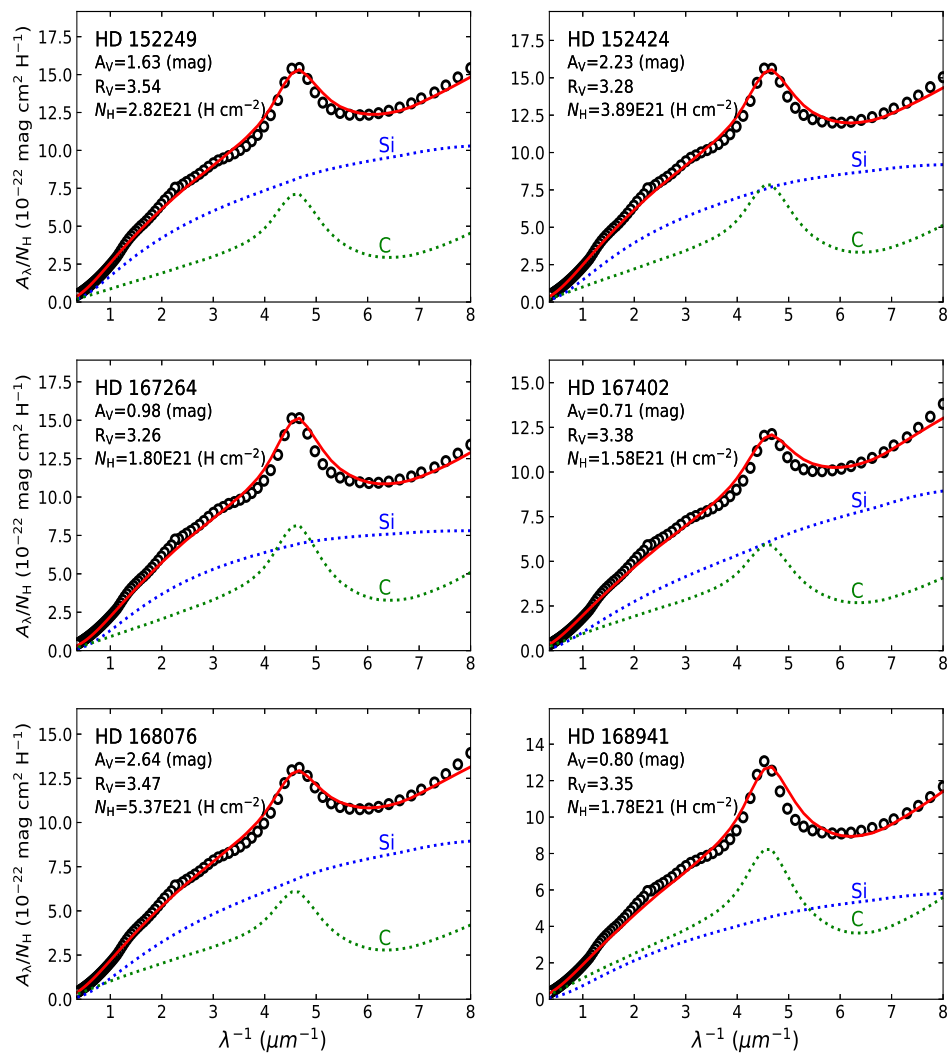}
	\caption{\footnotesize 
	\label{fig:extmod6}
	Same as Figure~\ref{fig:extmod1},
        but for HD\,152249, HD\,152424, HD\,167264,
        HD\,167402, HD\,168076, and HD\,168941.
	}
\end{figure}

\begin{figure}[htbp]
	\centering
	\includegraphics[width=0.9\textwidth,height=0.8\textheight]{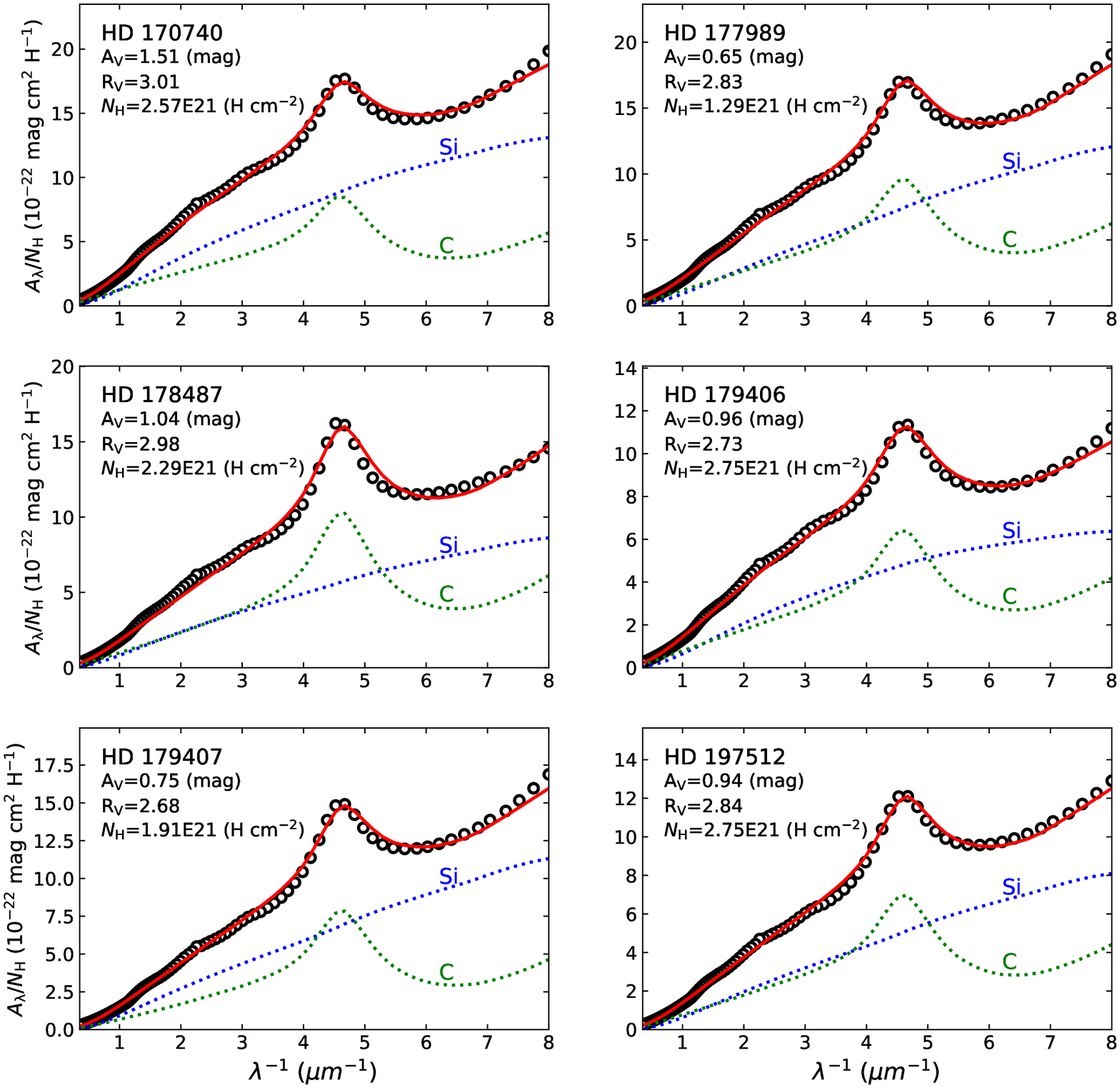}
	\caption{\footnotesize 
	\label{fig:extmod7}
	Same as Figure~\ref{fig:extmod1},
        but for HD\,170740, HD\,177989, HD\,178487,
        HD\,179406, HD\,179407, and HD\,197512.
	}
\end{figure}

\begin{figure}[htbp]
	\centering
	\includegraphics[width=0.49\textwidth,height=0.8\textheight]{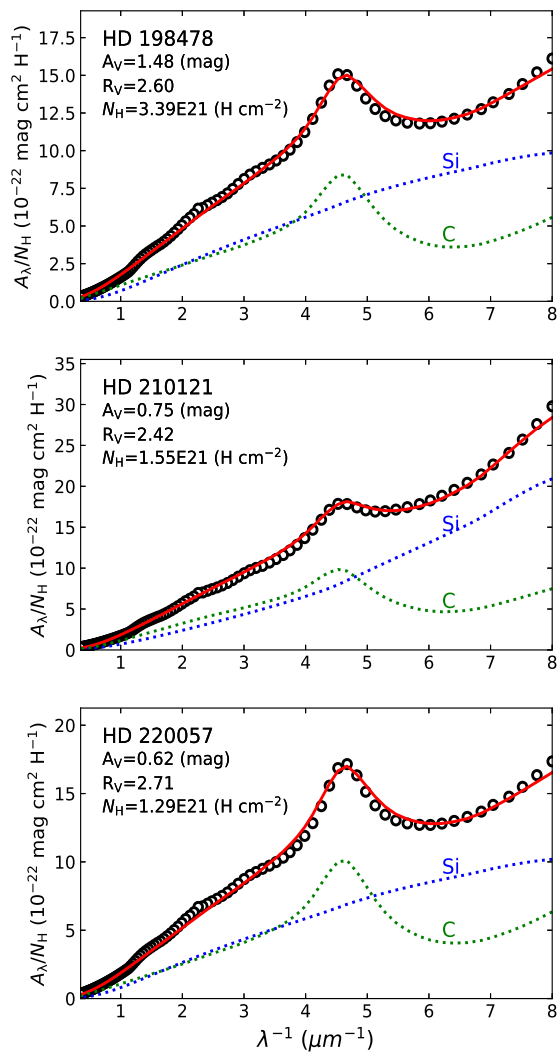}
	\caption{\footnotesize 
	\label{fig:extmod8}
	Same as Figure~\ref{fig:extmod1},
        but for HD\,198478, HD\,210121,
        and HD\,220057.
	}
\end{figure}

Among the 81 sight lines compiled in this work,
Mishra \& Li (2015, 2017) have already modeled
the extinction curves for 36 sightlines in the same
manner as described above.
In this work, we model the remaining 45 sight lines
and the best-fit model parameters are tabulated 
in Table~\ref{tab:modpara}. As shown in
Figures~\ref{fig:extmod1}--\ref{fig:extmod8},
a simple mixture of silicate and graphite is capable
of closely reproducing the observed extinction curves
of almost all sight lines from the UV to the near-IR.
The only exception is HD\,93222 for which the model
produces too broad a 2175$\Angstrom$ extinction bump
to be comparable to the observed bump
(see Figure~\ref{fig:extmod4}). Indeed, HD\,93222 is
rather unusual in the sense that, while its extinction curve
at $\lambda^{-1}<4\mum^{-1}$ is characteristic of
dense regions as reflected by its large $\RV=4.76$,
it exhibits a sharp 2175$\Angstrom$ bump and a steep
far-UV rise at $\lambda^{-1} > 5.9\mum^{-1}$
which are both characteristic of a small $\RV$ for diffuse
regions (i.e., $\RV<3.1$). A more thorough exploration of
HD\,93222 will be presented in a separate paper.

It is gratifying that the simple silicate-graphite model
also successfully fits the ``anomalous'' extinction curves
of HD\,62542 (see Figure~\ref{fig:extmod3})
and HD\,210121 (see Figure~\ref{fig:extmod8}).
The line of sight toward the B5V star HD\,62542
passes through a dense cloud in the Gum nebula region.
Its extinction curve shows an exceedingly 
anomalous 2175$\Angstrom$ extinction bump
of which the central wavelength shifts from 
the canonical 2175$\Angstrom$ to 2110$\Angstrom$,
and its width ($\simali$1.3$\mum^{-1}$) is substantially
broader than the Galactic average of
$\simali$0.99$\mum^{-1}$
(Cardelli \& Savage 1988).
The high-latitude cirrus cloud toward
the B3V star HD\,210121 also exhibits
an anomalous extinction curve which
deviates considerably from the CCM
parameterization expected from its small
$\RV\approx2.1$ (Larson et al.\ 1996).
Its extinction curve is characterized by
an extremely steep far-UV rise and by
a weak and broad 2200$\Angstrom$ hump
(Welty \& Fowler 1992, Li \& Greenberg 1998).
The robustness of the silicate-graphite model
in modeling the extinction curves of individual
sight lines is demonstrated in 
Figures~\ref{fig:extmod1}--\ref{fig:extmod8}
where a wide range of extinction-curve shapes
are accounted for, including the extremely anomalous
extinction curves seen in the lines of sight
toward HD\,62542 and HD\,210121.

\section{Discussion\label{sec:discussion}}
\begin{figure*}
	\centering	
	\vspace{-20mm}
	\includegraphics[width=0.9\textwidth,height=0.8\textheight]{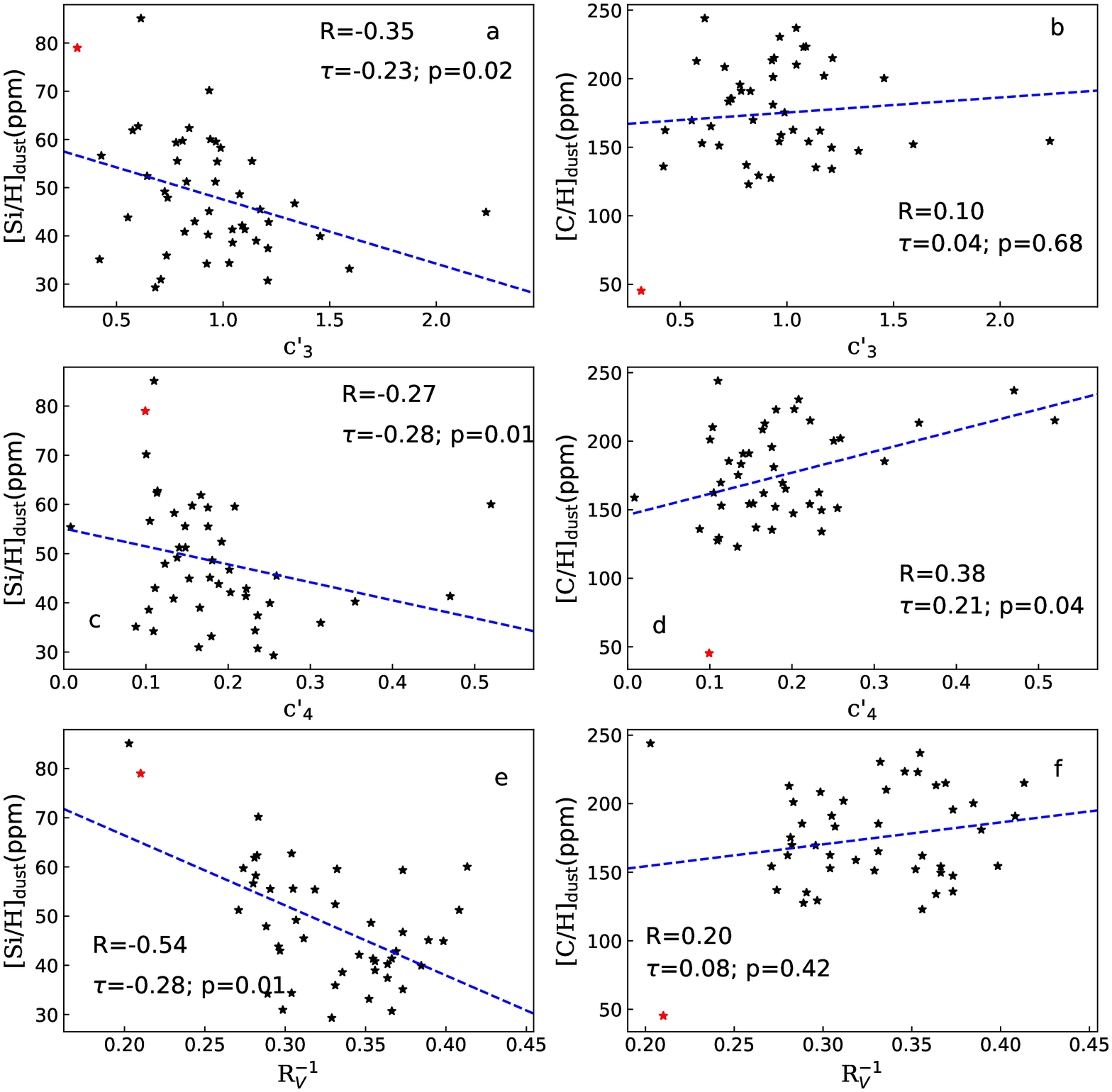}
	\vspace{-5mm}
	\caption{\footnotesize
        \label{fig:c3_c2h_si2h_mod}
         Upper panel: Correlation diagrams between
         the 2175$\Angstrom$ bump ($c_3^{\prime}$) 
         and the silicon depletion $\sidust$ (a) 
         or the carbon depletion $\cdust$ (b) 
         derived from fitting the extinction curve 
         of each sightline with a mixture of
         amorphous silicate dust and graphite dust.
	 Middle panel: The correlations between
	 the far-UV nonlinear extinction 
	 rise ($c_4^{\prime}$)
         and $\sidust$ (c) or $\cdust$ (d).
	 Lower panel: The correlations between
         $R_V^{-1}$ and $\sidust$ (e) or $\cdust$ (f). 
         In each panel, HD\,93222 is shown as a red star
         for which neither the 2175$\Angstrom$ bump
         nor the far-UV rise is well reproduced
         (see Figure~\ref{fig:extmod4}).
         }
\end{figure*}

As described by eqs.\,\ref{eq:Si2H} and \ref{eq:C2H}, 
we derive and tabulate in Table~\ref{tab:modpara} for
each sight line the silicon depletion ($\sidust$) and
carbon depletion ($\cdust$) in dust 
from modeling its extinction curve.
In Figure~\ref{fig:c3_c2h_si2h_mod}a
we examine the correlation between $\sidust$
and the strength of the 2175$\Angstrom$ 
extinction bump ($c_3^{\prime}$). 
With a Pearson correlation coefficient 
of $R\approx-0.35$
and a Kendall $\tau\approx-0.23$ 
and $p\approx0.02$,
it is clear that the silicon depletion does not 
correlate with the 2175$\Angstrom$ bump. 
The possible relation between $\cdust$
and the extinction bump is evaluated
in Figure~\ref{fig:c3_c2h_si2h_mod}b
and no correlation is found 
($R\approx0.10$, $\tau\approx0.04$ 
and $p\approx0.68$).
This can be understood in the context that,
although the silicate-graphite model assigns 
the 2175$\Angstrom$ bump to graphite,
the bulk carbon depletion is not consumed
by the small graphite grains which are responsible
for the extinction bump but by the submicron-sized
graphite grains which, together with the submicron-sized
silicate grains, account for the optical extinction. 
We have also explored the relation
between $\sidust$ and the strength of 
the nonlinear far-UV extinction rise 
($c_4^{\prime}$).
As shown in Figure~\ref{fig:c3_c2h_si2h_mod}c,
no correlation is found.
Similarly, Figure~\ref{fig:c3_c2h_si2h_mod}d
compares $\cdust$ with $c_4^{\prime}$
and also reveals no correlation.
%
We have also investigated how $\sidust$
and $\cdust$ vary with $\RV^{-1}$. 
As shown in Figures~\ref{fig:c3_c2h_si2h_mod}e and f, 
neither $\sidust$ nor $\cdust$ exhibits
strong correlation with $\RV^{-1}$.

We now assess whether the dust depletions are generally
consistent with the interstellar abundance constraints.
For an interstellar dust model to be considered viable,
the total abundance of element X ($\xtot$) implied by
the dust model---the abundance of this element
required to be tied up in dust ($\xdust$)
plus the gas-phase abundance ($\xgas$)---should not
exceed the interstellar abundance ($\xism$). 
Following Zuo et al.\ (2021) and Hensley \& Draine (2021),
we adopt the GCE-augmented protosolar abundances
as the interstellar reference abundances
(i.e., $\cism=339\pm39\ppm$,
$\oism=589\pm68\ppm$,
$\mgism=47.9\pm4.4\ppm$,
$\siism=42.7\pm4.0\ppm$, and
$\feism=47.9\pm4.4\ppm$
for the major dust-forming elements
C, O, Mg, Si and Fe;
see Table~1 in Zuo et al.\ 2021).

\begin{figure*}
\hspace*{-10mm}
\includegraphics[width=0.9\textheight,height=0.5\textwidth]{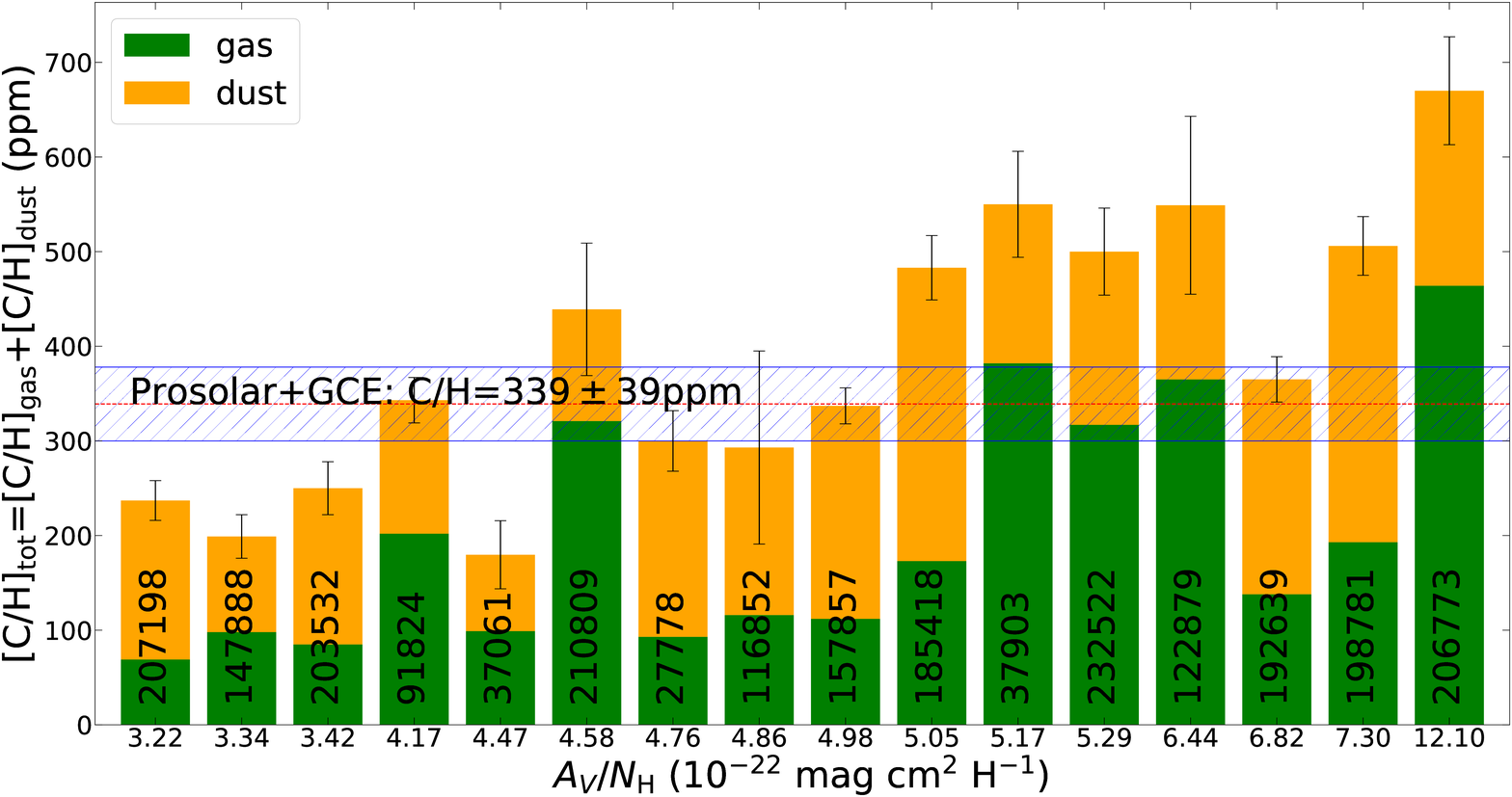}
\hspace*{-10mm}
\includegraphics[width=0.9\textheight,height=0.5\textwidth]{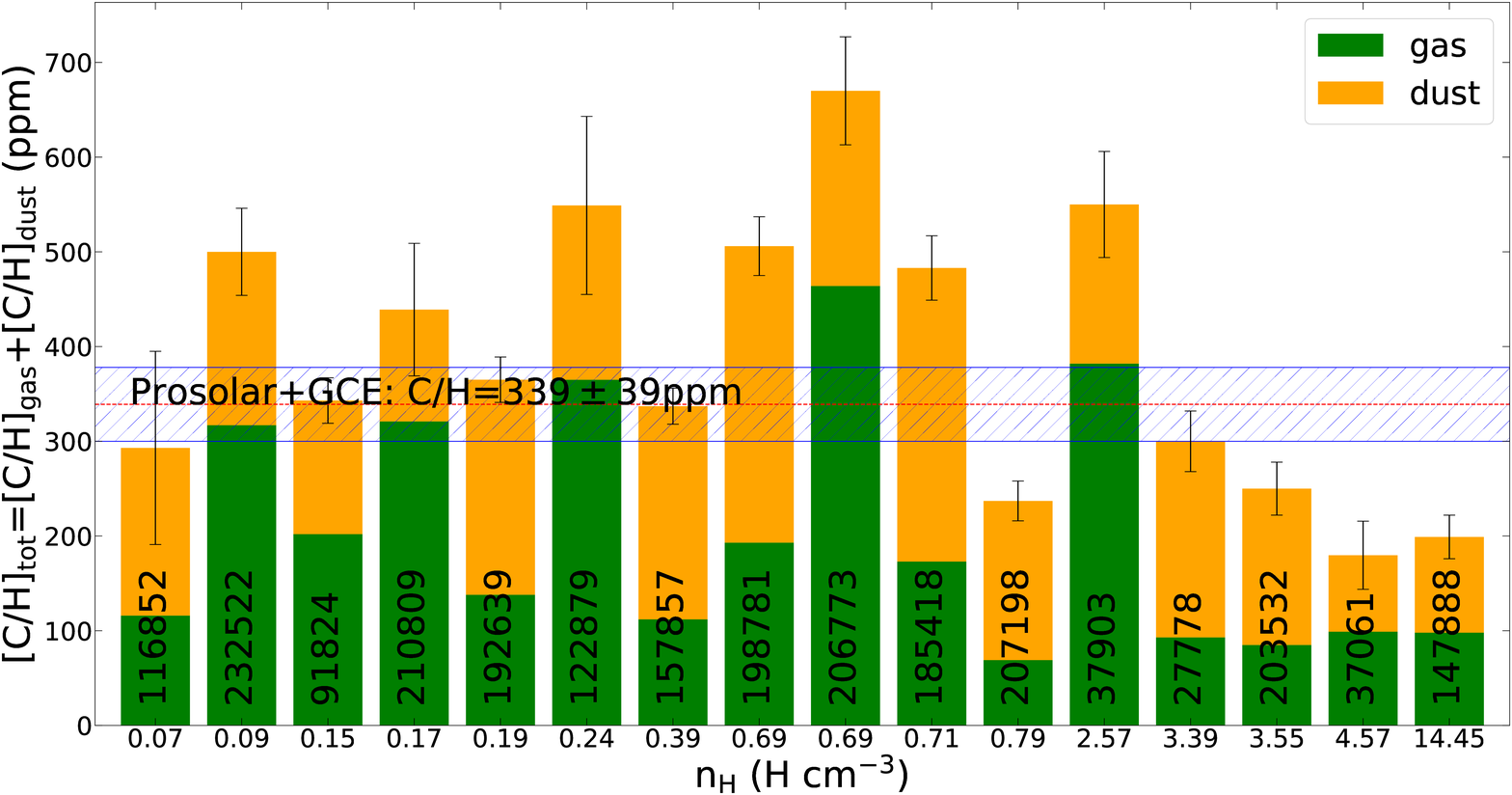}
\caption{\label{fig:C2H}\footnotesize
        Comparison of the GCE-augmented protosolar 
          C/H abundance ($\cism=339\pm39\ppm$; 
          blue horizontal shaded box)
          with $\ctot$ for each sight line,
          the gas-phase C/H abundance 
          (green vertical boxes) plus the C/H depletion
          required by the dust extinction modeling
          to be locked up in graphitic dust
          (orange vertical boxes).
          The (vertical) error bars are for $\ctot$,
          resulting from the uncertainties in $\cgas$.
          The sources are ordered either
          in the extinction-to-gas ratio $\AV/\NH$
          (upper panel) or in the hydrogen number
          density (bottom panel).
          }
\end{figure*}

Figure~\ref{fig:C2H} displays the measured
gas-phase $\cgas$ abundances and the model-derived
dust-phase $\cdust$ abundances of all the 16 sight lines
for each of which $\cgas$ has been observationally determined.
The inferred total dust-plus-gas abundances $\ctot$
are compared with the interstellar abundance $\cism$
which is represented by the GCE-augmented protosolar 
C/H abundance. Although the majorities (10/16) of
the sight lines are consistent with the interstellar abundance
constraints (i.e., $\ctot$ does not surpass $\cism$
for 10 sight lines within the observational uncertainties
of $\cgas$), six sight lines require $\ctot$ to exceed $\cism$.
For these six sight lines with $\ctot > \cism$,
the amount of C atoms available for making carbon dust
are insufficient to account for that required
by the observed extinction.
This was known as the ``C crisis'' in the mid-1990s
when the B star abundances were considered
as the interstellar reference abundances
(Snow \& Witt 1995, 1996; also see Li 2005).
Such a ``C crisis'' holds for these six sight lines
even if we assume the GCE-augmented protosolar
C abundance to be the interstellar C abundance.
All these six sight lines are characterized by a relatively
higher extinction-to-gas ratio of
$\AV/\NH\simgt5\times10^{-22}\magni\cm^2\HH^{-1}$,
while those sight lines with $\ctot\simlt\cism$ all
exhibit a lower extinction-to-gas ratio 
(i.e., $\AV/\NH < 5\times10^{-22}\magni\cm^2\HH^{-1}$).\footnote{%
  HD\,192639, the only sight line
  which has a high extinction-to-gas ratio
  ($\AV/\NH\approx6.8\times10^{-22}\magni\cm^2\HH^{-1}$)
  {\it and} does not violate the abundance constraints
  (i.e., $\ctot\simlt\cism$), has a relatively low gas-phase
  abundance of $\cgas\approx125\ppm$.
  }
Figure~\ref{fig:C2H} also reveals a rough trend
of increasing $\cgas$ with $\AV/\NH$.
However, $\cgas$ does not appear to show any systematic
variations with the hydrogen number density $\nH$.
%

\begin{figure}
\hspace*{-10mm}
\includegraphics[width=0.9\textheight,height=0.5\textwidth]{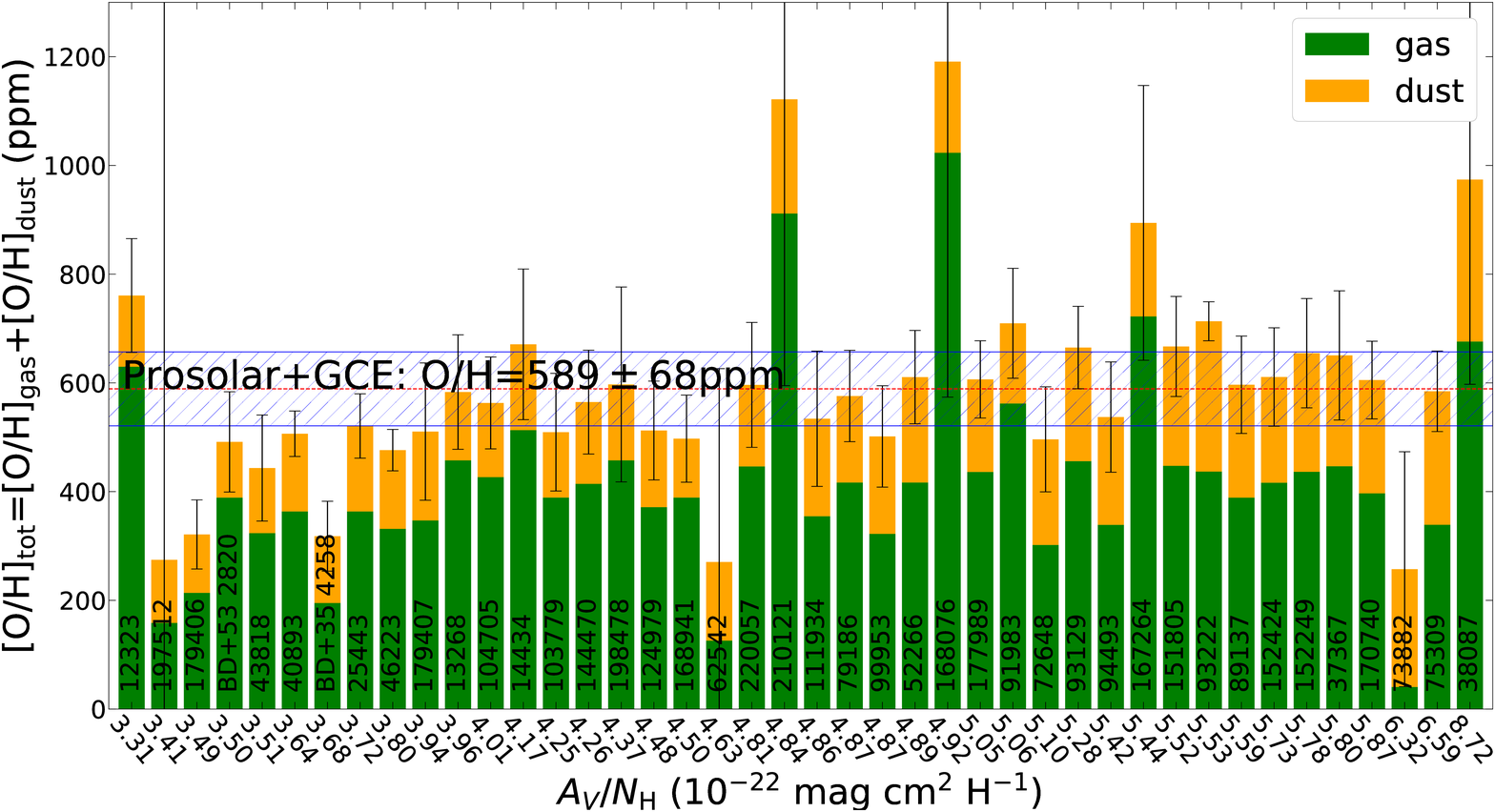}
\hspace*{-10mm}
\includegraphics[width=0.9\textheight,height=0.5\textwidth]{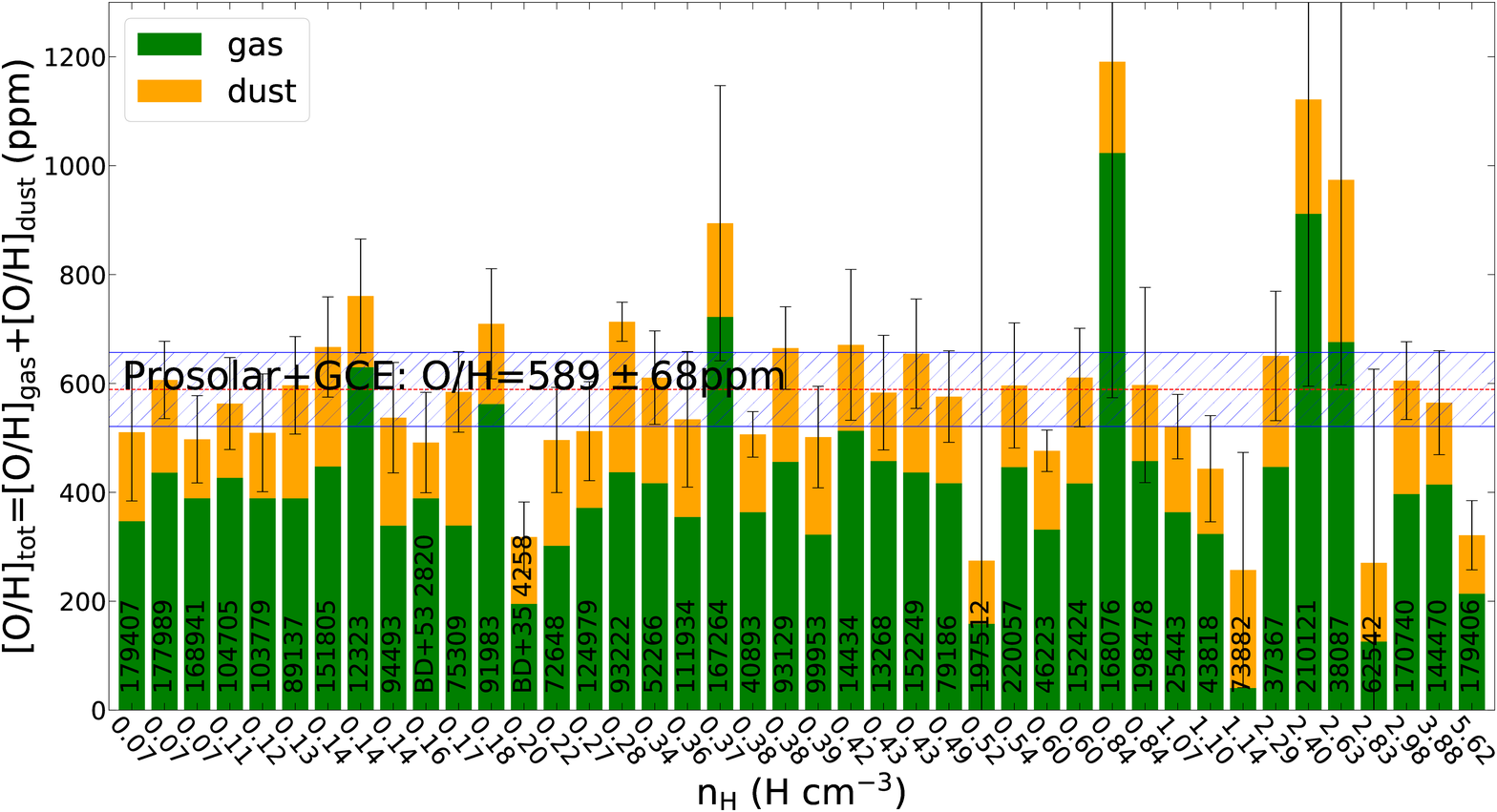}
		\caption{\label{fig:O2H}\footnotesize
                Same as Figure~\ref{fig:C2H} but for oxygen.  
              }
\end{figure}

The O/H depletion in dust is estimated from
$\odust\approx4\times\sidust$ on the basis
of the assumption of a stoichiometric composition of
MgFeSiO$_4$ for silicate dust. 
Figure~\ref{fig:O2H} displays the measured
gas-phase $\ogas$ abundances and the model-derived
dust-phase $\odust$ abundances of all the 42 sight lines
for each of which $\ogas$ has been observationally determined.
While $\ogas$ varies from one sight line to another,
there is no systematic variation of $\ogas$ with
$\AV/\NH$ or $\nH$.
Most prominently, within the observational uncertainties
of $\ogas$, $\otot$ is in general agreement with $\oism$
for almost all sight lines. The only exceptions are the three
sight lines toward HD\,179406, BD+35\,4258, and HD\,73882
for which $\otot$ is somewhat lower than $\oism$.
This clearly shows that the vast majority
(39/42) of the sight lines have no problem
in accommodating the oxygen atoms missing from
the gas phase, supporting our earlier finding
based on a ``gold'' sample of 10 sight lines
for which the gas-phase abundances of all
the major dust-forming elements have been
observationally measured (see Zuo et al.\ 2021).
This is in stark contrast to the so-called ``O crisis''
-- it has long been thought that in the ISM
oxygen is heavily depleted from the gas phase and far
exceeds (by as much as $\simali$160$\ppm$ of O/H)
that can be accounted for by the main oxygen-containing
refractory dust component such as silicates and oxides
(see Jenkins 2009, Whittet 2010, Poteet et al.\ 2015). 
While Wang et al.\ (2015b) attributed the excess O/H
to $\mu$m-sized H$_2$O ice grains
(which are large enough to suppress the 3.1$\mum$
absorption band of H$_2$O ice)
and Potapov et al.\ (2020) attributed it to the trapping
of H$_2$O ice in silicate grains,
here in this work as well as in Zuo et al.\ (2021)
we find that the dust depletions $\odust$ inferred from
the extinction combined with the gas-phase $\ogas$
are sufficient in fully accommodating the interstellar O/H
for the vast majority (39/42) of the sight lines.
%

\begin{figure}
\hspace*{-10mm}
\includegraphics[width=0.9\textheight,height=0.5\textwidth]{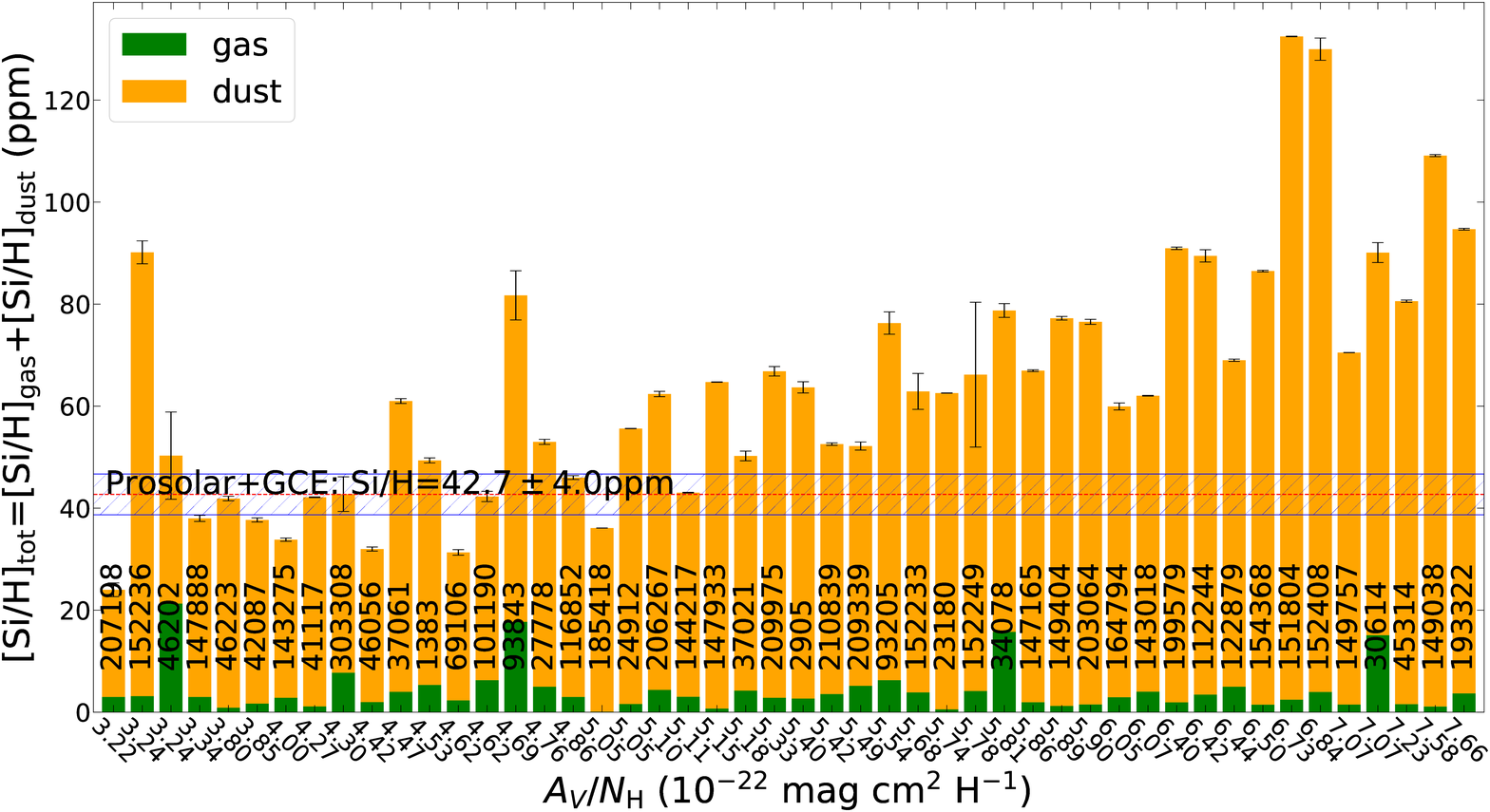}
\hspace*{-10mm}
\includegraphics[width=0.9\textheight,height=0.5\textwidth]{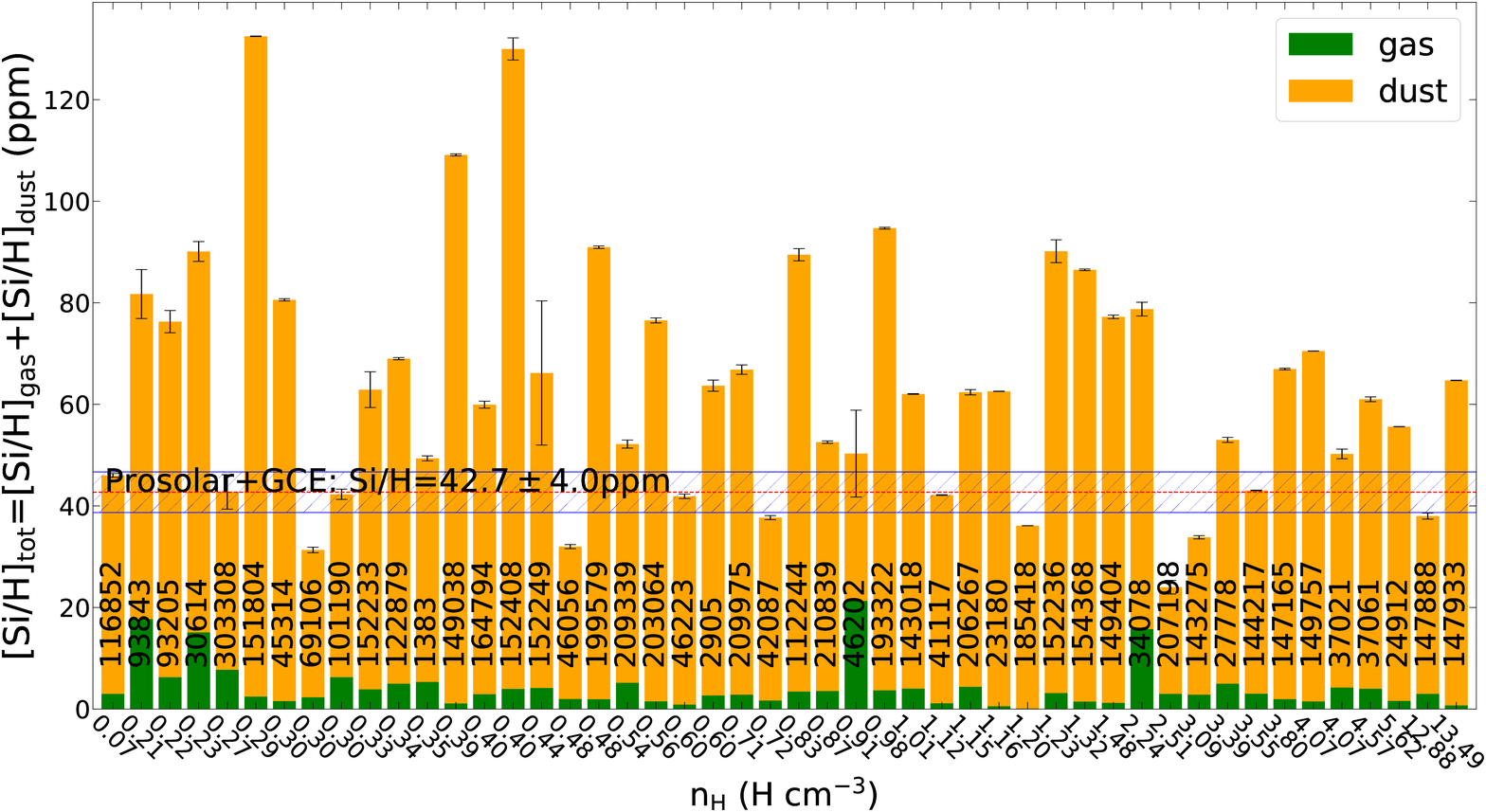}
		\caption{\label{fig:Si2H}\footnotesize
                Same as Figure~\ref{fig:C2H} but for silicon.  
              }
\end{figure}

Figure~\ref{fig:Si2H} compares the interstellar $\siism$
abundances with the measured gas-phase $\sigas$ abundances
and the model-derived dust-phase $\sidust$ abundances
and their combinations $\sitot$ of all the 48 sight lines
for each of which $\sigas$ has been observationally determined.
While most of the sight lines are highly depleted in silicon,
there are several sight lines in which $\sigas$ is rich and
accounts for $\simgt$1/3 of the interstellar $\siism$.
Also, there is no systematic variation of $\sigas$
with $\AV/\NH$ or $\nH$.
For most (43/48) of the sight lines, we find $\sitot$
substantially exceeds $\siism$, implying that there
are not enough silicon atoms to make the silicate
dust required to account for the observed extinction.
Figure~\ref{fig:Si2H} also shows that $\sitot$ tends
to increase with $\AV/\NH$, indicating that the shortage
of Si/H becomes more severe in sight lines with a higher
extinction-to-gas ratio.

\begin{figure}
\hspace*{-10mm}
\includegraphics[width=0.9\textheight,height=0.5\textwidth]{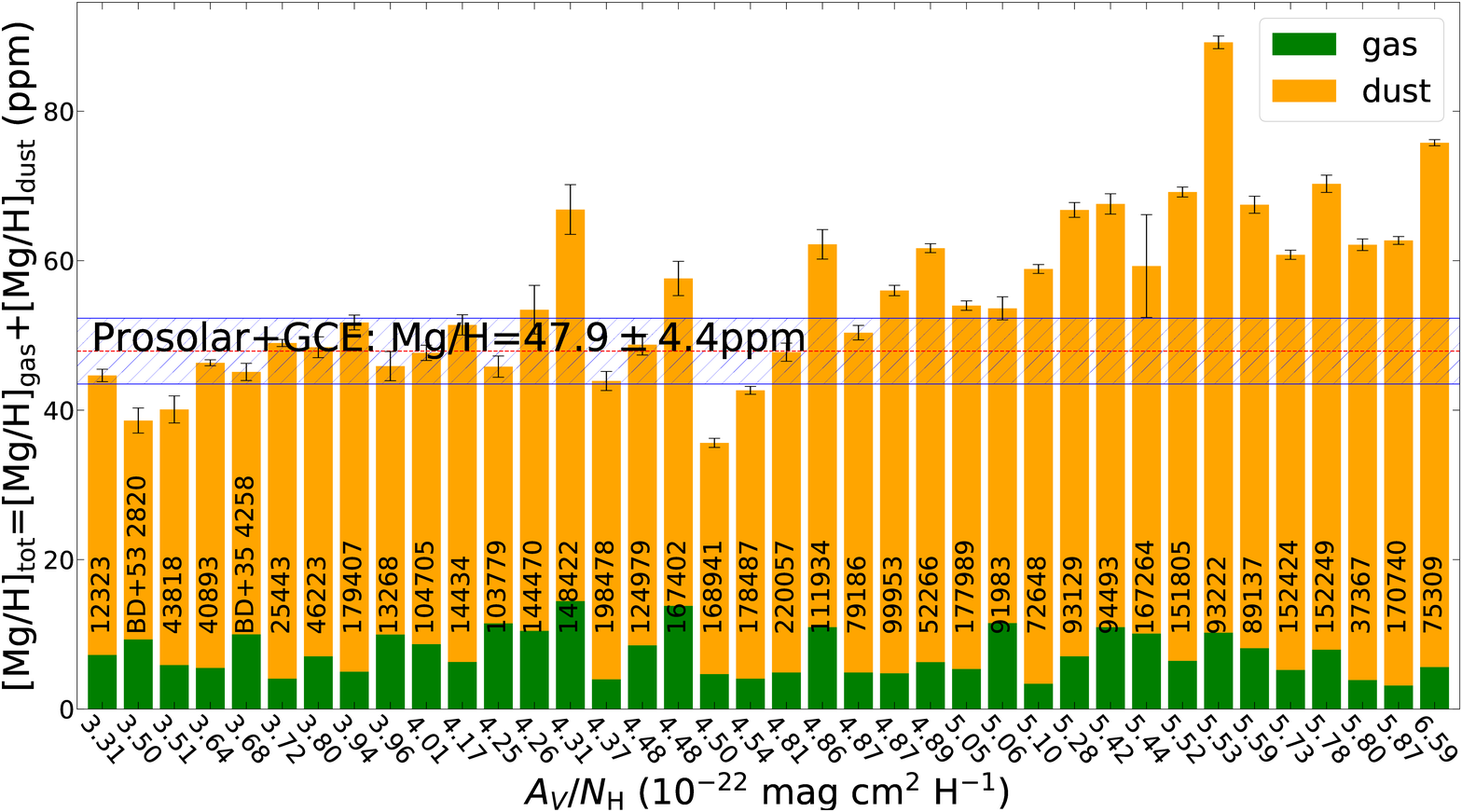}
\hspace*{-10mm}
\includegraphics[width=0.9\textheight,height=0.5\textwidth]{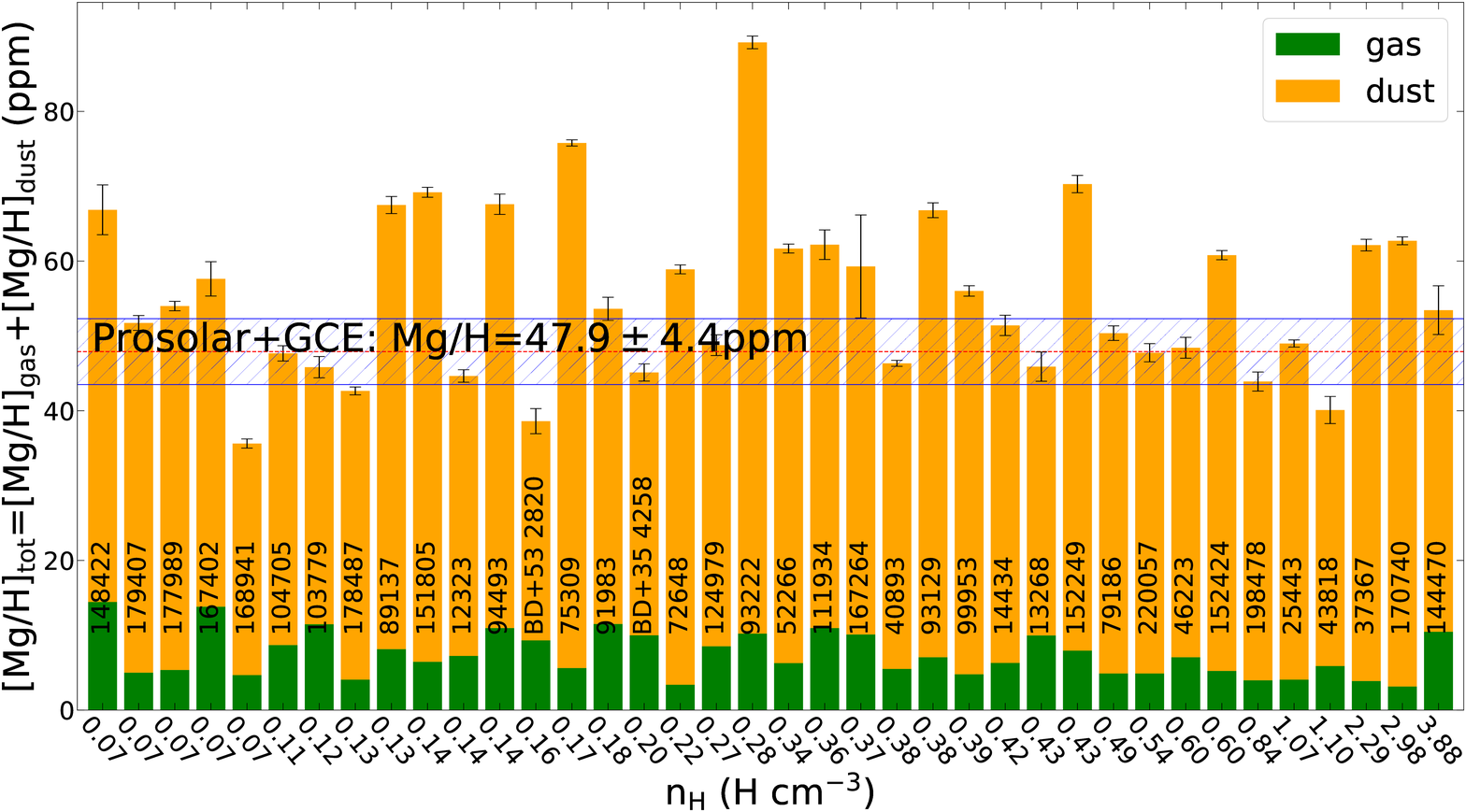}
\caption{\label{fig:Mg2H}\footnotesize
                 Same as Figure~\ref{fig:C2H} but for magnesium.
               }
\end{figure}

Figure~\ref{fig:Mg2H} displays the measured gas-phase
$\mggas$ abundances and the model-derived dust-phase
$\mgdust$ abundances of all the 38 sight lines for each
of which $\mggas$ has been observationally determined.
Although it is widely believed that magnesium is almost
fully depleted from the gas phase, there are appreciable
amounts ($\mggas\gtsim5\ppm$) of gaseous magnesium
atoms in essentially all the 38 sight lines.
The gas-phase $\mggas$ abundances vary
from one sight line to another, but does not
show any systematic variations with $\AV/\NH$ or $\nH$.
Figure~\ref{fig:Mg2H} also compares the interstellar
abundance $\mgism$ with the total dust-plus-gas
abundances $\mgtot$ inferred from the extinction modeling, 
revealing that $\simali$40\% of the sight lines require
$\mgtot>\mgism$ and most of these sight lines are
characterized by a higher extinction-to-gas ratio
(i.e., $\AV/\NH > 4.8\times10^{-22}\magni\cm^2\HH^{-1}$).

\begin{figure}
\hspace*{-10mm}
\includegraphics[width=0.9\textheight,height=0.5\textwidth]{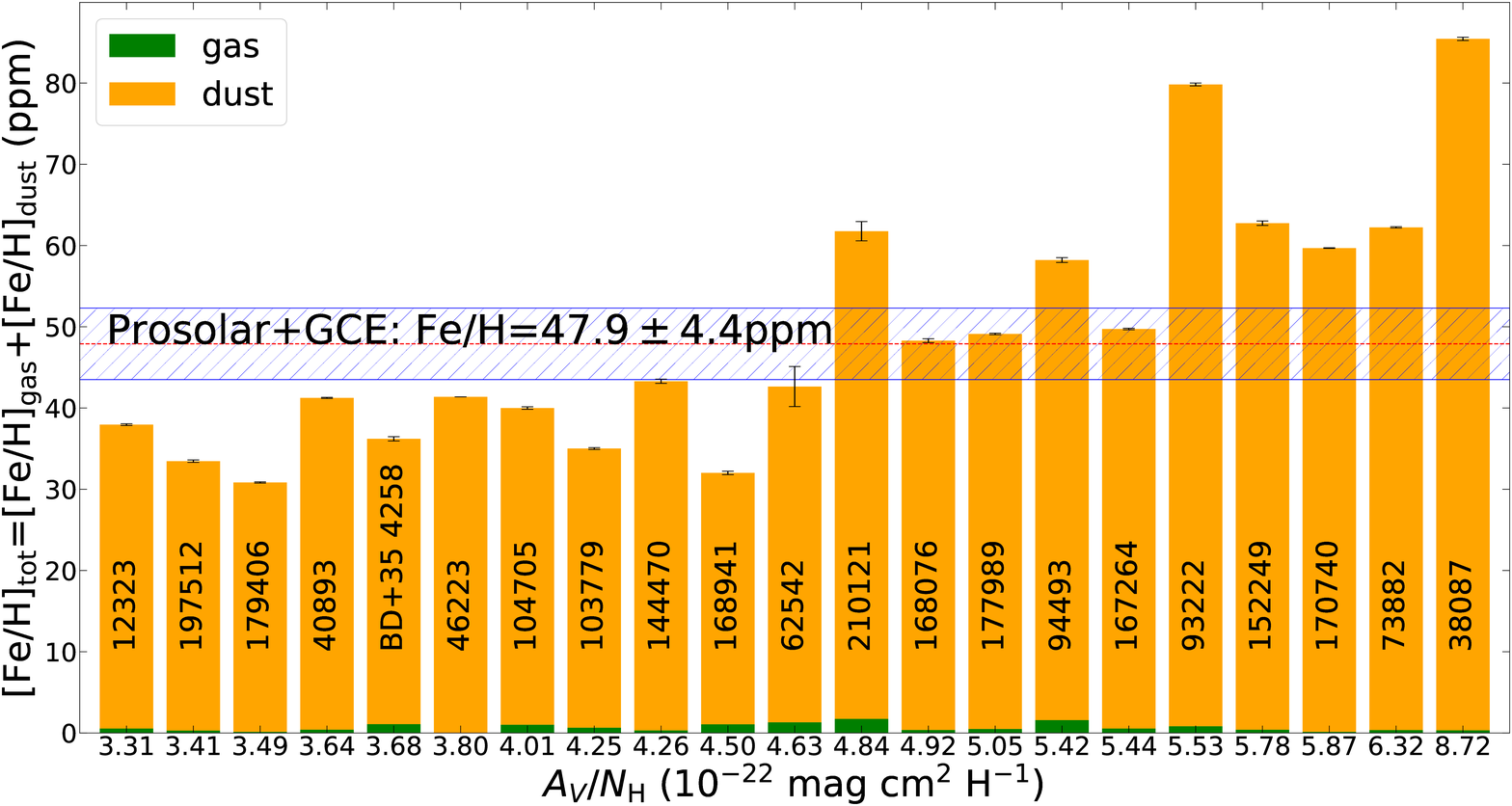}
\hspace*{-10mm}
\includegraphics[width=0.9\textheight,height=0.5\textwidth]{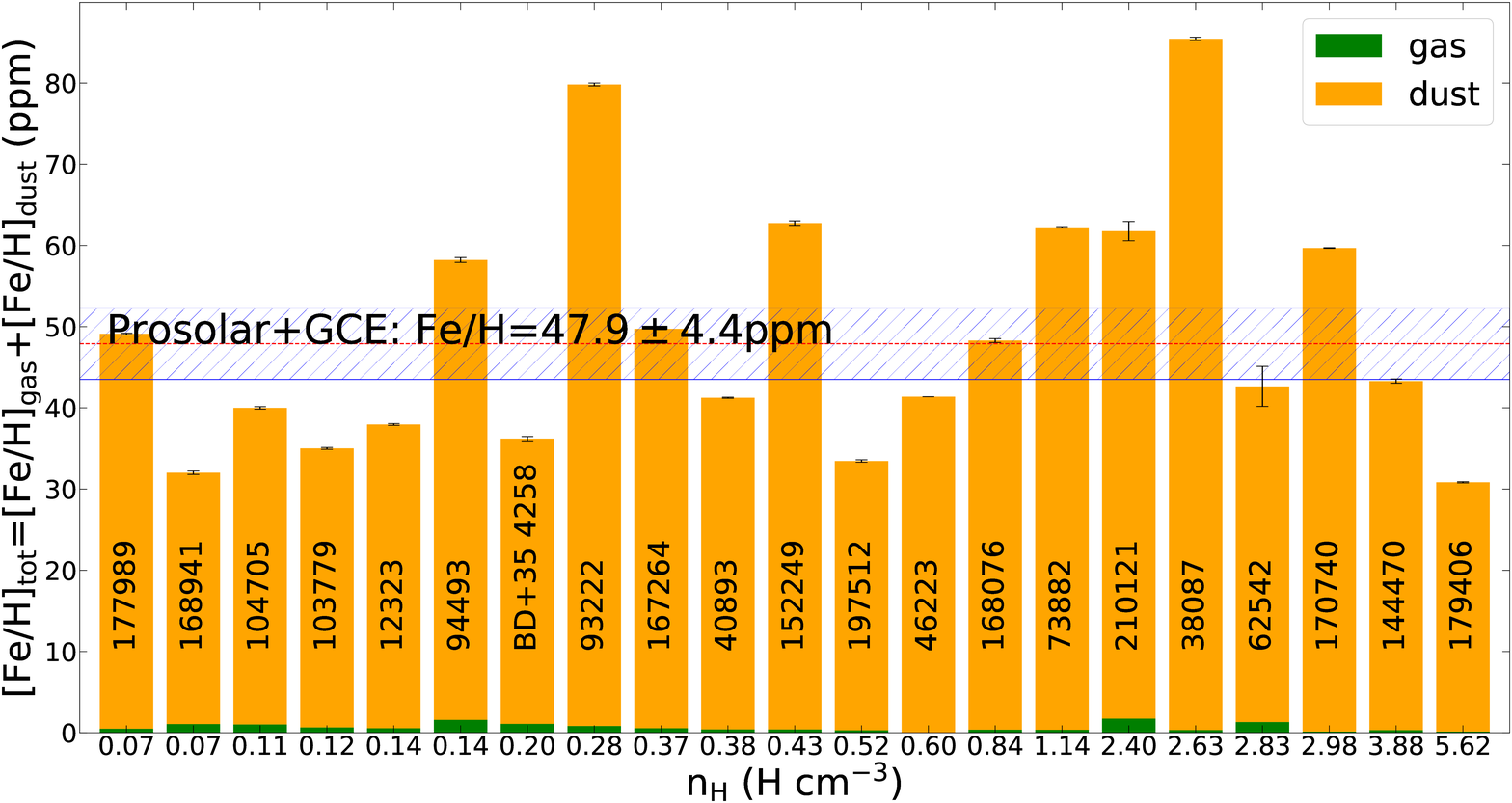}
\caption{\label{fig:Fe2H}\footnotesize
Same as Figure~\ref{fig:C2H} but for iron.
              }
\end{figure}

Figure~\ref{fig:Fe2H} compares the interstellar $\feism$
abundances with the measured gas-phase $\fegas$ abundances
and the model-derived dust-phase $\fedust$ abundances
and their combinations $\fetot$ of all the 21 sight lines
for each of which $\fegas$ has been observationally determined.
Unlike silicon and magnesium of which the gas-phase
abundances are unnegligible in the sight lines studied here,
iron is essentially completely depleted from the gas phase
in the 21 sight lines displayed in Figure~\ref{fig:Fe2H},
independent with $\AV/\NH$ or $\nH$.
While the majority of the sight lines have
$\fetot\simlt\feism$, $\simali$1/3 of the sight lines
require more iron than available to form silicate dust
in order to account for the observed extinction.

To summarize, for those sight lines with 
$\AV/\NH \simlt 4.8\times10^{-22}\magni\cm^2\HH^{-1}$,
the interstellar C/H, Si/H, Mg/H and Fe/H abundances
approximated by the GCE-augmented protosolar abundances
are sufficient to account for the gas-phase abundances
observationally measured and the dust-phase abundances
derived from the observed extinction.
However, for those sight lines with
$\AV/\NH \simgt 4.8\times10^{-22}\magni\cm^2\HH^{-1}$,
there are shortages of C, Si, Mg and Fe elements
for forming the dust to account for the observed extinction.
It is interesting to note that such an $\AV/\NH$ ratio
is close to that recently derived
for the Galactic ISM
(e.g., $\AV/\NH \approx 4.6\times10^{-22}\magni\cm^2\HH^{-1}$
for the ``gold'' sample of Zuo et al.\ (2021),
$\AV/\NH \approx 4.8\times10^{-22}\magni\cm^2\HH^{-1}$
of Zhu et al.\ (2017) for a large sample of sight lines toward
supernova remnants, planetary nebulae, and X-ray binaries),
although a nominal extinction-to-gas ratio of
$\AV/\NH \approx 5.3\times10^{-22}\magni\cm^2\HH^{-1}$
is commonly adopted for the diffuse ISM. 
A much lower ratio of 
$\AV/\NH\approx3.5\times10^{-22}\magni\cm^{2}\,\rmH^{-1}$
was derived by Lenz et al.\ (2017), 
for diffuse, low-column-density regions 
with $\NHI < 4\times 10^{20}\,\rmH\cm^{-2}$.
Perhaps the interstellar gas is not well mixed
as previously assumed so that the interstellar abundances
of heavy elements may actually have regional variations.
If this is true, a universal $\AV/\NH$ ratio is not expected.
Indeed, very recently De Cia et al.\ (2021) found large
variations in metallicity over a factor of 10 in the local
Galactic ISM. 
Finally, we explore
how the mean grain sizes derived 
in \S\ref{sec:extmod}
from fitting the observed extinction curves
vary with $R_V^{-1}$.
Let $\ameansil$ and $\ameangra$
respectively be the mean sizes 
of the silicate and graphite grains,
each weighted by grain area, grain mass,
or $V$-band extinction cross section
[$\Cext(a,\lambdaV)$].
We also derive the ``overall mean grain size'' 
as the average of $\ameansil$ and $\ameangra$,
weighted by the mass fraction 
of each dust component
in the same manner as described in
\S6 of Mishra \& Li (2017). 
Our results closely resemble that of Mishra \& Li
(2017; see their Figures 14, 15):
the area-, mass-, and $\Cext(a,\lambdaV)$-weighted 
mean grain sizes all anti-correlate with $R_V^{-1}$.
This clearly shows that {\it denser} regions
of {\it larger} $R_V$ values, on an average, are characterized 
by {\it larger} grains. The underlying physics could be related
to the rotational disruption of dust driven by radiative torques
which depends on the local conditions such as gas density
and starlight intensity (see Hoang 2019, 2021).


\section{Summary}\label{sec:summary}
We have synthesized and modeled the extinction curves
of a large number of Galactic sight lines from the near-IR
to the far-UV for which the gas-phase abundances of
at least one of the major dust-forming elements
(i.e., C, O, Si, Mg, and Fe) have been observationally
determined.
Our principal results are as follows:
\begin{enumerate}
\item The extinction curves of all sight lines 
  except HD\,93222 are closely reproduced
  from the far-UV to the near-IR by a simple mixture
  of silicate dust and graphite dust.
  The extinction curve of the sight line toward
  HD\,93222 is rather abnormal.
  Despite a large $\RV=4.76$,
  it exhibits a sharp 2175$\Angstrom$ extinction  bump
  and a steep far-UV rise at $\lambda^{-1} > 5.9\mum^{-1}$
  which are both characteristic of diffuse regions
  with $\RV<3.1$.
\item The gas-phase $\cgas$ and $\ogas$ abundances
  vary from one sight line to another, but do not
  show any systematic variations with the hydrogen
  number density $\nH$. While $\cgas$ appears to
  increase with the extinction-to-gas ratio $\AV/\NH$,
  $\ogas$ does not show any systematic variations
  with $\AV/\NH$.
\item While there are appreciable amounts of gas-phase
  Mg and Si atoms in the sight lines studied here,
  Fe is essentially completely depleted from the gas phase.
  Like $\ogas$, $\mggas$ and $\sigas$ vary from one
  sight line to another, but do not show any systematic
  variations with $\AV/\NH$ or $\nH$.
\item For those sight lines with
  $\AV/\NH \simlt 4.8\times10^{-22}\magni\cm^2\HH^{-1}$,
  the interstellar C/H, Si/H, Mg/H and Fe/H abundances
  approximated by the protosolar abundances
  augmented by Galactic chemical evolution
  are sufficient to account for the gas-phase abundances
  observationally measured and the dust-phase abundances
  derived from the observed extinction.
  In contrast, for those sight lines with
  $\AV/\NH \simgt 4.8\times10^{-22}\magni\cm^2\HH^{-1}$,
  there are shortages of C, Si, Mg and Fe elements
  for making dust to account for the observed extinction.
\item While it is generally believed that in the diffuse ISM
  a substantial fraction of the oxygen atoms
  remain unaccounted for in interstellar gas
  and dust, it is found that, for the majority
  of the lines of sight studied here, there does not
  appear to be a ``missing oxygen'' problem,
  i.e., the interstellar oxygen atoms are fully
  accommodated by gas and dust.
\end{enumerate}

\acknowledgments
WBZ and GZ are supported in part 
by the NSFC grants No.\,11988101 and No.\,11890694 
as well as the National Key R\&D Program of China 
(No.\,2019YFA0405502) and the CSST Milky Way Survey project.
We thank Dr.~B.T.~Draine, Dr.~A.~Mishra,
Dr.~A.N.~Witt and the anonymous referee
for very helpful discussions and suggestions.


\thispagestyle{empty}
\setlength{\voffset}{25mm}
\begin{deluxetable}{lcccccccccccccc}
\rotate 
\tablecolumns{15}
\tabletypesize{\tiny}
\tablewidth{0truein}
\center
\tablecaption{\footnotesize
                      \label{tab:extpara}
                      Extinction Parameters for
                      the 81 Interstellar Sight Lines
                      in Our Sample
                      }
\tablehead{
\colhead{Star}&
\colhead{$A_{V}$$^a$}&
\colhead{$E(B-V)$$^a$}&
\colhead{$R_{V}$$^a$}&
\colhead{$A_{U}$$^b$}&
\colhead{$A_{B}$$^b$}&
\colhead{$A_{J}$$^b$}&
\colhead{$A_{H}$$^b$}&
\colhead{$A_{K}$$^b$}&

\colhead{$c_{1}^{\prime}$$^a$}&
\colhead{$c_{2}^{\prime}$$^a$}&
\colhead{$c_{3}^{\prime}$$^a$}&
\colhead{$c_{4}^{\prime}$$^a$}&
\colhead{$x_{0}$$^a$}&
\colhead{$\gamma$$^a$}

\\
\colhead{}&

\colhead{(mag)}&
\colhead{(mag)}&
\colhead{}&

\colhead{(mag)}&
\colhead{(mag)}&
\colhead{(mag)}&
\colhead{(mag)}&
\colhead{(mag)}&

\colhead{}&
\colhead{}&
\colhead{}&
\colhead{}&
\colhead{($\um^{-1}$)}&
\colhead{($\um^{-1}$)}

}
\startdata
  BD+35\,4258$^1$&0.67$\pm$0.24&0.25$\pm$0.04&2.68$\pm$0.53&1.06&0.90&0.20&0.08&0.08&1.334$\pm$0.474&0.139$\pm$0.039&0.421$\pm$0.124&0.088$\pm$0.035&4.618$\pm$0.017&0.651$\pm$0.022\\
BD+53\,2820$^1$&0.88$\pm$0.15&0.29$\pm$0.03&3.04$\pm$0.32&1.36&1.18&0.26&0.08&0.10&1.372$\pm$0.621&0.156$\pm$0.026&0.682$\pm$0.107&0.255$\pm$0.063&4.591$\pm$0.015&0.732$\pm$0.041\\
HD\,1383$^2$&1.30$\pm$0.14&0.47$\pm$0.04&2.77$\pm$0.19&2.05&1.70&--&--&--&1.140$\pm$0.247&0.226$\pm$0.024&1.235$\pm$0.168&0.187$\pm$0.033&4.604$\pm$0.007&0.910$\pm$0.030\\
HD\,12323$^1$&0.63$\pm$0.14&0.23$\pm$0.04&2.75$\pm$0.41&0.99&0.85&0.15&0.13&0.07&0.669$\pm$0.188&0.403$\pm$0.081&1.211$\pm$0.279&0.236$\pm$0.071&4.577$\pm$0.015&0.883$\pm$0.030\\
HD\,13268$^2$&1.09$\pm$0.15&0.36$\pm$0.04&3.02$\pm$0.24&1.71&1.45&--&--&--&0.959$\pm$0.740&0.284$\pm$0.039&0.735$\pm$0.143&0.313$\pm$0.061&4.577$\pm$0.011&0.756$\pm$0.025\\
HD\,14434$^2$&1.23$\pm$0.11&0.48$\pm$0.04&2.57$\pm$0.16&1.93&1.68&--&--&--&0.632$\pm$0.154&0.408$\pm$0.044&0.935$\pm$0.137&0.178$\pm$0.042&4.600$\pm$0.017&0.862$\pm$0.030\\
HD\,24912 $$&1.00$\pm$0.21&0.35$\pm$0.04&2.86$\pm$0.51&--&--&--&--&--&1.187$\pm$0.728&0.270$\pm$0.054&0.942$\pm$0.219&0.050$\pm$0.024&4.541$\pm$0.016&0.846$\pm$0.028\\
HD\,25443$^1$&1.35$\pm$0.13&0.54$\pm$0.04&2.51$\pm$0.16&2.18&1.86&0.24&0.17&0.06&0.481$\pm$0.293&0.383$\pm$0.068&2.234$\pm$0.524&0.153$\pm$0.023&4.584$\pm$0.013&1.081$\pm$0.081\\
HD\,27778$^1$&0.91$\pm$0.14&0.35$\pm$0.04&2.59$\pm$0.24&1.58&1.25&0.19&0.11&0.07&1.421$\pm$0.252&0.232$\pm$0.038&0.878$\pm$0.180&0.386$\pm$0.062&4.603$\pm$0.012&0.974$\pm$0.032\\
HD\,30614 $$&0.87$\pm$0.16&0.29$\pm$0.04&3.01$\pm$0.33&--&--&--&--&--&1.099$\pm$0.288&0.200$\pm$0.032&0.748$\pm$0.146&0.127$\pm$0.030&4.570$\pm$0.011&0.900$\pm$0.030\\
HD\,37021 $$&2.80$\pm$0.17&0.48$\pm$0.02&5.84$\pm$0.26&--&--&--&--&--&1.063$\pm$1.047&0.020$\pm$0.008&0.235$\pm$0.043&0.007$\pm$0.007&4.584$\pm$0.047&1.081$\pm$0.036\\
HD\,37061$^1$&2.40$\pm$0.21&0.56$\pm$0.04&4.29$\pm$0.21&3.33&2.93&0.72&0.45&0.30&1.544$\pm$0.145&0.000$\pm$0.100&0.310$\pm$0.042&0.050$\pm$0.012&4.574$\pm$0.014&0.901$\pm$0.029\\
HD\,37367$^1$&1.49$\pm$0.24&0.42$\pm$0.04&3.55$\pm$0.44&2.33&1.94&0.42&0.28&0.16&1.424$\pm$0.304&0.153$\pm$0.024&0.989$\pm$0.172&0.134$\pm$0.025&4.598$\pm$0.005&0.865$\pm$0.023\\
HD\,37903$^1$&1.31$\pm$0.18&0.35$\pm$0.04&3.74$\pm$0.31&1.88&1.64&0.38&0.29&0.14&1.540$\pm$0.332&0.082$\pm$0.015&0.652$\pm$0.121&0.183$\pm$0.035&4.616$\pm$0.007&0.923$\pm$0.028\\
HD\,38087$^1$&1.48$\pm$0.24&0.30$\pm$0.04&4.93$\pm$0.44&1.92&1.74&0.60&0.35&0.21&1.632$\pm$0.254&0.007$\pm$0.002&0.615$\pm$0.114&0.110$\pm$0.024&4.572$\pm$0.007&0.870$\pm$0.029\\
HD\,40893$^1$&1.32$\pm$0.15&0.47$\pm$0.04&2.81$\pm$0.19&2.14&1.81&0.35&0.23&0.14&1.155$\pm$0.331&0.275$\pm$0.027&0.820$\pm$0.105&0.134$\pm$0.024&4.600$\pm$0.006&0.821$\pm$0.024\\
HD\,42087 $$&1.17$\pm$0.22&0.37$\pm$0.04&3.16$\pm$0.45&--&--&--&--&--&1.078$\pm$0.683&0.260$\pm$0.040&1.409$\pm$0.205&0.299$\pm$0.048&4.609$\pm$0.003&1.070$\pm$0.020\\
HD\,43818 $$&1.80$\pm$0.16&0.52$\pm$0.04&3.46$\pm$0.19&--&--&--&--&--&0.892$\pm$0.149&0.256$\pm$0.021&0.924$\pm$0.088&0.109$\pm$0.012&4.575$\pm$0.004&0.849$\pm$0.015\\
HD\,46056$^1$&1.30$\pm$0.15&0.49$\pm$0.04&2.66$\pm$0.21&2.06&1.75&0.35&0.23&0.13&0.759$\pm$0.240&0.330$\pm$0.039&1.194$\pm$0.185&0.205$\pm$0.034&4.581$\pm$0.010&0.911$\pm$0.028\\
HD\,46202$^1$&1.43$\pm$0.17&0.48$\pm$0.04&2.98$\pm$0.23&2.19&1.89&0.36&0.22&0.13&0.833$\pm$0.261&0.281$\pm$0.032&1.075$\pm$0.162&0.170$\pm$0.027&4.580$\pm$0.006&0.911$\pm$0.029\\
HD\,46223 $$&1.48$\pm$0.14&0.54$\pm$0.04&2.73$\pm$0.15&--&--&--&--&--&0.751$\pm$0.138&0.331$\pm$0.029&1.102$\pm$0.125&0.222$\pm$0.030&4.607$\pm$0.008&0.939$\pm$0.025\\
HD\,52266$^1$&0.91$\pm$0.16&0.29$\pm$0.04&3.14$\pm$0.37&1.36&1.19&0.22&0.10&0.09&0.748$\pm$0.203&0.310$\pm$0.049&0.973$\pm$0.193&0.008$\pm$0.007&4.598$\pm$0.015&0.944$\pm$0.031\\
HD\,62542$^1$&0.99$\pm$0.14&0.35$\pm$0.06&2.82$\pm$0.24&1.73&1.35&0.21&0.13&0.08&0.517$\pm$0.185&0.470$\pm$0.071&1.044$\pm$0.280&0.470$\pm$0.074&4.543$\pm$0.029&1.304$\pm$0.044\\
HD\,69106 $$&0.61$\pm$0.15&0.20$\pm$0.04&3.05$\pm$0.44&--&--&--&--&--&1.267$\pm$0.650&0.095$\pm$0.029&0.612$\pm$0.188&0.132$\pm$0.053&4.588$\pm$0.023&0.957$\pm$0.032\\
HD\,72648$^1$&1.31$\pm$0.17&0.38$\pm$0.04&3.44$\pm$0.27&1.96&1.68&0.33&0.22&0.11&1.480$\pm$0.429&0.102$\pm$0.017&1.136$\pm$0.174&0.176$\pm$0.035&4.585$\pm$0.009&0.970$\pm$0.031\\
HD\,73882$^1$&2.46$\pm$0.17&0.69$\pm$0.04&3.56$\pm$0.13&3.69&3.17&0.61&0.34&0.26&1.163$\pm$0.199&0.192$\pm$0.013&0.576$\pm$0.061&0.167$\pm$0.014&4.599$\pm$0.006&1.037$\pm$0.032\\
HD\,75309$^1$&1.02$\pm$0.18&0.29$\pm$0.04&3.53$\pm$0.40&1.54&1.32&0.28&0.19&0.12&0.997$\pm$0.100&0.216$\pm$0.034&0.935$\pm$0.198&0.100$\pm$0.032&4.598$\pm$0.014&0.926$\pm$0.031\\
HD\,79186 $$&1.28$\pm$0.26&0.40$\pm$0.04&3.21$\pm$0.56&--&--&--&--&--&1.132$\pm$0.663&0.259$\pm$0.060&1.174$\pm$0.302&0.259$\pm$0.075&4.569$\pm$0.016&0.998$\pm$0.034\\
HD\,89137 $$&0.72$\pm$0.13&0.27$\pm$0.04&2.68$\pm$0.28&--&--&--&--&--&1.531$\pm$0.617&0.279$\pm$0.081&0.780$\pm$0.255&0.175$\pm$0.072&4.636$\pm$0.027&0.841$\pm$0.028\\
HD\,91824$^1$&0.80$\pm$0.28&0.24$\pm$0.07&3.35$\pm$0.62&1.18&1.06&0.15&0.10&0.09&1.554$\pm$0.724&0.180$\pm$0.054&0.750$\pm$0.229&0.228$\pm$0.084&4.622$\pm$0.021&0.871$\pm$0.029\\
HD\,91983$^1$&0.84$\pm$0.15&0.29$\pm$0.04&2.89$\pm$0.34&1.23&1.10&--&--&--&1.123$\pm$0.480&0.207$\pm$0.035&1.090$\pm$0.233&0.203$\pm$0.055&4.612$\pm$0.015&0.978$\pm$0.033\\
HD\,93129 $$&1.75$\pm$0.39&0.48$\pm$0.10&3.65$\pm$0.42&--&--&--&--&--&1.865$\pm$0.300&0.237$\pm$0.035&0.810$\pm$0.139&0.156$\pm$0.029&4.606$\pm$0.009&0.990$\pm$0.031\\
HD\,93205 $$&1.23$\pm$0.16&0.38$\pm$0.04&3.25$\pm$0.24&--&--&--&--&--&0.933$\pm$0.505&0.254$\pm$0.031&0.744$\pm$0.157&0.171$\pm$0.058&4.614$\pm$0.043&0.959$\pm$0.033\\
HD\,93222$^1$&1.71$\pm$0.33&0.36$\pm$0.06&4.76$\pm$0.48&2.23&2.05&0.55&0.31&0.20&1.199$\pm$0.514&0.123$\pm$0.018&0.316$\pm$0.048&0.099$\pm$0.026&4.580$\pm$0.013&0.734$\pm$0.043\\
HD\,93843 $$&1.05$\pm$0.20&0.27$\pm$0.05&3.89$\pm$0.41&--&--&--&--&--&1.369$\pm$0.504&0.149$\pm$0.035&0.446$\pm$0.115&0.175$\pm$0.058&4.572$\pm$0.026&0.780$\pm$0.027\\
HD\,94493 $$&0.82$\pm$0.18&0.23$\pm$0.04&3.57$\pm$0.45&--&--&--&--&--&1.043$\pm$0.705&0.121$\pm$0.031&0.429$\pm$0.131&0.105$\pm$0.034&4.595$\pm$0.017&0.835$\pm$0.028\\
HD\,99953 $$&1.77$\pm$0.27&0.48$\pm$0.06&3.69$\pm$0.30&--&--&--&--&--&1.139$\pm$0.402&0.169$\pm$0.022&0.964$\pm$0.148&0.148$\pm$0.034&4.615$\pm$0.014&1.005$\pm$0.033\\
HD\,101190$^2$&0.92$\pm$0.12&0.37$\pm$0.04&2.48$\pm$0.21&1.40&1.25&--&--&--&0.405$\pm$0.094&0.399$\pm$0.057&1.295$\pm$0.231&0.208$\pm$0.047&4.625$\pm$0.015&1.078$\pm$0.035\\
HD\,103779$^2$&0.69$\pm$0.15&0.21$\pm$0.04&3.29$\pm$0.43&0.95&0.90&--&--&--&1.473$\pm$0.408&0.153$\pm$0.052&1.029$\pm$0.312&0.233$\pm$0.069&4.540$\pm$0.016&0.886$\pm$0.030\\
HD\,104705$^1$&0.65$\pm$0.24&0.23$\pm$0.07&2.81$\pm$0.57&0.91&0.80&0.18&0.10&0.07&1.037$\pm$2.074&0.217$\pm$0.050&1.155$\pm$0.316&0.166$\pm$0.067&4.569$\pm$0.017&0.943$\pm$0.031\\
HD\,111934 $$&1.25$\pm$0.18&0.51$\pm$0.06&2.45$\pm$0.20&--&--&--&--&--&1.263$\pm$0.346&0.178$\pm$0.031&0.829$\pm$0.154&0.141$\pm$0.029&4.593$\pm$0.005&0.817$\pm$0.021\\
HD\,116852 $$&0.51$\pm$0.12&0.21$\pm$0.04&2.42$\pm$0.37&--&--&--&--&--&0.518$\pm$0.249&0.376$\pm$0.103&0.633$\pm$0.173&0.010$\pm$0.015&4.548$\pm$0.041&0.782$\pm$0.069\\
HD\,122879$^2$&1.13$\pm$0.20&0.36$\pm$0.05&3.15$\pm$0.30&1.69&1.45&--&--&--&1.321$\pm$0.299&0.233$\pm$0.040&1.243$\pm$0.230&0.190$\pm$0.039&4.581$\pm$0.004&0.831$\pm$0.021\\
HD\,124979$^2$&1.05$\pm$0.10&0.38$\pm$0.03&2.75$\pm$0.19&1.81&1.43&--&--&--&1.308$\pm$0.368&0.211$\pm$0.038&0.930$\pm$0.194&0.355$\pm$0.071&4.579$\pm$0.017&0.824$\pm$0.027\\
HD\,144470$^1$&0.74$\pm$0.09&0.22$\pm$0.02&3.37$\pm$0.29&1.08&0.94&0.25&--&0.09&1.418$\pm$0.554&0.099$\pm$0.021&0.867$\pm$0.185&0.111$\pm$0.043&4.555$\pm$0.011&0.808$\pm$0.028\\
HD\,147165$^1$&1.47$\pm$0.23&0.38$\pm$0.03&3.86$\pm$0.52&2.21&1.88&0.40&0.25&0.15&1.562$\pm$0.302&0.042$\pm$0.011&0.633$\pm$0.129&0.023$\pm$0.007&4.612$\pm$0.011&0.887$\pm$0.029\\
HD\,147888$^1$&1.99$\pm$0.18&0.51$\pm$0.04&3.89$\pm$0.20&2.85&2.48&0.60&0.35&0.22&1.471$\pm$0.267&0.037$\pm$0.012&0.665$\pm$0.100&0.087$\pm$0.022&4.587$\pm$0.013&0.879$\pm$0.029\\
HD\,148422$^2$&0.88$\pm$0.16&0.29$\pm$0.04&3.02$\pm$0.33&1.35&1.17&--&--&--&0.401$\pm$0.114&0.391$\pm$0.070&0.644$\pm$0.143&0.192$\pm$0.048&4.601$\pm$0.014&0.776$\pm$0.025\\
HD\,149757$^1$&0.82$\pm$0.13&0.32$\pm$0.04&2.55$\pm$0.24&1.26&1.09&0.27&0.14&0.10&1.002$\pm$0.100&0.286$\pm$0.037&1.872$\pm$0.313&0.215$\pm$0.052&4.552$\pm$0.010&1.186$\pm$0.042\\
HD\,151805$^2$&1.42$\pm$0.21&0.43$\pm$0.05&3.29$\pm$0.30&2.28&1.84&--&--&--&1.193$\pm$0.453&0.178$\pm$0.023&0.602$\pm$0.081&0.114$\pm$0.028&4.614$\pm$0.011&0.807$\pm$0.041\\
HD\,152236 $$&2.24$\pm$0.26&0.60$\pm$0.03&3.73$\pm$0.39&--&--&--&--&--&0.764$\pm$0.665&0.258$\pm$0.049&1.291$\pm$0.369&0.150$\pm$0.029&4.610$\pm$0.023&1.104$\pm$0.094\\
HD\,152249$^2$&1.63$\pm$0.40&0.46$\pm$0.10&3.54$\pm$0.45&2.49&2.09&--&--&--&1.205$\pm$0.435&0.185$\pm$0.029&0.841$\pm$0.152&0.113$\pm$0.031&4.588$\pm$0.010&0.913$\pm$0.030\\
HD\,152424 $$&2.23$\pm$0.17&0.68$\pm$0.04&3.28$\pm$0.15&--&--&--&--&--&1.478$\pm$0.268&0.133$\pm$0.020&0.785$\pm$0.117&0.148$\pm$0.027&4.587$\pm$0.011&0.865$\pm$0.028\\
HD\,154368 $$&2.53$\pm$0.20&0.76$\pm$0.05&3.33$\pm$0.15&--&--&--&--&--&1.091$\pm$0.099&0.217$\pm$0.014&1.046$\pm$0.095&0.218$\pm$0.021&4.578$\pm$0.003&0.998$\pm$0.023\\
HD\,157857$^2$&1.48$\pm$0.17&0.43$\pm$0.04&3.45$\pm$0.23&2.41&2.04&--&--&--&1.548$\pm$0.269&0.057$\pm$0.021&1.115$\pm$0.203&0.263$\pm$0.049&4.563$\pm$0.011&0.848$\pm$0.028\\
HD\,167264 $$&0.98$\pm$0.15&0.30$\pm$0.04&3.26$\pm$0.31&--&--&--&--&--&1.514$\pm$0.510&0.085$\pm$0.022&0.726$\pm$0.141&0.138$\pm$0.041&4.596$\pm$0.016&0.819$\pm$0.033\\
HD\,167402$^2$&0.71$\pm$0.17&0.21$\pm$0.04&3.38$\pm$0.49&1.08&0.93&--&--&--&1.169$\pm$0.907&0.228$\pm$0.043&0.554$\pm$0.114&0.189$\pm$0.065&4.596$\pm$0.018&0.774$\pm$0.044\\
HD\,168076$^2$&2.64$\pm$0.17&0.76$\pm$0.04&3.47$\pm$0.12&3.61&3.33&--&--&--&0.977$\pm$0.483&0.164$\pm$0.019&0.740$\pm$0.101&0.123$\pm$0.017&4.604$\pm$0.007&0.972$\pm$0.021\\
HD\,168941$^2$&0.80$\pm$0.16&0.24$\pm$0.04&3.35$\pm$0.41&1.21&1.04&--&--&--&1.438$\pm$0.712&0.113$\pm$0.020&0.708$\pm$0.128&0.164$\pm$0.058&4.535$\pm$0.012&0.758$\pm$0.040\\
HD\,170740$^1$&1.51$\pm$0.46&0.50$\pm$0.13&3.01$\pm$0.49&2.43&2.03&0.43&0.29&0.17&1.144$\pm$0.225&0.253$\pm$0.045&0.966$\pm$0.193&0.208$\pm$0.039&4.595$\pm$0.005&0.942$\pm$0.025\\
HD\,177989$^1$&0.65$\pm$0.15&0.23$\pm$0.04&2.83$\pm$0.45&1.02&0.88&0.22&0.14&0.08&0.811$\pm$0.378&0.301$\pm$0.068&1.078$\pm$0.304&0.181$\pm$0.058&4.565$\pm$0.016&0.949$\pm$0.031\\
HD\,178487$^2$&1.04$\pm$0.15&0.35$\pm$0.04&2.98$\pm$0.27&1.61&1.35&--&--&--&0.902$\pm$0.691&0.224$\pm$0.038&1.045$\pm$0.222&0.103$\pm$0.033&4.576$\pm$0.016&0.842$\pm$0.028\\
HD\,179406$^1$&0.96$\pm$0.14&0.35$\pm$0.04&2.73$\pm$0.25&1.55&1.29&0.31&0.26&0.08&1.530$\pm$0.333&0.164$\pm$0.027&1.210$\pm$0.209&0.236$\pm$0.044&4.607$\pm$0.008&0.948$\pm$0.030\\
HD\,179407$^2$&0.75$\pm$0.14&0.28$\pm$0.04&2.68$\pm$0.33&1.23&1.03&--&--&--&0.568$\pm$0.157&0.369$\pm$0.066&1.336$\pm$0.279&0.202$\pm$0.054&4.581$\pm$0.016&0.970$\pm$0.035\\
HD\,185418$^1$&1.27$\pm$0.14&0.50$\pm$0.04&2.54$\pm$0.20&2.07&1.78&0.21&0.11&0.07&1.817$\pm$0.265&0.100$\pm$0.018&1.156$\pm$0.170&0.158$\pm$0.029&4.604$\pm$0.005&0.819$\pm$0.024\\
HD\,192639$^2$&1.91$\pm$0.16&0.61$\pm$0.04&3.14$\pm$0.16&2.98&2.50&--&--&--&1.248$\pm$0.238&0.190$\pm$0.020&1.008$\pm$0.130&0.148$\pm$0.027&4.575$\pm$0.010&0.866$\pm$0.029\\
HD\,197512$^1$&0.94$\pm$0.30&0.33$\pm$0.09&2.84$\pm$0.50&1.62&1.33&0.16&0.06&0.08&0.773$\pm$0.248&0.339$\pm$0.067&1.593$\pm$0.345&0.180$\pm$0.046&4.573$\pm$0.006&1.029$\pm$0.028\\
HD\,198478 $$&1.48$\pm$0.17&0.57$\pm$0.04&2.60$\pm$0.19&--&--&--&--&--&1.257$\pm$0.419&0.259$\pm$0.027&1.456$\pm$0.182&0.251$\pm$0.039&4.567$\pm$0.011&1.014$\pm$0.032\\
HD\,198781$^1$&0.75$\pm$0.13&0.35$\pm$0.04&2.14$\pm$0.28&1.22&1.04&0.10&0.07&0.06&0.663$\pm$0.381&0.474$\pm$0.090&1.912$\pm$0.506&0.271$\pm$0.054&4.590$\pm$0.013&1.125$\pm$0.067\\
HD\,199579$^1$&1.14$\pm$0.28&0.36$\pm$0.04&3.17$\pm$0.69&1.82&1.51&0.27&0.14&0.11&1.099$\pm$0.552&0.279$\pm$0.065&0.807$\pm$0.211&0.202$\pm$0.059&4.593$\pm$0.013&0.986$\pm$0.033\\
HD\,203532$^1$&0.94$\pm$0.11&0.28$\pm$0.03&3.37$\pm$0.24&1.56&1.26&0.27&0.12&0.08&0.758$\pm$0.165&0.267$\pm$0.034&1.502$\pm$0.246&0.204$\pm$0.036&4.599$\pm$0.011&1.266$\pm$0.040\\
HD\,206267$^2$&1.47$\pm$0.14&0.52$\pm$0.04&2.82$\pm$0.16&2.38&2.01&--&--&--&1.170$\pm$0.362&0.274$\pm$0.025&1.021$\pm$0.134&0.224$\pm$0.036&4.590$\pm$0.011&0.906$\pm$0.028\\
HD\,206773$^2$&1.99$\pm$0.21&0.45$\pm$0.04&4.42$\pm$0.26&2.72&2.50&--&--&--&1.018$\pm$0.283&0.154$\pm$0.014&0.504$\pm$0.078&0.049$\pm$0.015&4.583$\pm$0.016&0.893$\pm$0.030\\
HD\,207198$^2$&1.50$\pm$0.29&0.54$\pm$0.08&2.77$\pm$0.35&2.43&1.96&--&--&--&0.811$\pm$0.259&0.344$\pm$0.050&0.976$\pm$0.169&0.277$\pm$0.045&4.596$\pm$0.006&0.883$\pm$0.024\\
HD\,209339$^1$&1.00$\pm$0.20&0.36$\pm$0.07&2.78$\pm$0.34&1.52&1.32&--&--&--&1.156$\pm$0.304&0.238$\pm$0.039&0.989$\pm$0.191&0.080$\pm$0.021&4.603$\pm$0.007&0.875$\pm$0.027\\
HD\,210121$^1$&0.75$\pm$0.15&0.31$\pm$0.05&2.42$\pm$0.29&--&1.12&0.15&0.07&0.06&0.061$\pm$0.025&0.716$\pm$0.142&0.940$\pm$0.339&0.520$\pm$0.107&4.516$\pm$0.031&0.929$\pm$0.033\\
HD\,210809 $$&1.05$\pm$0.17&0.31$\pm$0.04&3.39$\pm$0.32&--&--&--&--&--&0.971$\pm$0.456&0.273$\pm$0.039&0.710$\pm$0.149&0.181$\pm$0.050&4.568$\pm$0.019&0.844$\pm$0.030\\
HD\,210839 $$&1.15$\pm$0.18&0.57$\pm$0.04&2.02$\pm$0.26&--&--&--&--&--&0.663$\pm$0.389&0.454$\pm$0.079&1.552$\pm$0.372&0.175$\pm$0.046&4.599$\pm$0.019&0.964$\pm$0.059\\
HD\,220057$^1$&0.62$\pm$0.20&0.23$\pm$0.06&2.71$\pm$0.49&1.06&0.89&0.08&0.05&0.07&1.090$\pm$1.933&0.215$\pm$0.041&1.214$\pm$0.246&0.222$\pm$0.080&4.617$\pm$0.018&0.938$\pm$0.071\\
HD\,232522 $$&0.82$\pm$0.16&0.27$\pm$0.04&3.05$\pm$0.41&--&--&--&--&--&0.594$\pm$0.123&0.378$\pm$0.067&1.063$\pm$0.223&0.229$\pm$0.058&4.555$\pm$0.009&0.934$\pm$0.031\\
HD\,303308 $$&1.36$\pm$0.17&0.45$\pm$0.05&3.02$\pm$0.21&--&--&--&--&--&0.865$\pm$0.213&0.263$\pm$0.027&0.905$\pm$0.132&0.155$\pm$0.027&4.588$\pm$0.008&0.945$\pm$0.030\\
\enddata 
\\
\flushleft{$^{a}$ Data taken from Valencic et al.\ (2004). \\
$^{b}$ U, B, J, H and K extinction data taken from
Fitzpatrick \& Massa (2007) for those sight lines
marked by ``1'' and from Gordon et al.\ (2009)
for those marked by ``2''. }
\end{deluxetable}

\thispagestyle{empty}
\setlength{\voffset}{25mm}
\begin{deluxetable}{llclcccccc}
	\rotate 
	\tablecolumns{10}
	\tabletypesize{\tiny}
	\tablewidth{0truein}
	\center
	\tablecaption{Hydrogen Densities and
         Gas-Phase C, O, Mg, Si and Fe Abundances
         of the 81 Interstellar Sight Lines in Our Sample
	\label{tab:GasAbund}
	}
	\tablehead{
		\colhead{Star}&
		\colhead{$N_{\rm H}$}&
		\colhead{$f({\rm H_{2}})$}&
		\colhead{$N({\rm H I})$}&
		\colhead{$N({\rm H_{2}})$}&
		\colhead{[C/H]$\rm_{gas}$}&
		\colhead{[O/H]$\rm_{gas}$}&
		\colhead{[Mg/H]$\rm_{gas}$}&
		\colhead{[Si/H]$\rm_{gas}$}&
		\colhead{[Fe/H]$\rm_{gas}$}
		\\
		\colhead{}&
		\colhead{($10^{21}\cm^{-2}$)}&
		\colhead{}&
		\colhead{($10^{21}\cm^{-2}$)}&
		\colhead{($10^{21}\cm^{-2}$)}&
		\colhead{(ppm)}&
		\colhead{(ppm)}&
		\colhead{(ppm)}&
		\colhead{(ppm)}&
		\colhead{(ppm)}
	        }
	\startdata
	
		BD+35\,4258 &1.82$_{-0.29}^{+0.13}$(1)&0.04&1.74$_{-0.28}^{+0.12}$(1)&0.04$\pm${0.01}(1)&--&195.1$\pm$64.3(1)&10.00$\pm$1.15(1)&--&1.10$\pm$0.26(6)\\
	BD+53\,2820 &2.51$_{-0.35}^{0.29}$(1)&0.1&2.24$_{-0.36}^{+0.26}$(1)&0.13$_{-0.04}^{+0.03}$(1)&--&389.0$\pm$92.2(1)&9.32$\pm$1.68(1)&--&--\\
	HD\,1383 &3.47$_{-0.40}^{+0.32}$(1)&0.18&2.88$_{-0.40}^{+0.33}$(1)&3.09$\pm${0.05}(1)&--&407.2$\pm$71.2(3)&7.59$\pm$0.78(1)&4.07$\pm$374.92(4)&0.15$\pm$49.03(4)\\
	HD\,12323 &1.91$\pm${0.18}(1)&0.19&1.55$_{-0.18}^{+0.14}$(1)&0.18$_{-0.05}^{+0.03}$(1)&--&629.8$\pm$104.7(1)&7.24$\pm$0.83(1)&--&0.55$\pm$0.11(6)\\
	HD\,13268 &2.75$\pm${0.38}(1)&0.21&2.19$\pm${0.35}(1)&0.29$\pm${0.05}(1)&--&457.5$\pm$105.3(1)&9.99$\pm$1.95(1)&--&--\\
	HD\,14434 &2.95$\pm${0.52}(23)&0.2&2.34$\pm${0.54}(3)&0.30$\pm${0.05}(3)&--&513.0$\pm$138.6(3)&6.31$\pm$1.36(23)&--&--\\
	HD\,24912 &1.98$\pm${0.54}(2)&0.34&1.20$\pm${0.18}(3)&0.34$\pm${0.07}(3)&163.1$\pm$86.6(34)&326.1$\pm$97.5(3)&1.99$\pm$0.73(2)&1.61$\pm$0.44(2)&0.92$\pm$0.25(13)\\
	HD\,25443 &3.63$\pm${0.42}(1)&0.46&1.95$_{-0.23}^{+0.27}$(1)&0.83$\pm${0.15}(1)&--&363.6$\pm$59.2(1)&4.08$\pm$0.48(1)&--&--\\
	HD\,27778 &2.51$\pm${0.44}(7)&0.49&0.89(5)&0.62(10)&79.2$\pm$27.6(14)&269.1$\pm$53.8(3)&1.05$\pm$0.22(7)&3.43$\pm$0.60(9)&0.10$\pm$0.02(9)\\
	HD\,30614 &1.23$\pm${0.28}(25)&0.36&0.93$\pm${0.21}(27)&0.22$\pm${0.09}(13)&--&723.6$\pm$236.0(24)&18.63$\pm$7.74(13)&12.59$\pm$16.26(4)&0.66$\pm$0.40(4)\\
	HD\,37021 &4.79$\pm${1.50}(7)&--&4.79$\pm${1.50}(3)&--&90.9$\pm$37.9(14)&257.0$\pm$82.6(3)&1.78$\pm$0.57(7)&2.99$\pm$1.15(9)&0.58$\pm$0.19(9)\\
	HD\,37061 &5.37$\pm${1.20}(7)&--&5.37$\pm${1.20}(3)&--&98.1$\pm$35.5(14)&316.2$\pm$72.2(3)&1.12$\pm$0.26(7)&0.19$\pm$0.17(4)&0.06$\pm$0.08(4)\\
	HD\,37367 &2.57$\pm${0.45}(23)&0.32&1.91$\pm${0.44}(3)&0.41$\pm${0}(29)&--&446.7$\pm$118.9(3)&3.89$\pm$0.78(23)&--&--\\
	HD\,37903 &3.16$\pm${0.47}(6)&0.53&1.45$\pm${0.33}(3)&0.83$\pm${0.12}(17)&332.1$\pm$49.4(15)&239.9$\pm$37.4(3)&1.12$\pm$0.26(7)&--&0.17$\pm$0.06(6)\\
	HD\,38087 &1.70$\pm${0.81}(6)&0.51&0.81$\pm${0.81}(6)&0.44$\pm${0.08}(17)&--&676.1$\pm$376.5(19)&--&--&0.33$\pm$0.20(6)\\
	HD\,40893 &3.63$_{-0.42}^{+0.25}$(1)&0.21&2.88$_{-0.40}^{+0.27}$(1)&0.39$\pm${0.03}(1)&--&363.6$\pm$41.8(1)&5.51$\pm$0.40(1)&--&0.42$\pm$0.07(6)\\
	HD\,42087 &3.09$\pm${0.71}(6)&0.21&0.98$\pm${0.22}(27)&0.33$\pm${0.11}(17)&--&1096.4$\pm$605.1(19)&--&1.41$\pm$3.25(4)&0.14$\pm$0.13(4)\\
	HD\,43818 &5.13$\pm${1.50}(23)&--&3.98$\pm${1.30}(3)&--&--&323.7$\pm$97.5(3)&5.89$\pm$1.81(23)&--&--\\
	HD\,46056 &3.39$\pm${0.98}(19)&0.28&1.38$\pm${0.48}(27)&0.48$\pm${0.07}(17)&--&467.8$\pm$181.5(19)&--&1.66$\pm$32.47(4)&0.32$\pm$0.12(6)\\
	HD\,46202 &4.79$\pm${1.70}(19)&0.2&0.692$\pm${0.18}(27)&0.48$\pm${0.08}(17)&--&363.1$\pm$206.2(19)&--&17.78$\pm$71.32(4)&0.22$\pm$9.82(4)\\
	HD\,46223 &3.89$_{-0.18}^{+0.27}$(1)&0.24&2.88$_{-0.13}^{+0.27}$(1)&0.47$_{-0.05}^{+0.07}$(1)&--&331.6$\pm$38.1(1)&7.07$\pm$1.39(1)&0.89$\pm$0.44(4)&0.03$\pm$0.00(4)\\
	HD\,52266 &1.86$_{-0.21}^{+0.17}$(1)&0.11&1.66$_{-0.19}^{+0.15}$(1)&0.10$\pm${0.02}(1)&--&416.7$\pm$85.9(1)&6.28$\pm$0.60(1)&--&--\\
	HD\,62542 &2.14$\pm${1.00}(19)&0.6&0.79$\pm${0}(5)&0.65$\pm${0.40}(16)&--&125.9$\pm$355.7(19)&--&--&1.32$\pm$2.46(28)\\
	HD\,69106 &1.29$\pm${0.12}(1)&0.09&1.17$\pm${0.11}(1)&0.06$\pm${0.01}(1)&--&315.9$\pm$46.6(1)&6.02$\pm$0.62(1)&1.99$\pm$0.49(4)&0.40$\pm$0.27(4)\\
	HD\,72648 &2.57$\pm${0.41}(1)&0.39&1.55$_{-0.21}^{+0.25}$(1)&0.50$\pm${0.17}(1)&--&301.9$\pm$96.6(1)&3.39$\pm$0.59(1)&--&--\\
	HD\,73882 &3.89$\pm${0.68}(19)&0.66&1.29(57)&1.29$\pm${0.26}(33)&--&40.7$\pm$216.0(19)&--&--&0.36$\pm$0.08(6)\\
	HD\,75309 &1.55$_{-0.18}^{+0.11}$(1)&0.19&1.26$_{-0.17}^{+0.09}$(1)&0.15$_{-0.03}^{+0.02}$(1)&--&339.0$\pm$74.0(1)&5.62$\pm$0.41(1)&--&--\\
	HD\,79186 &2.63$\pm${0.46}(23)&0.4&1.59$\pm${0.36}(3)&0.53$\pm${0.12}(3)&--&416.7$\pm$84.0(3)&4.90$\pm$0.97(23)&--&--\\
	HD\,89137 &1.29$_{-0.06}^{+0.18}$(1)&0.16&1.07$_{-0.05}^{+0.17}$(1)&0.11$\pm${0.02}(1)&--&388.9$\pm$89.6(1)&8.15$\pm$1.14(1)&--&--\\
	HD\,91824 &1.45$\pm${0.13}(1)&0.09&1.32$\pm${0.12}(1)&0.07$_{-0.02}^{+0.01}$(1)&145.3$\pm$14.0(15)&691.8$\pm$104.6(3)&10.93$\pm$1.13(1)&--&--\\
	HD\,91983 &1.66$_{-0.15}^{+0.19}$(1)&0.15&1.41$_{-0.16}^{+0.20}$(1)&0.13$\pm${0.02}(1)&--&562.2$\pm$101.1(1)&11.51$\pm$1.54(1)&--&--\\
	HD\,93129 &3.31$_{-0.31}^{+0.46}$(1)&0.1&2.95$_{-0.27}^{+0.48}$(1)&0.16$_{-0.02}^{+0.03}$(1)&--&456.0$\pm$75.8(1)&7.07$\pm$0.99(1)&--&--\\
	HD\,93205 &2.40$\pm${0.28}(1)&0.04&2.29$\pm${0.26}(1)&0.05$_{-0.01}^{+0.02}$(1)&--&375.2$\pm$49.9(20)&10.46$\pm$1.30(1)&4.90$\pm$1.80(4)&0.55$\pm$0.37(4)\\
	HD\,93222 &3.09$_{-0.29}^{+0.21}$(1)&0.04&2.95$_{-0.27}^{+0.20}$(1)&0.059$\pm${0.01}(1)&--&436.9$\pm$35.9(20)&10.23$\pm$0.85(1)&--&0.83$\pm$0.18(6)\\
	HD\,93843 &2.09$_{-0.19}^{+0.24}$(1)&0.04&2.00$_{-0.18}^{+0.23}$(1)&0.04$\pm${0.001}(1)&--&407.3$\pm$96.6(1)&8.52$\pm$1.00(1)&15.85$\pm$430.77(4)&0.93$\pm$76.58(4)\\
	HD\,94493 &1.51$_{-0.139}^{+0.174}$(1)&0.16&1.26$_{-0.15}^{+0.17}$(1)&0.12$_{-0.03}^{+0.02}$(1)&--&338.9$\pm$101.4(1)&10.97$\pm$1.36(1)&--&1.59$\pm$0.30(6)\\
	HD\,99953 &3.63$\pm${0.502}(1)&0.22&2.82$\pm${0.45}(1)&0.40$\pm${0.13}(1)&--&322.2$\pm$93.3(1)&4.79$\pm$0.70(1)&--&--\\
	HD\,101190 &2.24$_{-0.26}^{+0.16}$(1)&0.24&1.77$_{-0.25}^{+0.12}$(1)&0.27$_{-0.05}^{+0.03}$(1)&--&446.7$\pm$69.0(1)&8.71$\pm$1.00(1)&4.57$\pm$71.47(4)&0.54$\pm$36.63(4)\\
	HD\,103779 &1.62$\pm${0.19}(1)&0.1&1.48$\pm${0.17}(1)&0.08$\pm${0.02}(1)&--&389.1$\pm$108.2(1)&11.47$\pm$1.42(1)&--&0.66$\pm$0.11(6)\\
	HD\,104705 &1.62$_{-0.15}^{+0.19}$(1)&0.13&1.41$_{-0.16}^{+0.20}$(1)&0.11$\pm${0.02}(1)&--&426.7$\pm$84.5(1)&8.69$\pm$1.02(1)&--&1.02$\pm$0.15(6)\\
	HD\,111934 &2.57$_{-0.65}^{+0.36}$(1)&0.18&2.09$_{-0.67}^{+0.29}$(1)&0.23$\pm${0.08}(1)&--&354.8$\pm$124.4(1)&10.97$\pm$1.97(1)&--&--\\
	HD\,116852 &1.02$\pm${0.09}(1)&0.11&0.91$\pm${0.08}(1)&0.06$\pm${0.01}(1)&117.3$\pm$103.2(15)&537.5$\pm$133.2(1)&7.76$\pm$0.74(1)&2.57$\pm$0.39(4)&0.98$\pm$0.35(4)\\
	HD\,122879 &2.45$_{-0.23}^{+0.28}$(1)&0.17&2.04$_{-0.24}^{+0.28}$(1)&0.21$_{-0.03}^{+0.04}$(1)&324.3$\pm$38.2(15)&448.1$\pm$72.8(1)&6.44$\pm$0.76(1)&4.97$\pm$0.61(2)&0.53$\pm$0.10(6)\\
	HD\,124979 &2.34$\pm${0.38}(1)&0.21&1.86$\pm${0.39}(1)&0.25$_{-0.05}^{+0.03}$(1)&--&371.6$\pm$90.9(1)&8.53$\pm$1.39(1)&--&--\\
	HD\,144470 &1.74$\pm${0.300}(22)&0.13&1.51$\pm${0.31}(22)&0.11$\pm${0.02}(22)&--&414.3$\pm$95.5(24)&10.47$\pm$3.25(13)&--&0.32$\pm$0.26(13)\\
	HD\,147165 &2.51$\pm${0.510}(21)&0.05&2.40$\pm${0.48}(21)&0.06$\pm${0.01}(21)&--&371.5$\pm$121.7(21)&3.54$\pm$0.73(2)&1.96$\pm$0.41(2)&0.36$\pm$24.68(4)\\
	HD\,147888 &5.37$_{-2.10}^{+0.87}$(1)&0.11&4.79$_{-2.10}^{+0.88}$(1)&0.28$_{-0.05}^{+0.03}$(1)&105.4$\pm$22.6(14)&301.7$\pm$50.6(1)&1.86$\pm$0.33(1)&2.44$\pm$0.60(9)&0.15$\pm$0.08(6)\\
	HD\,148422 &2.04$_{-0.24}^{+0.38}$(1)&0.14&1.74$_{-0.24}^{+0.36}$(1)&0.14$_{-0.03}^{+0.04}$(1)&--&--&14.45$\pm$3.33(1)&--&--\\
	HD\,149757 &1.40$\pm${0.03}(2)&0.64&0.51$\pm${0.02}(3)&0.45$\pm${0.06}(3)&100.9$\pm$48.6(11)&307.1$\pm$29.1(8)&1.86$\pm$0.16(2)&1.50$\pm$0.04(2)&0.32$\pm$0.09(4)\\
	HD\,151805 &2.57$\pm${0.24}(1)&0.17&2.14$\pm${0.25}(1)&0.22$_{-0.05}^{+0.04}$(1)&--&447.4$\pm$92.0(1)&6.46$\pm$0.67(1)&--&--\\
	HD\,152236 &6.92$\pm${2.00}(6)&0.16&5.89(79)&0.54$\pm${0.17}(17)&--&2291.1$\pm$1034.8(19)&--&2.63$\pm$187.92(4)&0.18$\pm$0.07(6)\\
	HD\,152249 &2.82$\pm${0.39}(1)&0.14&2.40$\pm${0.39}(1)&0.19$\pm${0.04}(1)&--&436.4$\pm$100.5(1)&7.95$\pm$1.16(1)&4.17$\pm$14.20(4)&0.41$\pm$0.28(4)\\
	HD\,152424 &3.89$_{-0.54}^{+0.45}$(1)&0.24&3.02$_{-0.49}^{+0.42}$(1)&0.47$_{-0.12}^{+0.05}$(1)&--&416.4$\pm$90.5(1)&5.24$\pm$0.62(1)&--&--\\
	HD\,154368 &3.89$\pm${0.47}(18)&0.75&1.00$\pm${0.11}(16)&1.45$\pm${0.25}(16)&--&338.7$\pm$179.5(19)&2.67$\pm$0.43(2)&1.49$\pm$0.24(2)&0.38$\pm$0.16(28)\\
	HD\,157857 &2.75$\pm${0.48}(23)&0.36&1.82$\pm${0.42}(3)&0.49(29)&109.3$\pm$19.1(15)&467.7$\pm$92.4(3)&5.63$\pm$1.29(23)&--&--\\
	HD\,167264 &1.80$\pm${0.30}(24)&0.21&1.41$\pm${0.49}(27)&0.19(26)&--&722.2$\pm$252.7(24)&10.11$\pm$6.88(13)&--&0.56$\pm$0.09(13)\\
	HD\,167402 &1.58$\pm${0.15}(1)&0.19&1.35$_{-0.12}^{+0.16}$(1)&0.15$_{-0.02}^{+0.03}$(1)&--&--&13.82$\pm$2.29(1)&--&--\\
	HD\,168076 &5.37$\pm${3.10}(18)&0.18&4.47$\pm${3.10}(16)&0.48$\pm${0.10}(16)&--&1023.3$\pm$617.2(19)&--&--&0.38$\pm$0.24(28)\\
	HD\,168941 &1.78$_{-0.25}^{+0.16}$(1)&0.14&1.51$_{-0.24}^{+0.17}$(1)&0.13$\pm${0.02}(1)&--&389.1$\pm$80.1(1)&4.68$\pm$0.61(1)&--&1.07$\pm$0.21(6)\\
	HD\,170740 &2.57$_{-0.59}^{+0.36}$(1)&0.56&1.23$_{-0.48}^{+0.17}$(1)&0.72$_{-0.20}^{+0.13}$(1)&--&396.8$\pm$71.5(1)&3.16$\pm$0.53(1)&--&0.14$\pm$0.04(28)\\
	HD\,177989 &1.29$\pm${0.15}(1)&0.22&0.98$_{-0.14}^{+0.11}$(1)&0.15$\pm${0.04}(1)&--&436.3$\pm$71.0(1)&5.37$\pm$0.63(1)&--&0.50$\pm$0.09(6)\\
	HD\,178487 &2.29$_{-0.42}^{+0.21}$(1)&0.28&1.66$_{-0.38}^{+0.15}$(1)&0.32$\pm${0.07}(1)&--&--&4.07$\pm$0.53(1)&--&--\\
	HD\,179406 &2.75$\pm${0.71}(6)&0.39&1.70$\pm${0.70}(32)&0.54$\pm${0.09}(17)&--&213.8$\pm$63.5(19)&--&--&0.14$\pm$0.07(6)\\
	HD\,179407 &1.91$_{-0.35}^{+0.22}$(1)&0.18&1.58$_{-0.36}^{+0.22}$(1)&0.17$_{-0.04}^{+0.35}$(1)&--&346.9$\pm$126.2(1)&5.01$\pm$0.99(1)&--&--\\
	HD\,185418 &2.63$_{-0.18}^{+0.24}$(1)&0.4&1.55$_{-0.14}^{+0.18}$(1)&0.53$_{-0.07}^{+0.09}$(1)&167.7$\pm$22.8(15)&380.2$\pm$43.8(1)&4.07$\pm$0.42(1)&0.06$\pm$0.01(12)&0.32$\pm$0.09(12)\\
	HD\,192639 &3.09$\pm${0.71}(18)&0.35&1.95$\pm${0.45}(3)&0.54$\pm${0.14}(3)&125.2$\pm$29.4(15)&446.6$\pm$110.9(3)&5.13$\pm$1.23(7)&--&0.32$\pm$0.11(28)\\
	HD\,197512 &2.75$\pm${0.79}(18)&0.33&1.82$\pm${0.75}(31)&0.46$\pm${0.06}(16)&--&158.5$\pm$1888.7(19)&--&--&0.30$\pm$0.15(28)\\
	HD\,198478 &3.39$_{-0.94}^{+1.01}$(1)&0.34&2.09$_{-0.63}^{+0.58}$(1)&0.58$_{-0.46}^{+0.44}$(1)&--&457.4$\pm$179.3(1)&3.98$\pm$1.28(1)&--&--\\
	HD\,198781 &1.48$\pm${0.17}(1)&0.41&0.85$_{-0.06}^{+0.14}$(1)&0.30$_{-0.07}^{+0.06}$(1)&135.9$\pm$19.8(15)&501.0$\pm$73.9(1)&3.72$\pm$0.44(1)&--&--\\
	HD\,199579 &1.78$\pm${0.31}(18)&0.38&1.10(79)&0.34$\pm${0.03}(16)&--&46.8$\pm$22.4(19)&--&1.62$\pm$0.34(4)&0.30$\pm$0.08(28)\\
	HD\,203532 &2.75$\pm${0.48}(23)&0.32&1.86$\pm${0.43}(3)&0.44$\pm${0.09}(3)&82.4$\pm$28.3(15)&257.0$\pm$46.4(3)&1.38$\pm$0.25(23)&--&--\\
	HD\,206267 &3.09$_{-0.29}^{+0.36}$(1)&0.46&1.66$_{-0.15}^{+0.23}$(1)&0.71$_{-0.11}^{+0.13}$(1)&--&407.7$\pm$54.7(1)&3.56$\pm$0.44(1)&4.08$\pm$0.68(2)&0.22$\pm$0.06(28)\\
	HD\,206773 &1.74$_{-0.12}^{+0.20}$(1)&0.3&1.23$_{-0.09}^{+0.20}$(1)&0.26$_{-0.05}^{+0.04}$(1)&426.4$\pm$50.2(15)&446.5$\pm$59.9(1)&5.50$\pm$0.65(1)&--&--\\
	HD\,207198 &3.16$_{-0.44}^{+0.36}$(1)&0.39&1.91$_{-0.40}^{+0.31}$(1)&0.62$\pm${0.07}(1)&102.1$\pm$25.1(14)&445.9$\pm$59.9(1)&3.64$\pm$0.43(1)&2.19$\pm$1.07(9)&0.43$\pm$0.08(9)\\
	HD\,209339 &1.86$\pm${0.17}(1)&0.15&1.58$\pm${0.15}(1)&0.14$_{-0.03}^{+0.02}$(1)&--&355.0$\pm$52.3(1)&6.93$\pm$0.80(1)&5.06$\pm$0.88(2)&0.48$\pm$0.06(6)\\
	HD\,210121 &1.55$\pm${0.40}(6)&0.93&0.43$\pm${0.18}(29)&0.72(29)&--&911.6$\pm$526.8(19)&--&--&1.74$\pm$1.19(6)\\
	HD\,210809 &2.24$_{-0.26}^{+0.31}$(1)&0.08&2.04$_{-0.24}^{+0.28}$(1)&0.09$_{-0.02}^{+0.03}$(1)&295.3$\pm$48.6(15)&339.0$\pm$78.0(1)&10.94$\pm$1.69(1)&--&--\\
	HD\,210839 &3.02$\pm${0.28}(1)&0.42&1.74$\pm${0.20}(1)&0.63$\pm${0.07}(1)&--&490.1$\pm$56.4(1)&3.64$\pm$0.35(1)&3.48$\pm$0.40(2)&0.32$\pm$0.06(28)\\
	HD\,220057 &1.29$_{-0.45}^{+0.29}$(1)&0.29&0.89$_{-0.47}^{+0.29}$(1)&0.19$_{-0.03}^{+0.04}$(1)&--&446.3$\pm$114.9(1)&4.90$\pm$1.22(1)&--&--\\
	HD\,232522 &1.62$\pm${0.15}(1)&0.19&1.32$_{-0.15}^{+0.12}$(1)&0.15$_{-0.04}^{+0.03}$(1)&296.6$\pm$36.4(15)&647.4$\pm$202.2(1)&8.69$\pm$0.83(1)&--&--\\
	HD\,303308 &2.88$_{-0.47}^{+0.20}$(1)&0.11&2.57$_{-0.47}^{+0.18}$(1)&0.16$_{-0.03}^{+0.02}$(1)&--&423.0$\pm$35.9(20)&8.50$\pm$0.71(1)&7.08$\pm$658.80(4)&1.00$\pm$0.14(6)
	\enddata 
	\\
\flushleft{
(1) Jenkins (2019);
(2) Gnaci\'nski \& Krogulec (2006);
(3) Cartledge et al.\ (2004);
(4) van Steenberg \& Shull (1988);
(5) Welty \& Crowther (2010);
(6) Jensen \& Snow (2007a);
(7) Jensen \& Snow (2007b);
(8) Knauth et al.\ (2006);
(9) Miller et al.\ (2007);
(10) Sheffer et al.\ (2007);
(11) Sofia et al.\ (1994); 
(12) Sonnentrucker et al.\ (2003);
(13) Jenkins et al.\ (1986);
(14) Sofia et al.\ (2011);
(15) Parvathi et al.\ (2012);
(16) Rachford et al.\ (2002);
(17) Rachford et al.\ (2009);
(18) Jensen et al.\ (2007);
(19) Jensen et al.\ (2005);
(20) Andr\'e et al.\ (2003);
(21) Cartledge et al.\ (2008); 
(22) Cartledge et al.\ (2003);
(23) Cartledge et al.\ (2006);
(24) Meyer et al.\ (1998);
(25) Crinklaw et al.\ (1994);
(26) Federman et al.\ (1994); 
(27) Diplas \& Savage (1994);
(28) Snow et al.\ (2002);
(29) Sheffer et al.\ (2008);
(30) Welty \& Crowther (2010);
(31) Fitzpatrick \& Massa (1990); 
(32) Hanson et al.\ (1992);
(33) Cui et al.\ (2005);
(34) Sofia et al.\ (2004).
}
\end{deluxetable}


\begin{deluxetable}{lcccccccc}
	\tablecolumns{9}
	\tabletypesize{\scriptsize}
	\tablewidth{0truein}
	\center
	\tablecaption{Model Parameters for Fitting
          the UV/Optical/Near-IR Extinction
          with a Mixture of Silicate and Graphite Grains
	  \label{tab:modpara}
	  }
	\tablehead{
		\colhead{Star}&
		\colhead{$\AV/\NH$}&
		\colhead{$\alphaS$}&
		\colhead{$\acS$}&
		\colhead{$\alphaC$}&
		\colhead{$\acC$}&
		\colhead{$\chi^{2}$}&
		\colhead{$\cdust$}&
		\colhead{$\sidust$}
		\\
		\colhead{}&
		\colhead{($10^{-22}\magni\cm^2\HH^{-1}$)}&
		
		\colhead{}&
		\colhead{($\mu$m)}&
		\colhead{}&
		\colhead{($\mu$m)}&
		\colhead{}&
		
		\colhead{(ppm)}&
		\colhead{(ppm)}
	}
	\startdata
	BD+35\,4258&3.68&2.4&0.11&3.7&0.25&0.002&136&35\\
	BD+53\,2820&3.5&2.9&0.17&3.7&0.28&0.003&151&29\\
	HD\,12323&3.31&3.15&0.18&3.9&0.34&0.001&134&37\\
	HD\,13268&3.96&3.4&0.29&3.55&0.19&0.002&185&36\\
	HD\,14434&4.17&3.1&0.14&3.65&0.22&0.001&181&45\\
	HD\,25443&3.72&3.24&0.22&4.02&0.14&0.001&154&45\\
	HD\,37367&5.8&2.4&0.14&3.89&0.4&0.002&175&58\\
	HD\,38087&8.72&1.26&0.1&3.63&0.54&0.001&244&85\\
	HD\,40893&3.64&2.8&0.15&3.85&0.38&0.001&123&41\\
	HD\,43818&3.51&2.88&0.18&3.75&0.29&0.002&127&34\\
	HD\,46223&3.8&3.1&0.17&3.75&0.25&0.001&154&41\\
	HD\,52266&4.89&2.82&0.16&3.88&0.44&0.001&159&55\\
	HD\,62542&4.63&3.54&0.12&2.18&0.06&0.001&237&41\\
	HD\,72648&5.1&2.25&0.14&3.88&0.24&0.001&135&56\\
	HD\,73882&6.32&2.8&0.16&2.92&0.15&0.001&213&62\\
	HD\,75309&6.59&2.7&0.17&3.83&0.35&0.002&201&70\\
	HD\,79186&4.87&3&0.18&3.52&0.25&0.001&202&45\\
	HD\,89137&5.59&2.64&0.12&3.5&0.21&0.001&196&59\\
	HD\,91983&5.06&2.78&0.13&3.36&0.14&0.001&223&42\\
	HD\,93129&5.28&2.6&0.16&3.3&0.26&0.001&137&60\\
	HD\,93222&5.53&2.38&0.19&3.4&0.1&0.001&45&79\\
	HD\,94493&5.42&2.4&0.13&3.6&0.45&0.001&162&57\\
	HD\,99953&4.87&2.52&0.15&3.75&0.54&0.001&154&51\\
	HD\,103779&4.25&2.78&0.17&3.25&0.1&0.003&163&34\\
	HD\,104705&4.01&2.84&0.15&3.83&0.29&0.001&162&39\\
	HD\,111934&4.86&2.8&0.15&3.9&0.28&0.002&191&51\\
	HD\,124979&4.48&3.2&0.21&3.73&0.3&0.002&213&40\\
	HD\,144470&4.26&2.2&0.14&3.9&0.2&0.002&129&43\\
	HD\,148422&4.31&3.35&0.26&3.78&0.35&0.001&165&52\\
	HD\,151805&5.52&2.61&0.16&3.8&0.34&0.001&153&63\\
	HD\,152249&5.78&2.54&0.15&3.8&0.49&0.001&170&62\\
	HD\,152424&5.73&2.32&0.12&3.72&0.39&0.001&191&56\\
	HD\,167264&5.44&2.05&0.1&3.85&0.48&0.003&183&49\\
	HD\,167402&4.48&2.83&0.16&3.6&0.42&0.002&170&44\\
	HD\,168076&4.92&2.55&0.13&3.6&0.5&0.001&185&48\\
	HD\,168941&4.5&2.48&0.12&3.6&0.3&0.003&208&31\\
	HD\,170740&5.87&2.75&0.13&3.66&0.45&0.001&230&60\\
	HD\,177989&5.05&2.9&0.13&3.7&0.32&0.001&223&49\\
	HD\,178487&4.54&2.83&0.14&3.95&0.5&0.003&210&39\\
	HD\,179406&3.49&2.42&0.1&3.7&0.32&0.001&150&31\\
	HD\,179407&3.94&2.98&0.15&3.93&0.3&0.001&147&47\\
	HD\,197512&3.41&2.88&0.13&3.78&0.33&0.001&152&33\\
	HD\,198478&4.37&2.66&0.1&3.66&0.3&0.001&200&40\\
	HD\,210121&4.84&3.55&0.17&2.8&0.06&0.002&215&60\\
	HD\,220057&4.81&2.7&0.11&3.8&0.32&0.002&215&43

\enddata
\end{deluxetable}

\end{document}